\documentclass[letterpaper,onecolumn,onecolappendix,appendixfloats]{emulateapj}
\usepackage{float}
\usepackage{hyperref}
\usepackage{epsfig}
\usepackage{graphicx}
\usepackage{graphics}
\usepackage{csquotes}
\usepackage{latexsym}
\usepackage{amsmath}
\newcommand{\nd}{$\dots$}
\newcommand{\kms}{\rm km \ s^{-1}}
\newcommand{\kmsmpc}{\rm km \ s^{-1} \ Mpc^{-1}} 
\newcommand{\msun}{\rm M$_{\odot}$}
\shorttitle{SN 2017cbv}
\shortauthors{Wang et al.}

\begin{document}

\title{Optical and Near-Infrared Observations of the Nearby Type Ia Supernova 2017\lowercase {cbv}}
%\author{Lingzhi Wang\altaffilmark{1,2}, et al.}
\author{Lingzhi Wang}
\affiliation{CAS Key Laboratory of Optical Astronomy, National Astronomical Observatories, Chinese Academy of Sciences, Beijing 100101, China wanglingzhi@bao.ac.cn} 
\affiliation{Chinese Academy of Sciences South America Center for Astronomy, National Astronomical Observatories, CAS, Beijing 100101, China}

\author{Carlos Contreras}
\affiliation{Carnegie Observatories, Las Campanas Observatory, Colina El Pino, Casilla 601, Chile}

\author{Maokai Hu}
\affiliation{Purple Mountain Observatory, Nanjing, 201008, Jiangsu, People’s Republic of China}

\author{Mario A. Hamuy}
\affiliation{Departamento de Astronom\'ia, Universidad de Chile, Camino El Observatorio 1515, Las Condes, Santiago, Chile}
\affiliation{Millennium Institute of Astrophysics, Santiago, Chile}

\author{Eric Y. Hsiao }
\affiliation{Department of Physics, Florida State University, Tallahassee, FL 32306, USA}

\author{David J. Sand}
\affiliation{Department of Astronomy and Steward Observatory, University of Arizona, 933 North Cherry Avenue, Tucson, AZ 85721-0065, USA}

\author{Joseph P. Anderson}
\affiliation{European Southern Observatory, Alonso de C\'ordova 3107, Casilla 19, Santiago, Chile}

\author{Chris Ashall}
\affiliation{Department of Physics, Florida State University, Tallahassee, FL 32306, USA}

\author{Christopher R. Burns}
%\affiliation{}
%\altaffiliation{Carnegie Fellow}
\affiliation{Observatories of the Carnegie Institution for Science, 813 Santa Barbara St., Pasadena, CA 91101, USA}

\author{Juncheng Chen}
\affiliation{Chinese Academy of Sciences South America Center for Astronomy, National Astronomical Observatories, CAS, Beijing 100101, China}
\affiliation{School of Electronics and Information Engineering, Wuzhou University, Wuzhou 543002, China}

\author{Tiara R. Diamond}
\affiliation{ }
%\affiliation{Laboratory of Observational Cosmology, Code 665, NASA Goddard Space Flight Center, Greenbelt, MD 20771, USA}

\author{Scott Davis}
\affiliation{Department of Physics, Florida State University, Tallahassee, FL 32306, USA}

\author{Francisco F\"orster}
\affiliation{Departamento de Astronom\'ia, Universidad de Chile, Camino El Observatorio 1515, Las Condes, Santiago, Chile}
\affiliation{Millennium Institute of Astrophysics, Santiago, Chile}

\author{Llu\'is Galbany}
\affiliation{Departamento de F\'isica Te\'orica y del Cosmos, Universidad de Granada, E-18071 Granada, Spain}

\author{Santiago Gonz\'alez-Gait\'an}
\affiliation{CENTRA/COSTAR, Instituto Superior T\'ecnico, Universidade de Lisboa, Av. Rovisco Pais 1, 1049-001 Lisboa, Portugal}

\author{Mariusz Gromadzki}
\affiliation{Astronomical Observatory, University of Warsaw, Al. Ujazdowskie 4, 00-478 Warszawa, Poland}

\author{Peter Hoeflich}
\affiliation{Department of Physics, Florida State University, Tallahassee, FL 32306, USA}

\author{Wenxiong Li}
\affiliation{Physics Department/Tsinghua Center for Astrophysics, Tsinghua University, Beijing, 100084, China}

\author{G. H. Marion}
\affiliation{Department of Astronomy, The University of Texas at Austin, 1 University Station C1400, Austin, TX 78712-0259, USA}

\author{Nidia Morrell}
\affiliation{Carnegie Observatories, Las Campanas Observatory, Colina El Pino, Casilla 601, Chile}

\author{Giuliano Pignata}
\affiliation{Departamento de Ciencias F\'isicas, Universidad Andres Bello, Avda. Rep\'ublica 252, Santiago, 8320000, Chile}
\affiliation{Millennium Institute of Astrophysics, Santiago, Chile}

\author{Jose L. Prieto}
\affiliation{N\'ucleo de Astronom\'ia de la Facultad de Ingenier\'ia, Universidad Diego Portales, Av. Ej\'ercito 441, Santiago, Chile}
\affiliation{Millennium Institute of Astrophysics, Santiago, Chile}

\author{Mark M. Phillips}
\affiliation{Carnegie Observatories, Las Campanas Observatory, Colina El Pino, Casilla 601, Chile}

%\author{Alessandro Razza}
%\affiliation{Departamento de Astronom\'ia, Universidad de Chile, Camino El Observatorio 1515, Las Condes, Santiago, Chile}
%\affiliation{European Southern Observatory, Alonso de C\'ordova 3107, Casilla 19, Santiago, Chile}

\author{Melissa Shahbandeh}
\affiliation{Department of Physics, Florida State University, Tallahassee, FL 32306, USA}

\author{Nicholas Suntzeff}
\affiliation{George P. and Cynthia Woods Mitchell Institute for Fundamental Physics and Astronomy, Department of Physics and Astronomy,Texas A\&M University, College Station, TX, 77843, USA}

\author{Stefano Valenti}
\affiliation{Department of Physics, University of California, 1 Shields Avenue, Davis, CA 95616-5270, USA}

\author{Lifan Wang}
\affiliation{George P. and Cynthia Woods Mitchell Institute for Fundamental Physics and Astronomy, Department of Physics and Astronomy,Texas A\&M University, College Station, TX, 77843, USA}

\author{Xiaofeng Wang}
\affiliation{Physics Department/Tsinghua Center for Astrophysics, Tsinghua University, Beijing, 100084, China}

\author{D. R. Young}
\affiliation{Astrophysics Research Centre, School of Mathematics and Physics, Queens University Belfast, Belfast BT7 1NN, UK}

\author{Lixin Yu}
\affiliation{CAS Key Laboratory of Optical Astronomy, National Astronomical Observatories, Chinese Academy of Sciences，Beijing 100101, China}
\affiliation{Chinese Academy of Sciences South America Center for Astronomy, National Astronomical Observatories, CAS, Beijing 100101, China}

\author{Jujia Zhang}
\affiliation{Yunnan Observatories, Chinese Academy of Sciences, Kunming 650216, China}
\affiliation{Key Laboratory for the Structure and Evolution of Celestial Objects, Chinese Academy of Sciences, Kunming 650216, China}

\begin{abstract}
  Supernova (SN) 2017cbv in NGC 5643 is one of a handful of type Ia supernovae (SNe~Ia) reported to have excess blue emission at early times. This paper presents extensive $BVRIYJHK_s$-band light curves of SN~2017cbv, covering the phase from $-16$ to $+125$ days relative to $B$-band maximum light. SN~2017cbv reached a $B$-band maximum of 11.710$\pm$0.006~mag, with a post-maximum magnitude decline $\Delta m_{15}(B)$=0.990$\pm$0.013 mag. The supernova suffered no host reddening based on Phillips intrinsic color, Lira-Phillips relation, and the CMAGIC diagram. By employing the CMAGIC distance modulus $\mu=30.58\pm0.05$~mag and assuming $H_0=72~\kmsmpc$, we found that 0.73~\msun $^{56}$Ni was synthesized during the explosion of SN~2017cbv, which is consistent with estimates using reddening-free and distance-free methods via the phases of the secondary maximum of the NIR-band light curves. We also present 14 near-infrared spectra from $-18$ to $+49$~days relative to the $B$-band maximum light, providing constraints on the amount of swept-up hydrogen from the companion star in the context of the single degenerate progenitor scenario. No $Pa{\beta}$ emission feature was detected from our post-maximum NIR spectra, placing a hydrogen mass upper limit of 0.1 $M_{\odot}$. The overall optical/NIR photometric and NIR spectral evolution of SN~2017cbv is similar to that of a normal SN~Ia, even though its early evolution is marked by a flux excess no seen in most other well-observed normal SNe~Ia. We also compare the exquisite light curves of SN~2017cbv with some $M_{ch}$ DDT models and sub-$M_{ch}$ double detonation models.
  
\end{abstract}

\keywords{supernovae: general -- supernovae: individual: SN~2017cbv}

\clearpage

%%%%%%%%%%%%%%%%%%%%%%%%%%%%%%%%%%%%%%%%%%%%%%%%%%%%%%%%%%%%%%%
%Three methods (NIR-absolute calibration, SNooPy fit, and CMAGIC diagram) are used to estimate the distance modulus $\mu=30.42\pm0.16$~mag, which is consistent with an external distance measurement of SN~2013aa which exploded in the same host galaxy of SN~2017cbv.

\section{Introduction}
Type Ia supernovae (SNe~Ia) have served as cosmological distance indicators for the past three decades and led to the discovery of the accelerating expansion of the Universe \citep{Riess98,Perlmutter99}. After corrections for the light-curve/color parameters \citep[i.e., $\Delta m_{15}; $][]{Phillips93,Riess96,Tripp98,Wang05,Guy05}, the magnitude dispersion on the Hubble diagram of SNe~Ia can be brought down to below 0.1 mag rms \citep[e.g.,][]{Wang05,Wang09a,Burns18,He18}. Increasing evidence suggests that NIR light curves of SNe~Ia are better standard candles \citep{Phillips12,Avelino19}, and are intrinsically less affected by dust extinction from the host galaxy \citep{Meikle00,Krisciunas04a,Krisciunas04c,Krisciunas07}. 
The scatter in the NIR Hubble diagram can reach 0.15 mag without applying any light-curve/color parameter corrections \citep{Krisciunas04a,WoodVasey08,Folatelli10}.

There is a general consensus that SNe~Ia are the thermonuclear explosion of carbon-oxygen white dwarfs, and many of them seem to explode near the Chandrasekhar mass \citep[$M_{ch}$; ][]{Hillebrandt00}, although they may originate from progenitors of other masses as well \citep{Scalzo14b}. There are two popular progenitor 
scenarios: the single degenerate (SD) and the double degenerate (DD), see recent reviews by \citet{wang12}, \citet{Maoz14} and \citet{Jha19}. In the SD model, a carbon-oxygen white dwarf accretes material from a non-degenerate companion star such as a red giant, subgiant, main-sequence, or a helium  star \citep{Whelan73,Livne90,Woosley94,Nomoto97}, while in the latter, the system comprises two white dwarfs \citep{Webbink84,Iben84}.

A very clear sign of the SD model lies in the very early light curves, when a \enquote{blue bump} may appear in the NUV-optical bands as a result of the collision of SN ejecta with the non-degenerate companion \citep{Kasen10,Marion16}. Excess emission in the early light curve has been reported in a number of SNe~Ia \citep{Cao15,Marion16,Hosseinzadeh17,Jiang18,Dimitriadis19a,Shappee19,Li19b}. Alternatively, these early light curve features may be associated with mixing of radioactive $^{56}$Ni \citep{Piro16,Miller18,Magee18,Jiang20,Magee20a,Magee20b}, He-shell detonation \citep{Jiang17,Maeda18,Polin19,Siebert20}, or circumstellar material interaction in DD scenario \citep{Levanon17,Levanon19}. Meanwhile, other studies have searched for narrow H$\alpha$/He emission lines in nebular phase spectra as a characteristic  signature of the SD scenario \citep{Marietta00,Mattila05,Pan10,Leonard07,Pan12,Liu12,Liu13,Lundqvist13,Lundqvist15,Shappee18,Graham15,Maguire16,Botyanszki18,Sand18,Sand19,Shappee18,Dimitriadis19b,Holmbo19,Tucker19b}. However, no definitive late-time narrow emission features of hydrogen (i.e., H$_{\alpha}$) have been detected among current samples of normal SNe~Ia. Recent observations for fast-declining, sub-luminous SNe~Ia have detected narrow H$_{\alpha}$ emission in two cases, SNe~2018fhw \citep{Kollmeier19} and 2018cqj \citep{Prieto20}. The H$_{\alpha}$ luminosity could be due to stripped hydrogen from a nondegenerate companion in the SD progenitor scenario \citep{Kollmeier19,Prieto20}. Or it could originate from ejecta-CSM interaction \citep{Kollmeier19,Vallely19,Dessart20}, a scenario similar to luminous Type Ia-CSM objects \citep{Hamuy03,Wang_2004,Prieto07,Aldering_2006,Dilday12,Silverman13b,Graham19}.

Additionally, recent observational and theoretical studies of SNe Ia have shown that the NIR spectra have several key physical diagnostics capable of discriminating potential progenitor systems and explosion mechanisms \citep{Hsiao19,Ashall19a,Ashall19b}. Unburned carbon C I $\lambda$1.069~$\mu$m can be used to probe the primordial material directly from the progenitor \citep{Hsiao13,Hsiao15}. Its abundance and distribution in the ejecta also provides strong constraints on explosion models. For instance, the turbulent deflagration and pure deflagration models predict a large amount of unburned carbon \citep{Gamezo03,Kozma05}. In contrast, delayed detonation models predict nearly complete carbon burning \citep{Kasen09}. On the other hand, substantial unburned carbon is not expected to survive in the explosions of sub-Chandrasekhar mass white dwarfs through the double detonation mechanism \citep{Fink10}. Furthermore, narrow P$\beta$ $\lambda$1.282~$\mu$m emission is expected in the single degenerate scenario (for a red giant companion) 1--2 months after maximum light \citep{Maeda14}.  %the companion signature P$\alpha {\beta}$ $\lambda$1.282~$\mu$m from red-giant model (RGa case) is expected to be detectable in the 1-2 months after maximum light \citep{Maeda14}. 
Searching for such emission has only been undertaken in a handful of objects but is another promising signature of the single degenerate scenario \citep[e.g.][]{Sand16}.
%Those desirable properties could help diagnose the possible detection of embedded hydrogen stripped from the companion star \citep{Sand16}.

SN~2017cbv (DLT17u) gained much attention as it shows a very clear \enquote{blue bump} as reported by \citet{Hosseinzadeh17}. Neither narrow emission lines H/He have been detected in the nebular phase spectra \citep{Sand18}, nor time variable narrow line features of Na I D and Ca II H\&K have been detected within high resolution spectra \citep{Ferretti17}. These observational signatures, together with the fact that SN~2017cbv was discovered so young, make it an interesting target to study with respect to its progenitor system and explosion physics. 

Here we present extensive optical and NIR observations of SN~2017cbv, including $BVRIYJHK_s$-band photometry lasting 140 days, using the same instrument, and 14 near-infrared spectra. In Section 2 we describe the observational data and data analyses of SN~2017cbv. In section 3 we present the physical properties of SN~2017cbv from our well-sampled light curves, including our light/color curves, color magnitude diagrams, host reddening and its distance determination, and its bolometric light curves. In section 4 we present theoretical perspectives of SN~2017cbv. We summarize our results in Section 5.

%%%%%%%%%%%%%%%%%%%%%%%%%%%%%%%Data%%%%%%%%%%%%%%%%%%%%%%%%%%%%
\section{Data and Data analyses}
Our data includes optical/NIR photometry from $-16$ to $+125$~days, and 14 NIR spectra from $-18$ to $+49$~days after $B$-band maximum, making SN~2017cbv one of the earliest NIR ever taken for SNe Ia. 
SN 2017cbv was discovered on 2017 March 10.14 (UT dates are used throughout this paper) by the $D<$40 Mpc survey \citep[DLT40; ][]{Tartaglia18} and a confirmation 
image was obtained with the same telescope 30 minutes later \citep{Tartaglia17}. It has coordinates 
$\alpha = 14^h32^m34.42^s$ and $\delta = -44^{\circ}08'02.8''$(J2000.0). It is located 68 arcsec west and 145 arcsec north of the center of galaxy NGC 5643, which has a SAB(rs)c morphology \citep{deVaucouleurs91} and a Tully-Fisher distance modulus of 31.14$\pm$0.40 mag \citep{Tully88}. SN~Ia 2013aa also exploded in this galaxy \citep{Parker13,Parrent13} and a comprehensive comparison between SN~2017cbv and SN2013aa is described in \citet{Burns20}. A spectrum of SN~2017cbv was acquired shortly after its discovery, showing it 
to be a very young SN Ia with high-velocity features, similar to that of SN~1999aa at t $>$ 2 weeks before maximum light \citep{Hosseinzadeh17a,Hosseinzadeh17b}. 

Our photometric observations of SN~2017cbv started on 2017 March 13.17, $\sim 16$ days before $B-$band maximum. Data were collected in $BVRIYJHK_s$ bands with the CTIO 1.3 m telescope and dual-channel optical/near-infrared (NIR) camera ANDICam. This instrument has an optical FOV of $6.3'\times 6.3'$ ($0.37''$ pixel$^{-1}$) and a NIR FOV of $2.4'\times 2.4'$ ($0.27''$ pixel$^{-1}$).
%gfwhm=158.8   158.8/2.355/635
\begin{figure}[h]
\begin{center}
\includegraphics[trim=7cm 1cm 5cm 0cm, clip=true,width=0.85\textwidth]{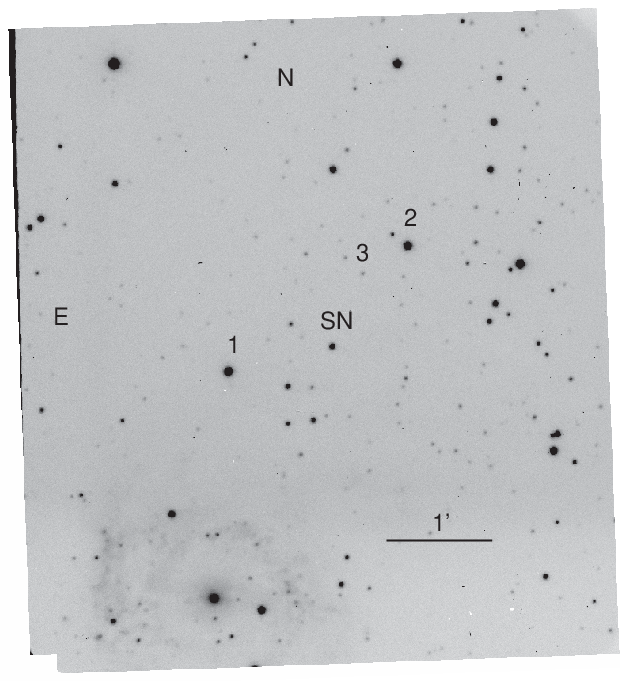}
\caption{SN~2017cbv in NGC~5643. This is a $B$-band image taken with the CTIO 1.3m on 2017 July 31.99. The supernova 2017cbv and 3 local reference stars (Tables~\ref{tab:localbvri} and~\ref{tab:localyjhk}) for $BVRIYJHK_s$ calibration are marked by numbers. Star \# 2 was used for optical calibration and \#1 was used for NIR calibration. \label{fig:chart}}
\end{center}
\end{figure}

The first NIR spectrum of SN~2017cbv was taken with FLAMINGOS-2 \citep{Eikenberry06} on Gemini South 8.2-m telescope, at only 2.30 days past the explosion, representing one of the earliest NIR ever taken for a SNe Ia. Similar early-phase NIR spectra were only obtained for SN 2011fe \citep{Hsiao13} and iPTF13ebh \citep{Hsiao15}.
 Five more NIR spectra were also obtained with FLAMINGOS-2, covering the phase from $-18$ days to $+36$ days from $B$-band maximum light. Two other spectra were provided by ePESSTO \citep{Smartt15} with the SOFI instrument \citep{Moorwood98} on the NTT. Five  NIR spectra were also acquired with FIRE \citep{Simcoe13} on the Magellan Baade telescope. One more spectrum was taken with the SpeX spectrograph \citep{Rayner03} on the NASA Infrared Telescope Facility (IRTF). A journal of the NIR spectroscopic observations is shown in  Table~\ref{tab:logspec}.
 
 \begin{deluxetable}{ccrrrr}[h]
%\rotate
%\tablewidth{0pc}
\tablewidth{0pt}
%\tablenum{5}
\tablecaption{ Log of the NIR spectroscopic observations. \label{tab:logspec}}
\tablehead{\colhead{UT Date }& \colhead{MJD} & \colhead{$t_{B}^{max}$} & \colhead{Instrument} &  \colhead{$t_{exp}$} & \colhead{Airmass}  \\
           &           &          &       &   Sec  &  }
\startdata
2017-03-11 & 57823.30 & $-$17.57 & FLAMINGOS-2 & 8$\times$150 & 1.03 \\
2017-03-13 & 57825.30 & $-$15.57 & FLAMINGOS-2 & 8$\times$150 & 1.04\\
2017-03-17 & 57829.40 & $-$11.47 & FLAMINGOS-2 & 8$\times$60 & 1.03 \\
2017-03-22 & 57834.30 &  $-$6.57 & FLAMINGOS-2 & 8$\times$20 & 1.08 \\
2017-03-26 & 57838.25 &  $-$2.62 & FIRE  & 8$\times$126.8 & 1.06 \\ %15.9
2017-04-02 & 57845.30 &   4.43 & FLAMINGOS-2 & 8$\times$15 & 1.06 \\
2017-04-14 & 57857.14 &  16.27 & FIRE  & 8$\times$126.8 & 1.16 \\ %16.9
2017-04-21 & 57864.27 &  23.40 & FIRE  & 4$\times$126.8 & 1.08\\ %8.5
2017-04-23 & 57866.20 &  25.33 & SOFI  & 4$\times$50 & 1.04 \\
2017-05-02 & 57875.27 &  34.40 & FIRE  & 4$\times$126.8 & 1.14 \\  %8.5
2017-05-04 & 57877.20 &  36.33 & FLAMINGOS-2 & 8$\times$25 & 1.05 \\
2017-05-04 & 57877.37 &  36.50 & SpeX  & 10$\times$150   &  2.47    \\
2017-05-17 & 57890.22 &  49.35 & FIRE  & 4$\times$126.8 & 1.12\\  %8.5
2017-05-17 & 57890.24 &  49.37 & SOFI  & 4$\times$50 & 1.16 

\enddata
\tablecomments{$t_B^{max}$: Relative to the epoch of $B$-band maximum, $t_B^{max}$ = 57840.87 MJD from Gaussian process regression \citep{Rasmussen06,scikit11}.}
\end{deluxetable}

\subsection{Optical Photometry}
%\subsubsection{Observations}
\subsubsection{Data Reduction and Astrometry}
The raw optical images were collected by the Yale SMARTS team and reduced through a data pipeline via the NOAO IRAF package\footnote{http://ast.noao.edu/data/software}. The reduction included the subtraction of an overscan region, a zero frame and the division by a normalized dome flat\footnote{http://www.astro.yale.edu/smarts/ANDICAM/data.html}. The reduced images were downloaded from the SMARTS ftp site. Then cosmic rays were detected and removed using the Laplacian Cosmic Ray Identification \citep{vanDokkum01}\footnote{\url{http://www.astro.yale.edu/dokkum/lacosmic/}}. The astrometric calibration for the optical images were carried out using Astrometry.net \citep{Lang10}.

\subsubsection{Differential Photometry}
As SN~2017cbv is located far away ($160''$) from the center of its host galaxy, light contamination from the host is negligible (see Figure~\ref{fig:chart}). Aperture and point-spread-function (PSF) photometry were performed on the optical 
images of SN~2017cbv.

The photometry of the stars in the field was not completely consistent when comparing Carnegie Supernova Project \citep[CSP; ][]{Burns20} and ANDICam images. We found a uncorrected illumination pattern in the ANDICam images, which is more evident when combining a large number of images. Unfortunately there are no available flat field images to make a proper correction, so instead we opted to use only stars in the neighborhood of the SN to produce relative photometry. In detail, we used 2 bright and isolated stars in the close neighbourhood of SN~2017cbv to measure differential photometry (we did not detect time-series variability for the two selected stars during our observing window). The Carnegie Supernova Project also observed the field in $ugriBV$ bands \citep{Burns20} and provided us their calibration, in order to derive the $BVRI$ values. We measured differential photometry using PSF and aperture photometry relative to star \#2 \citep[using PSFEx and SEP tools respectively; ][]{Bertin96,Bertin11,Barbary18} as star \#1 was saturated in all CSP $ri$-band images, but not saturated on our ANDICam images. We used the PSF photometry as it was also used in the NIR images. We took the aperture photometry as a sanity check, showing a 0.02~mag systematic deviation, which was added to the final error budget as a systematic error.
 %We found a 0.02 magnitudes systematic difference between aperture and PSF methods. So we averaged both and added 0.01 magnitudes to the total error bar for each individual data point because of this effect.

\subsubsection{Calibration of the Local Sequence Stars}
Due to the illumination problem (see above), we did not calibrate the stars in the SN field, instead we used calibrations from CSP photometry (Burns et al. 2020). The $BV$-band calibrations of ANDICam in the natural system were transformed from CSP $BV$ bands via its color term coefficients of $B-V$ in Table~\ref{tab:cal}. The ANDICam $R$ and $I$-band calibrations in the natural system were derived in a similar way and the standard magnitudes of CSP $RI$ bands were measured from the average values of $ri$ bands following the SDSS transformation equations \citep{Jester05,Lupton05,Jordi06}. The $BVRI$ magnitudes of the star \#2 are listed in Table~\ref{tab:localbvri}.

\subsubsection{Optical Color Terms}
In spite of having a calibrated set of local sequence stars, we still need
color terms to transform instrumental photometry into the standard system.
The color terms of ANDICam were determined from images
of the standard field Rubin 149, following the equations:

\begin{align}
  B &= b + Z_B - k_B\times X+C_B\times (B-V)\\
  V &= v + Z_V - k_V\times X+C_V\times (B-V)\\
  R &= r + Z_R - k_R\times X+C_R\times (V-R)\\
  I &= i + Z_I - k_I\times X+C_I\times (V-I)
\end{align}

\noindent where $bvri$ are the instrumental magnitudes, $BVRI$ are the standard magnitudes,
$Z_B, Z_V, Z_R, Z_I$ are the zero point magnitudes, $k_B$, $k_V$, $k_R$, $k_I$ are
the extinction coefficients, $X$ is the airmass, and $C_B, C_V , C_R, C_I$ are the color terms. We adopted the same extinction coefficients from CTIO's calibration pages in Table~\ref{tab:cal}. The Rubin~149 field was observed for 60 photometric nights and the color term parameters in Table~\ref{tab:cal} were determined calibrating the Landolt standard stars in Rubin~149.

\subsubsection{Comparison with Other Photometric Observations}
\begin{figure}[h]
\begin{center}
\includegraphics[width=0.85\textwidth]{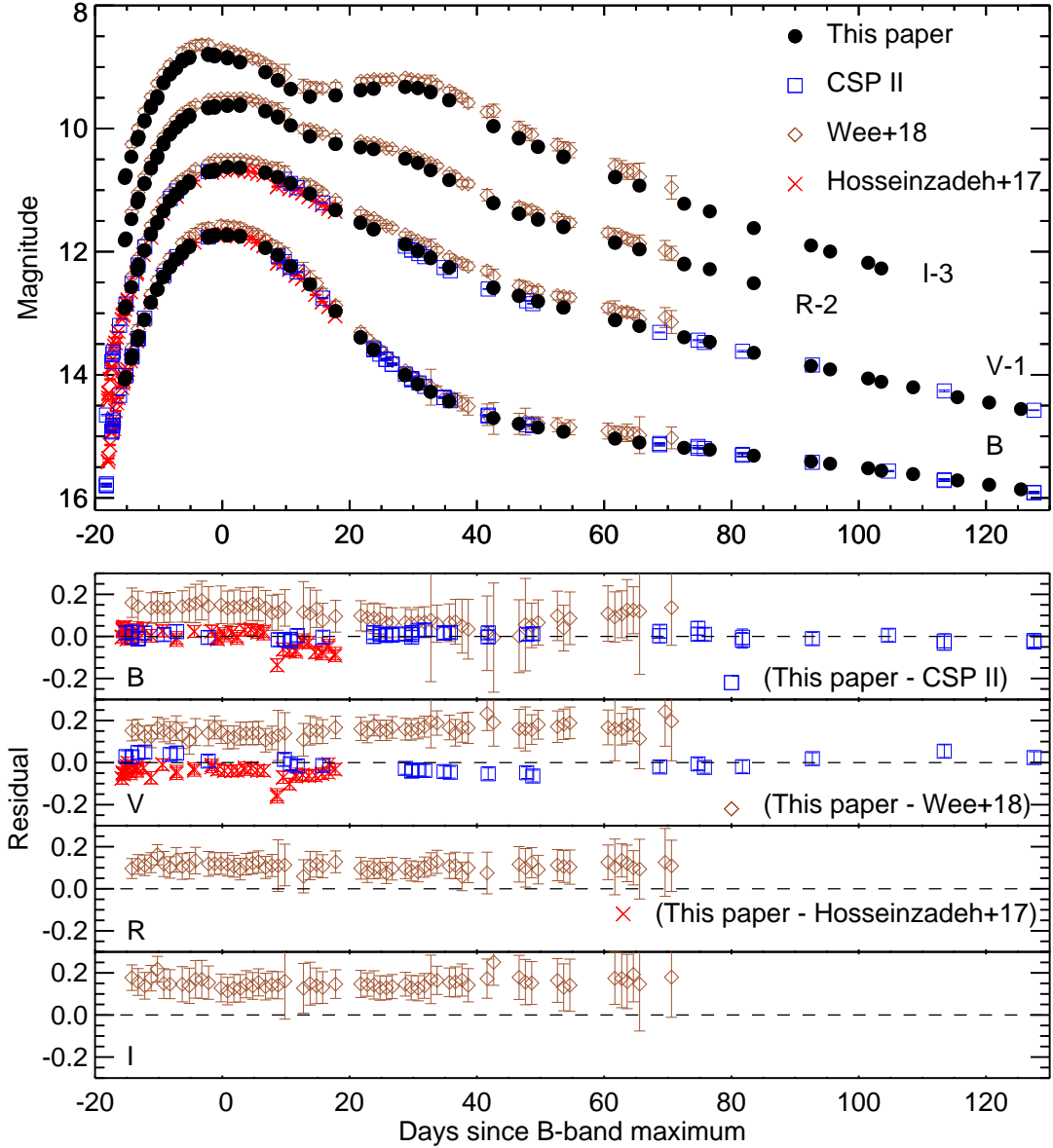}
\caption{Comparison of $BVRI$-band light curves of SN 2017cbv between ANDICam (this work), \citet{Wee18}, the CSP II program (Burns et al. 2020), and the Las Cumbres Observatory's 1-m telescope \citep{Hosseinzadeh17}. The bottom four panels show the differences between ANDICam and other surveys in $BVRI$ bands, respectively. \label{fig:cmpoptical}}
\end{center}
\end{figure}

We compared the $BVRI$-band photometry of SN 2017cbv observed with ANDICam to the data of \citet{Wee18}, CSP II program (Burns et al. 2020), and the 1-m data taken with the Las Cumbres Observatory Supernova Key Project and Global Supernova Project \citep{Hosseinzadeh17}, see Figure~\ref{fig:cmpoptical}. Our $BV$-band photometry are consistent with CSP II and \citet{Hosseinzadeh17} within 0.05~mag. While our $BVRI$-band light curves are systematically fainter by $0.101\pm0.042$, $0.156\pm0.027$, $0.108\pm0.018$, $0.152\pm0.023$~mag relative to the same filter in \citet{Wee18}. We double-checked our photometry by employing two independent methods simultaneously (aperture and PSF photometry) while only aperture photometry was used in \citet{Wee18}. Our aperture photometry is corrected by the aperture size.
The aperture correction was measured on bright isolated stars, between $15''$ and $7''$ and applied to $7''$ aperture photometry. This is the same method used to measure standard stars by \citet{Landolt09}. Our calibration involved independent observations of local sequence stars taken from the CSP II project, while \citet{Wee18} took observations of standard star fields for calibration and target fields with the same instrument. Our $BV$ band photometry is most consistent with those from CSP II (Burns et al. 2020) and \citet{Hosseinzadeh17}.

\  \par
\subsection{NIR Photometry}
\subsubsection{Data reduction and Astrometry}
The ANDICam NIR raw images have been binned on-site at CTIO using an IRAF script and uploaded to the Yale Repository. For the binned NIR images, we first applied flat-field correction and cosmic-ray rejection. Then a sky frame was subtracted from each image using neighboring images: if the three dithered science images a, b, c were taken one by one with the same exposure time, the difference of two neighboring images a-b, b-a, c-b were taken as the sky-subtracted images. We measured the dither offsets relative to the first frame by picking up a bright source on the dithered science images using {\it skycat}\footnote{\url{http://archive.eso.org/cms/tools-documentation/skycat.html}} and took the measured dither offsets as the initial values for all science frames. Stars were extracted on the image after the initial dither offset correction \citep{Bertin96} and their pixel positions were matched to get additional shifts. The images with corrections for the dither offset (initial + extra) corrected frames were then combined to create a co-added image, which can be used to perform photometry simultaneously via Source Extraction and Photometry \citep[SEP; ][]{Barbary18} and PSF Extractor \citep[PSFEx; ][]{Bertin96,Bertin11} after astrometric calibration. For the NIR images, WCS information was added manually to one reference image, and then all other images were aligned with the reference using the python module $Astroalign$ \citep{Beroiz19}. 

%Astrometry.net does not work on the NIR images due to its smaller FOV.

\subsubsection{Differential Photometry}
As we did for the optical images, we performed differential photometry of SN~2017cbv relative to neighbor stars 1 and 3 ($\sim$ 3 magnitudes fainter than 1, see Figure~\ref{fig:chart}). We did not take 2 as the local star because it is too close to the image's edge. We found that PSF photometry performs better than aperture photometry. We tested this by comparing the variance of the difference between stars 1 and 3, which is smaller for PSF than aperture photometry in all filters. We also noted an intrinsic dispersion of the order of 0.025 magnitudes for the PSF difference distribution, which we account for as an instrumental noise component, and add it to the photometry error budget. As star 1 is brighter and there is no time-series variability during our observing period, we finally adopted the differential PSF photometry of SN~2017cbv relative only to star 1.

\subsubsection{Calibration of the Local Sequence Stars}
We found systematic differences for the nightly zero points when calibrating the three standard fields (RU149, P9144, and LHS2397a) during photometric nights. So we took the $JHK_s$ photometry of the star \# 1 from the Two Micron All Sky Survey \citep[2MASS; ][]{Cutri03a,Cutri03b,2mass} to avoid a calibration problem. Its $Y$-band magnitude was derived from the relationship between $Y-J$ and $J-H$ colors in \citet{Hodgkin09}. The $YJHK_s$ magnitudes of star \#1 are listed in Table~\ref{tab:localyjhk}.

\subsubsection{Comparison with \citet{Wee18}}
\begin{figure}[h]
\begin{center}
\includegraphics[width=0.85\textwidth]{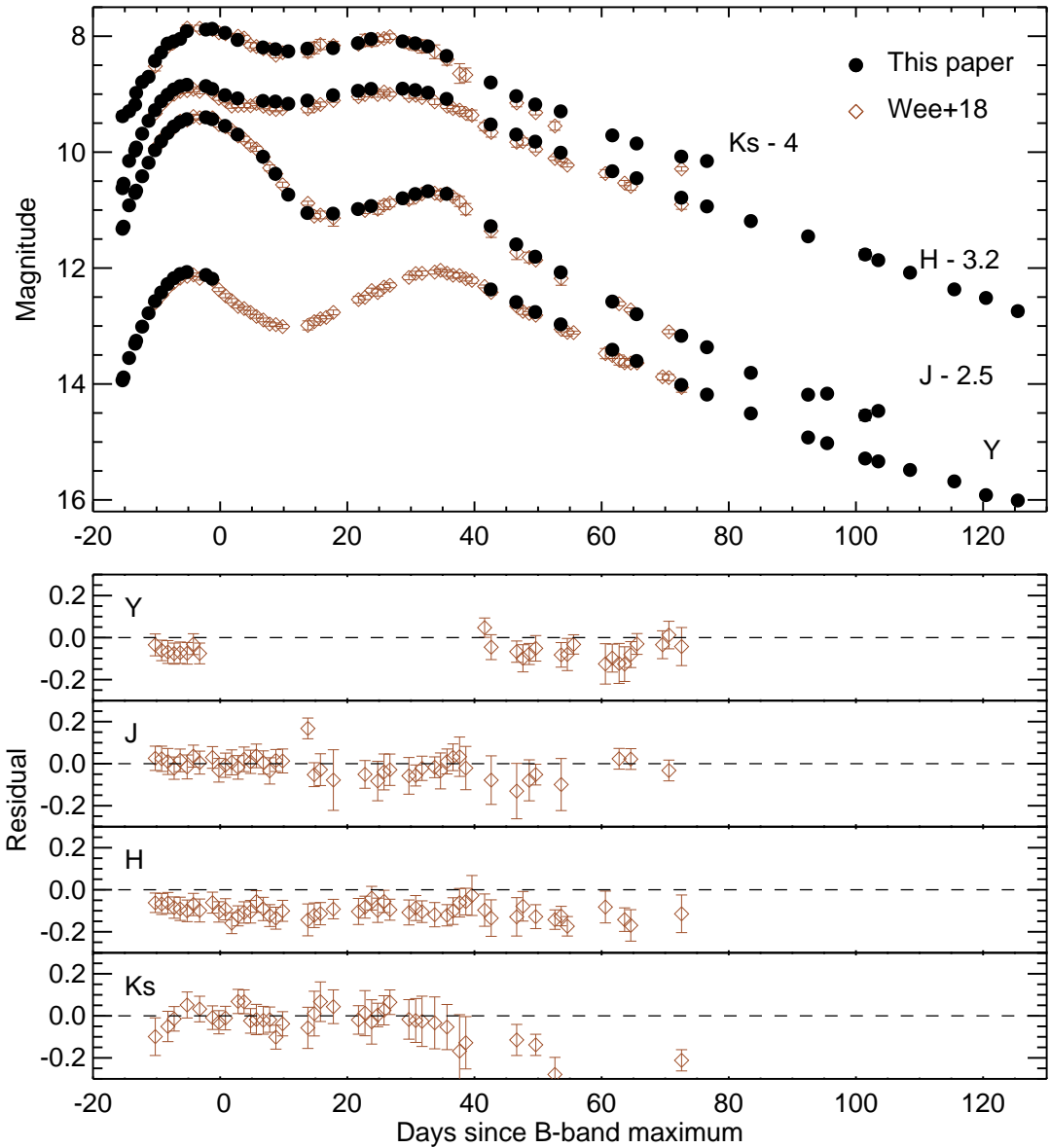}
\caption{Same as Figure~\ref{fig:cmpoptical}, but for $YJHK_s$ bands. \label{fig:cmpnir}}
\end{center}
\end{figure}

We also compared the $YJHK_s$-band photometry with that from \citet{Wee18} in Figure~\ref{fig:cmpnir}. The $YJK_s$-bands are consistent, the differences being $-0.063\pm0.040$, $-0.014\pm0.049$, $-0.035\pm0.078$~mag respectively. 
For $H$ band, we found a systematic difference of $-0.101\pm0.032$~mag.

\begin{deluxetable}{lrc}[thb]
\tablewidth{0pt}
%\tablenum{2}
\tablecaption{Transformation coefficients of color terms in the $BVRI$ bands. \label{tab:cal}}
\tablehead{\colhead{Filter } &  \colhead{Color term $C$} & \colhead{Extinction coefficient $K$}  \\}
\startdata

$B$ &   $ 0.059\pm0.014$  &    0.251 \\
$V$ &   $-0.041\pm0.007$  &    0.149 \\
$R$ &   $-0.015\pm0.027$  &    0.098 \\
$I$ &   $-0.079\pm0.015$  &    0.066 
%$Y$ & &0.100\\
%$J$ & &0.100\\
%$H$ & &0.055\\
%$K_s$ & &0.085
\enddata
%\tablecomments{Photometry of RU 149 were obtained in \citet{Landolt09}, and $JHK$ photometry were from \citet{Cutri03a,Cutri03b}}
\end{deluxetable}

\begin{deluxetable}{lrrllll}[h]
%\rotate
%\tablewidth{0pc}
\tablewidth{0pt}
%\tablenum{3}
\tablecaption{ Magnitudes of the optical photometry sequence for SN2017cbv. \label{tab:localbvri}}
\tablehead{\colhead{Star }& \colhead{R.A.} & \colhead{Dec.} & \colhead{$B$}  & \colhead{$V$} & \colhead{$R$} & \colhead{$I$} \\
                          & (J2000.0)      & (J2000.0)      &                &               &               &              }
%\tablehead{\colhead{ID }& \colhead{R.A.} & \colhead{Dec.}  & \colhead{$I$} & \colhead{$Y$} & \colhead{$J$} & \colhead{$H$} & \colhead{$K$} \\}
\startdata

2    & 14:32:30.4 &$-$44:07:04.6 & $14.378\pm0.013$ & $13.876\pm0.011$ & $13.546\pm0.023$ & $13.260\pm0.014$

 %17 & 14:32:32.813 & -44:07:23.62 &    \nd          &   \nd           &   \nd           &  \nd            & 14.521 (12) &    14.216 (7) &    13.027 (8) &    12.579 (28) 

\enddata
\tablecomments{Units of right ascension are hours, minutes, and seconds, and units of declination are degrees, arcminutes, and arcseconds. \\
See Figure~\ref{fig:chart} 1 for a chart of SN2017cbv.\\
%$^a$ Based on 16 nights for $BVRI$-band calibrations with the CTIO 1.3 m telescope.\\
}
\end{deluxetable}

\begin{deluxetable}{lrrlrrr}[h]
\tabletypesize{\footnotesize}
%\rotate
%\tablewidth{0pc}
\tablewidth{0pt}
%\tablenum{4}
\tablecaption{ Magnitudes of the NIR photometric sequence for SN2017cbv$^a$. \label{tab:localyjhk}}
\tablehead{\colhead{Star }& \colhead{R.A.} & \colhead{Dec.} & \colhead{$Y$}  & \colhead{$J$} & \colhead{$H$} & \colhead{$K_s$}  \\
                          & (J2000.0)      & (J2000.0)      &                &               &               &                 }
\startdata

1 & 14:32:40.0 & $-$44:08:17.3 &10.962$\pm$0.029 &10.496 $\pm$0.024 & 9.724 $\pm$ 0.022 & 9.606 $\pm$ 0.021 \\
3 & 14:32:32.8 & $-$44:07:20.5&14.166$\pm$0.035 &13.760 $\pm$0.029 & 13.108 $\pm$ 0.025& 12.815 $\pm$0.037

\enddata
\tablecomments{The bright star \# 1 was used to calibrate the NIR photometry of SN 2017cbv.\\
$^a$ The $JHK_s$ magnitudes are taken from 2MASS \citep{Skrutskie06,Cutri03a,Cutri03b}. \\
$^b$ The $Y$-band magnitudes are measured from 2MASS $JH$ magnitudes following equation 5 in \citet{Hodgkin09}. }
\end{deluxetable}

\subsection{NIR Spectroscopy}
\subsubsection{Data Reduction}

Six NIR spectra were taken with FLAMINGOS-2 \citep{Eikenberry06} on Gemini South 8.2-m telescope at Gemini South Observatory, including the earliest one at only 2.30 days past explosion. The FLAMINGOS-2 spectra were acquired in long-slit mode with the $JH$ grism and filter in place, along with a slit width of 0.72 arcsec. This setup yielded a wavelength coverage of 1.0 - 1.8 $\mu$m and a resolution of $R \sim 1000$. The long-slit spectra were acquired at the parallactic angle with a standard ABBA pattern, and were reduced in a standard way using the F2 PyRAF package provided by Gemini Observatory. The XTELLCOR pipeline was used to perform telluric corrections and flux calibrations.  More details of the data reduction process can be found in \citet{Brown19,Hsiao19}.

Two NIR spectra were provided by the ePESSTO collaboration \citep{Smartt15} with the SOFI instrument \citep{Moorwood98} on the 3.6-m NTT at La Silla Observatory. SOFI spectra were observed with blue and red grisms with a slit width of 1.0 arcsec, a wavelength coverage of 0.9 to 2.5 $\mu$m and a resolution of $R \sim 500$. The conventional ABBA nod-along the-slit technique was adopted. For each SOFI spectrum, a Vega-like or a Solar analogue was also observed for flux calibration. SOFI spectra were reduced by performing the following steps: cross-talk correction, flat field correction, wavelength calibration, sky subtraction and spectral extraction, telluric absorption correction and flux calibration \citep{Smartt15}.

Five NIR spectra were observed with FIRE \citep{Simcoe13} on the 6.5-m Magellan Baade telescope at Las Campanas Observatory. The FIRE spectra were acquired with the high-throughput prism mode with a slit of 0.6 arcsec, a wavelength coverage of 0.8 to 2.5 $\mu$m and a similar resolution as SOFI. For each epoch, the conventional ABBA nod-along-the-slit technique and the sampling-up-the-ramp readout mode were used. Meanwhile, an A0V star was observed for telluric correction  close in time, angular distance and airmass to our science target \citep{Vacca03,Hsiao15,Hsiao19}. An IDL pipeline {\it firehose} \citep{Simcoe13} was specifically developed to reduce FIRE spectra.

One SpeX \citep{Rayner03} spectrum was taken with the 3.0-m NASA Infrared Telescope Facility at the summit of Mauna Kea. The SpeX spectrum was obtained in the cross-dispersed mode with a slit of 0.5 arcsec, yielding a wavelength range of 0.8 to 2.5 $\mu$m \citep{Hsiao19}. The data was reduced with the IDL code Spextool \citep{Cushing04}, which is designed for handling the SpeX data. The flux calibration process for SpeX spectrograph is similar as other devices, such as FIRE, SOFI, and FlAMINGOS-2.

%\subsubsection{}
\subsubsection{NIR Spectral Diagnostics of SN~2017cbv}
\begin{figure}[htb]
\begin{center}
\includegraphics[width=0.9\textwidth]{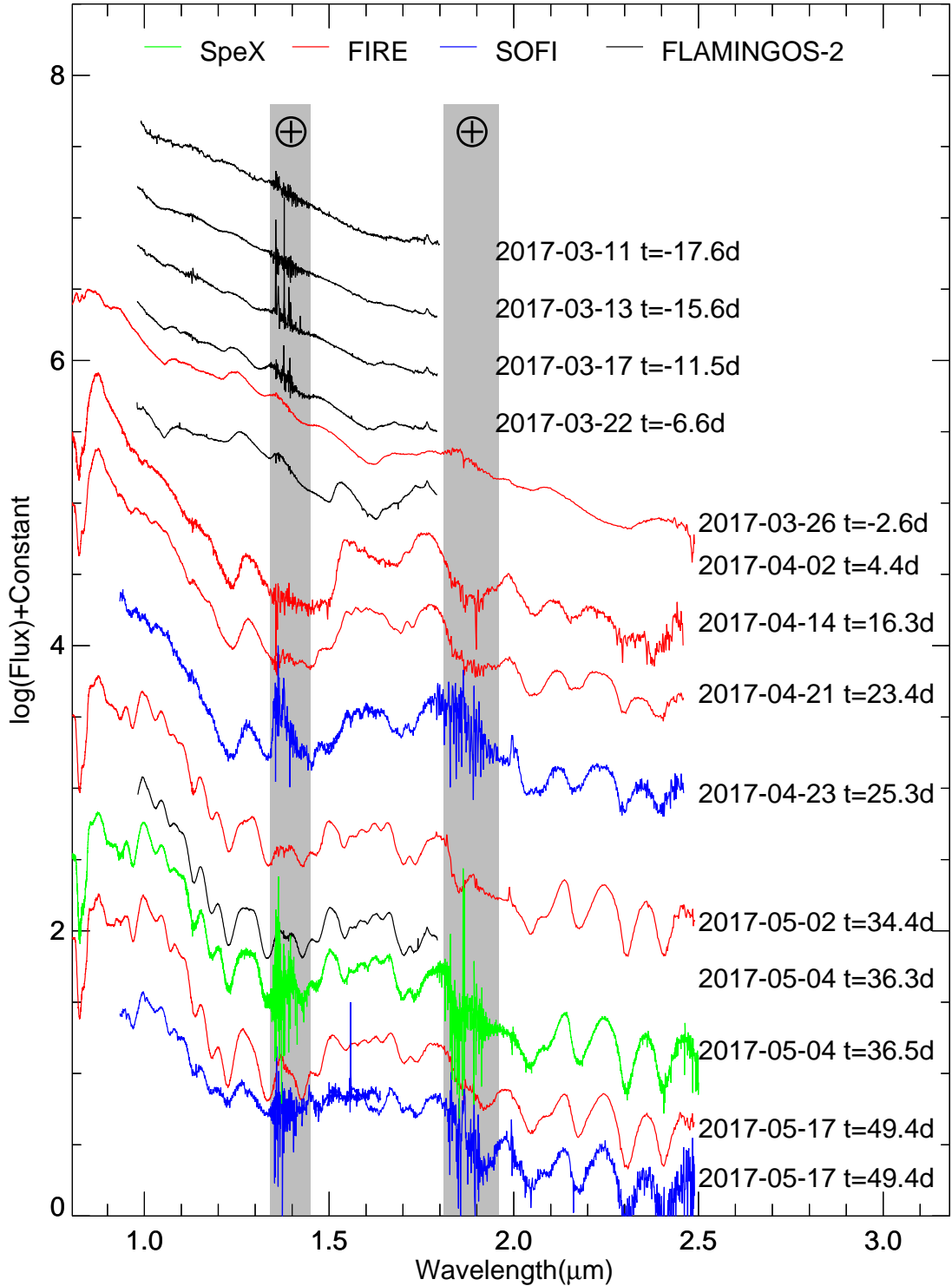}
\caption{Time series NIR spectra of SN~2017cbv from NIR spectrograph: SpeX, FIRE, SOFI, and FLAMINGOS-2. \label{fig:spec}}
\end{center}
\end{figure}

%\subsection{Optical spectra}
Figure~\ref{fig:spec} shows the NIR spectra obtained for SN~2017cbv, covering the phases from $-17.6$ to $+49.4$ days relative to $t_B^{max}$. The early NIR spectra are dominated by electron scattering with a well-defined photosphere. Thus the first three early spectra in Figure~\ref{fig:spec} show relatively featureless blue continua. Later, spectral features develop at 1.05, 1.25 and 1.65 $\mu$m when the SN is close to $B$ band maximum \citep{Wheeler98}. The most prominent features at those epochs are associated with intermediate-mass species i.e., O I, Mg II, and Si III, \citep[see also Figure 2 in][]{Hsiao13}. Specifically, the strong and relatively isolated absorption feature at 1.05 $\mu$m was identified as Mg II $\lambda 1.0927 \mu$m by \citet{Wheeler98} and also in SNe 1999ee \citep{Hamuy02b,Hamuy02a}, 2005cf \citep{Gall12} and 2011fe \citep{Hsiao13}. The emission feature at 1.25 $\mu$m was identified as Si III by \citet{Hsiao13}, and Fe III by \citet{Hamuy02a} and \citet{Rudy02}. A strong feature around 1.6 $\mu$m was identified as Fe III by \citet{Hsiao13}, Mg II/Si II/ Co II by \citet{Marion09}, and Si II by \citet{Wheeler98} and \citet{Gall12}. At longer wavelengths, the spectrum is also featureless, except for a weak emission feature around 2.05 $\mu$m perhaps due to Si III \citep{Wheeler98}. 

\ \par

Figure~\ref{fig:early} displays the comparison between SNe 2017cbv and 2011fe at $-17.6$, $-11.5$, $-2.6$ and $+4.4$ days relative to the $B$-band maximum \citep{Hsiao13}. Both spectra are matched very well except for the Mg II $\lambda$ $1.0927 \mu$m absorption feature, which is very weak in SN 2017cbv at $t= -17.6$ and $-11.5$ days. 

\ \par
Two weeks past maximum, the spectra show dramatic evolution dominated by strong emission/absorption features. The Mg II $\lambda$ $1.0927 \mu$m absorption feature disappeared and the most remarkable features were the strong and wide peaks around 1.5 to 1.7 $\mu$m, which can be attributed to blends of Co II, Fe II, and Ni II \citep{Wheeler98}. Meanwhile, new peaks at around 2.2 and 2.4 $\mu$m appeared and gradually developed, which are mainly attributed to iron-group element Co II \citep{Wheeler98}. Figure~\ref{fig:aftermax} shows the post-maximum comparisons of SNe 2017cbv, 2011fe \citep{Hsiao13}, 2012fr \citep{Hsiao19} and 2014J \citep{Sand16} at comparable phases. They are very consistent with each other at $+16.3$, $+34.4$, $+36.3$, $+36.5$, and $+49.4$ days after the maximum light. Two spectra were taken at $+49.4$~days on May 17, one by FIRE and the other by SOFI. We adopted the spectrum observed by FIRE as the colors from the FIRE spectrum are closer to the NIR-band photometry and the S/N is higher as well in Figure~\ref{fig:spec}.

\ \par
\subsubsection{Mg II velocity}

Mg II is a product of explosive carbon burning and thus a sensitive probe of the location of the inner edge of carbon burning \citep{Wheeler98,Marion01,Marion09,Hoeflich02,Hsiao13} in velocity space. The Mg II $\lambda 1.0927 \mu$m line is expected to be observed with decreasing velocity in the early spectral evolution and then is expected to remain at an almost constant velocity when the photosphere has receded below the inner edge of the Mg II distribution. We can only measure the absorption minimum of the Mg II $\lambda~1.0927 \mu$m at $t= -11.5$, $-6.6$, $-2.6$ and $+4.4$ days, with an almost constant velocity $\sim 10,500 $km s$^{-1}$. This suggests that its photosphere had receded below the inner edge of magnesium, according to the analysis by \citet{Wheeler98}. Alternatively, \citet{Meikle96} interpreted the constant velocity of Mg II as a detached feature. %From observations, the measured Mg velocity does not change when the photosphere recedes deeper into the core. 
%meaning that it generate well above the region of peak emissivity
\subsubsection{C I}
Unburned carbon provides the most direct diagnostic of the primordial material from the progenitor. The weak absorption carbon feature was mainly detected from early optical spectra via C II $\lambda$ 0.6580 $\mu$m \citep{Thomas07,Thomas11,Scalzo10,Silverman11,Silverman12d,Taubenberger11,Zheng13}. Alternatively, NIR C I $\lambda$~1.0693 $\mu$m can be taken as a superior carbon tracer compared with the optical C II $\lambda$ 0.6580 $\mu$m, as C I can be detected at maximum light \citep[e.g., ][]{Hoeflich02}. \citet{Hosseinzadeh17} detected a strong C II $\lambda$ 0.6580 $\mu$m feature at $t= -19$ day, similar to SN 2013dy at $t= -16$ day \citep{Zheng13}, although the carbon feature disappeared by day $-13$ for SN~2017cbv. Very strong carbon C II was also seen in iPTF~16abc \citep{Miller18}. We tried to detect the C I from our NIR spectra and we saw a notch close to $\lambda$ 1.0693 $\mu$m near 1.03$\mu$m, taken on Apr. 2, 2017, or $t\sim 4.4$~days after $B$-band maximum. We applied the automated spectrum synthesis code SYNAPPS \citep{Thomas11} to identify the C I $\lambda$ 1.0693 $\mu$m, as showed in Figure~\ref{fig:synapps}. The blueshift of the C I line (green bashed line in Figure~\ref{fig:synapps}) was observed at 11,000~$\kms$ at $t \sim 4.4$~days after $B$-band maximum. The velocity of the unburned carbon in the NIR spectrum was consistent with velocity of Mg II $\lambda~1.0927 \mu$m at the same epoch in \S 2.3.3. If the detected C I is real, SN~2017cbv could be the second case to support the hypothesis that a change in the ionization condition occurs as the temperature cools, indicating that the signature of C II $\lambda~0.6580 \mu$m appears in the very early phase before $B$-band maximum and the C I $\lambda~1.0693 \mu$m appears later, i.e., around maximum, as similarly reported by \citet{Hsiao13} for SN~2011fe. Note that the detection of C I $\lambda$ 1.0693 $\mu$m in Figure~\ref{fig:synapps} is interpreted by SYNAPPS \citep{Thomas11}, which is independent of the very early optical detection C II $\lambda$ 0.6580 $\mu$m \citep{Hosseinzadeh17}.

\subsubsection{Pa$\beta$}
\citet{Maeda14} has emphasized that the Pa$\beta$ in post-maximum NIR spectra can provide a powerful diagnostic of the presence of unbound hydrogen-rich matter expelled from a companion. Hydrodynamic and radiative transfer models in \citet{Maeda14} found that the post-maximum Pa$\beta$ is easily observed and this feature grows stronger at $\sim$ 1-2 months after maximum, covering a range of viewing angles between the observer, supernova and the companion star. \citet{Sand16} tried to identify the Pa$\beta$ line for the nearby Type Ia SN 2014J. They found no evidence for the presence of Pa$\beta$ emission after comparing the observed spectra around Pa$\beta~\lambda~1.282\mu$m with the red-giant scenario corresponding to 0.3, 0.1, 0.03\msun of hydrogen for the boundary cases: $\theta = 0^{\circ}$ and 180$^{\circ}$ \citep{Maeda14}. Thus \citet{Sand16} gave a rough hydrogen mass upper limit of 0.1 $M_{\odot}$ for all SN-companion star orientations and claimed that it was not distinguishable between the scenario with hydrogen masses of 0.03 \msun and observations. We have compared our post-maximum NIR spectra of SN 2017cbv with those of SNe 2011fe and 2014J at several phases of $t= +16.3$, $+34.4$, $+36.3$, $+36.5$, $+49.4$ days, as shown in Figure~\ref{fig:pbeta}. We can see our five post-maximum spectra are well-matched with that of SN~2014J and both have comparable signal-to-noise ratio in Figure~\ref{fig:pbeta}. No Pa$\beta$ lines were detected from our five post-maximum NIR spectra by visual inspection and this yields a similar hydrogen mass limit of less than 0.1 $M_{\odot}$ from the companion star of SN~2017cbv, although the limit depends on the viewing angles. Analysis of Pa$\beta$ line of SN 2017cbv using the same NIR spectrum at 34 days after maximum was performed in \citet{Hosseinzadeh17} and they drew similar conclusions. Nondetection of H$_\alpha$ from nebular spectroscopy gave an even lower hydrogen mass limit \citep{Sand18}.

\begin{figure}[htb]
\begin{center}
\includegraphics[width=0.9\textwidth]{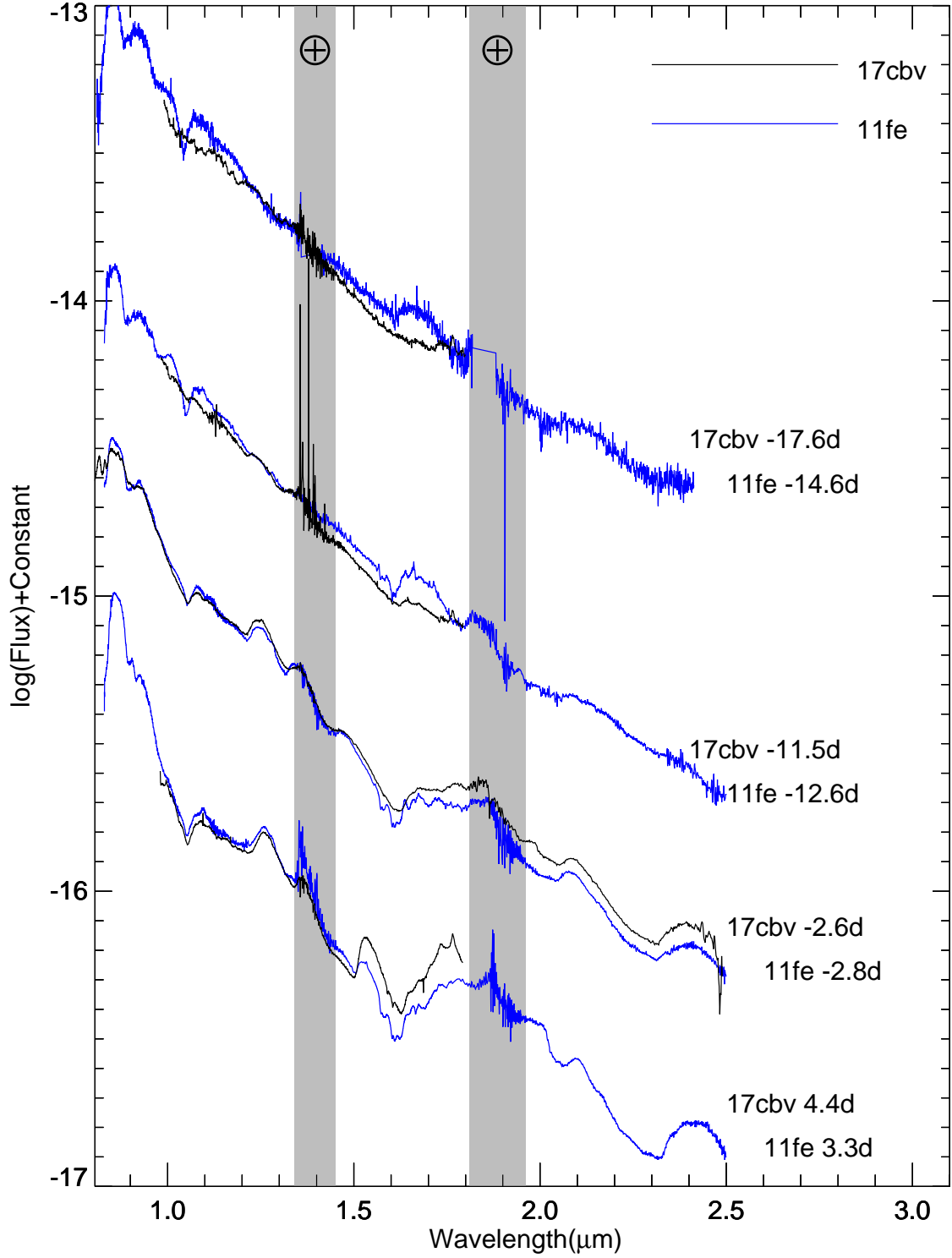}
\caption{Comparison NIR spectra of SN~2017cbv (black) and SN~2011fe (blue) from early epochs through roughly maximum light. \label{fig:early}}
\end{center}
\end{figure}

\begin{figure}[htb]
\begin{center}
\includegraphics[width=0.9\textwidth]{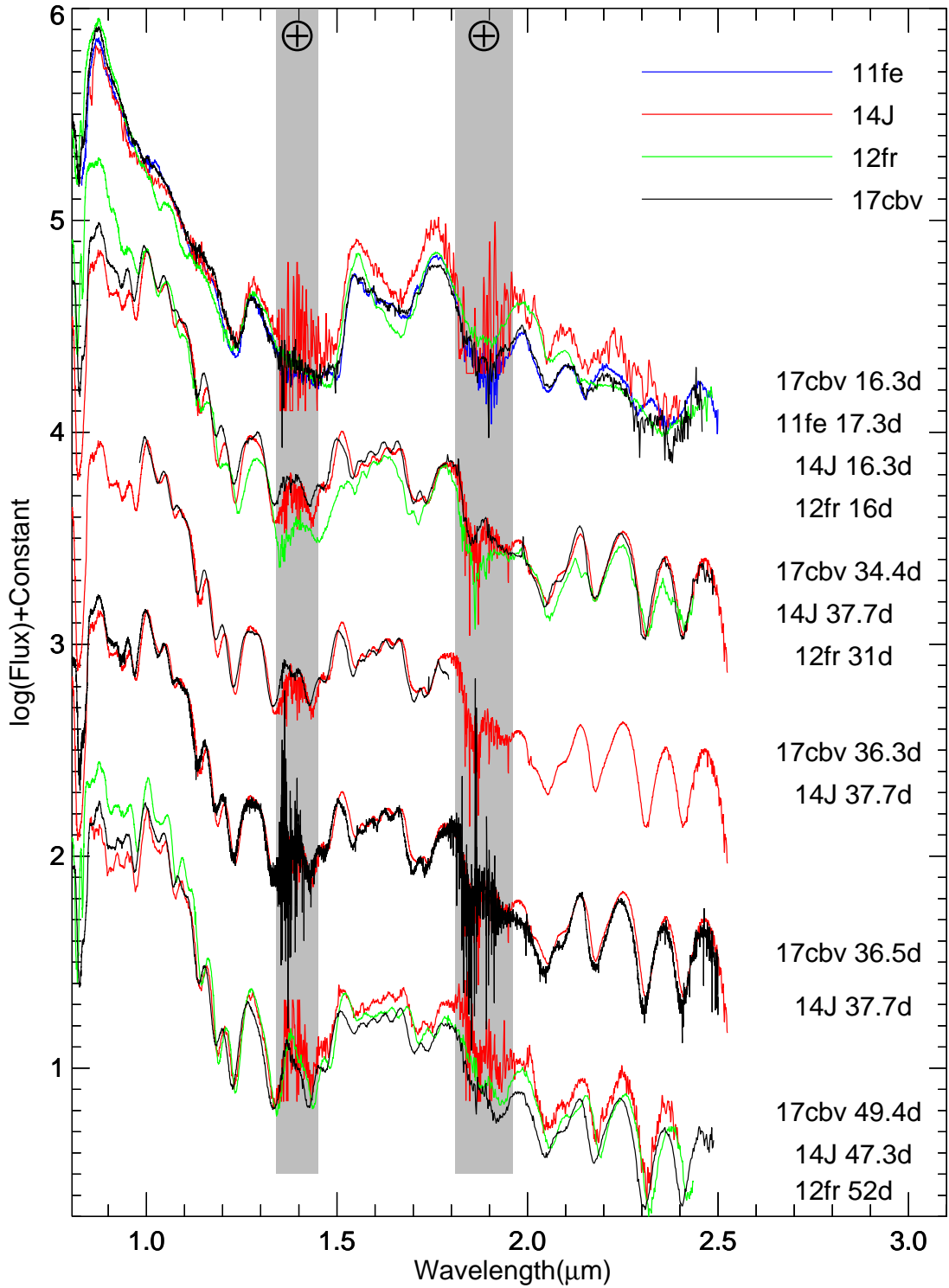}
\caption{Comparison spectra of SN~2017cbv (black), SN~2011fe (blue), SN~2012fr (green) and SN~2014J (red) after maximum light. \label{fig:aftermax}}
\end{center}
\end{figure}

\begin{figure}[htb]
\begin{center}
\includegraphics[width=0.9\textwidth]{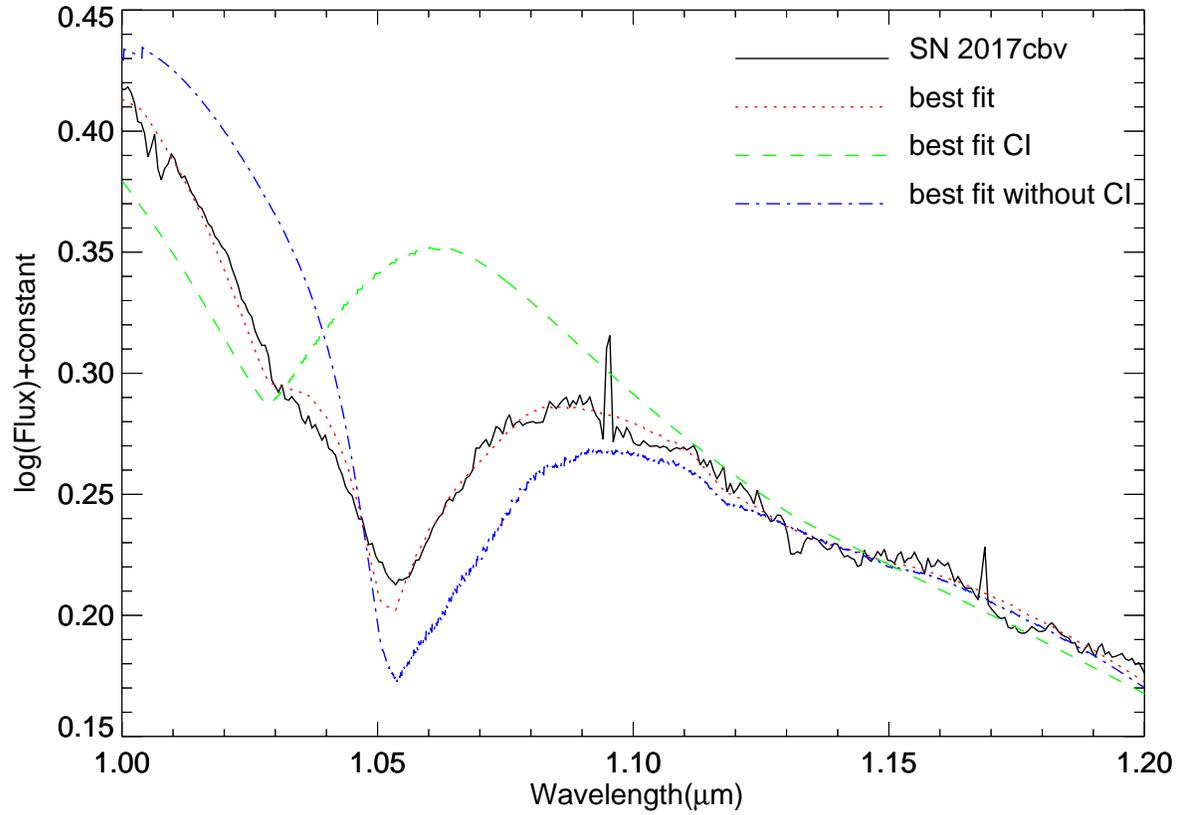}
\caption{SYNAPPS fit to the region of the C I $\lambda 1.069 \mu$m line of SN 2017cbv taken at $t\sim 4.4$~days. The spectrum is plotted as a solid black curve and the best-fit synthesized spectra are plotted with all ions (eg, C I $\lambda 1.069 \mu$m, Mg II $\lambda 1.093 \mu$m and other ions; red dotted), with only C I (green dashed), and with all ions except C I (blue dash-dotted curves). There is likely a detection of C I in the spectrum, with a clear notch seen in the blue wing of the Mg II line. \label{fig:synapps}}
\end{center}
\end{figure}

\begin{figure}[htb]
\begin{center}
\includegraphics[width=0.9\textwidth]{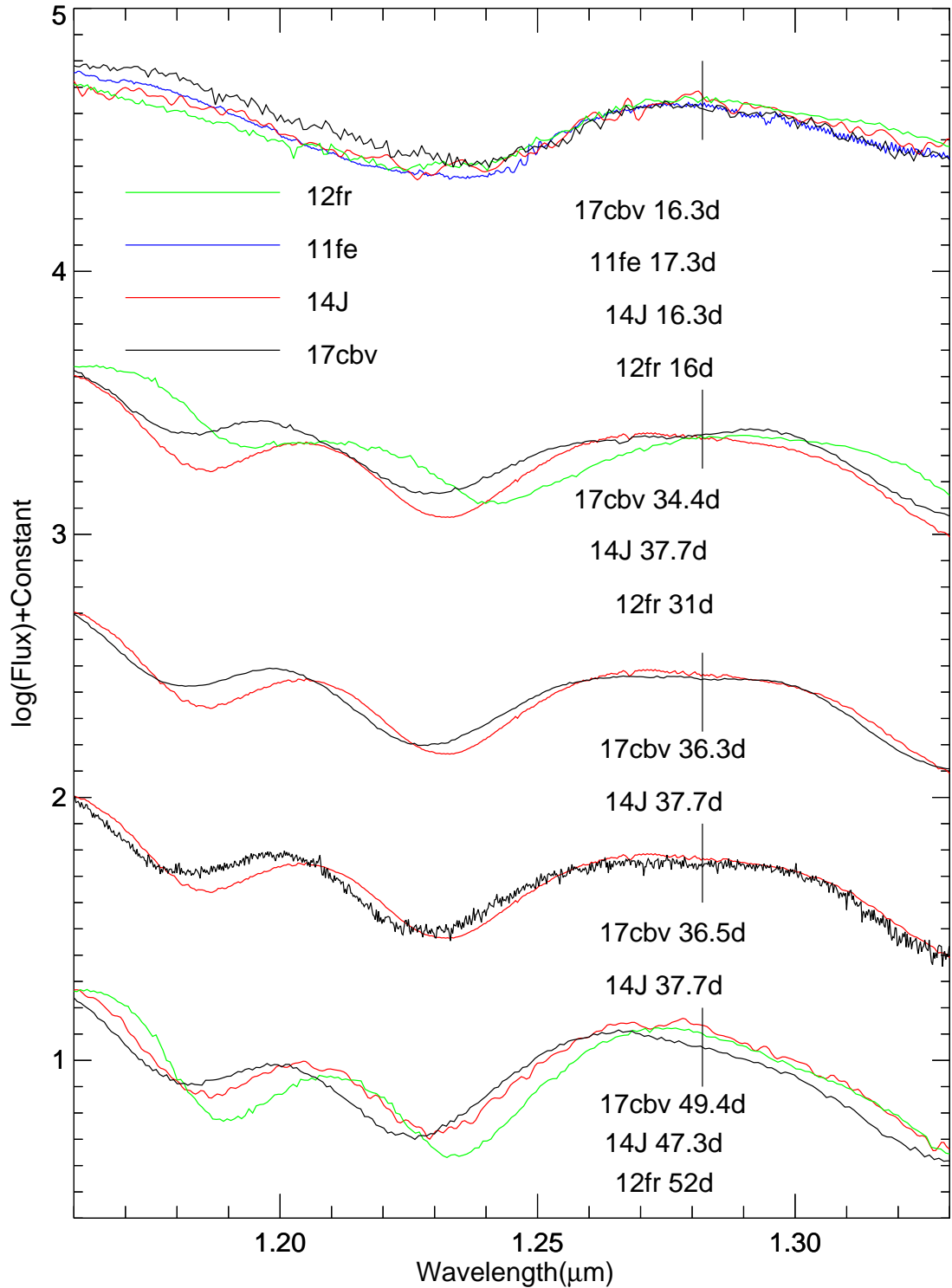}
\caption{Comparison spectra of $Pa\beta$ at 1.282 $\mu$m for SN~2017cbv in black, SN~2011fe in blue, SN~2012fr in green and SN~2014J in red after maximum light. \label{fig:pbeta}}
\end{center}
\end{figure}

\clearpage
\section{Further Analyses of Physical Properties of SN 2017\lowercase{cbv}}
\subsection{Light Curves and Color Evolution}
\subsubsection{Optical Light Curves}
Figure~\ref{fig:lc} shows the $BVRIYJHK_s$-band light curves of SN~2017cbv from our observations (also see Appendix Table~\ref{tab:bvri} for optical and Appendix Table~\ref{tab:yjhk} for near IR data). The light curves were sampled during the period $t\sim -16$ to $+125$ days relative to $B$-band maximum, making SN~2017cbv one of the best-observed SNe~Ia in optical and NIR bands simultaneously. The morphology of the light curves resembles those of normal SNe~Ia, showing a shoulder in the $R$ band and a pronounced secondary-maximum in $I$ and NIR bands. The NIR light curves of SN~2017cbv reached their first peak $\sim$ 4 days earlier than the $B$-band curve, consistent with the statistical analysis of a SNe Ia sample \citep{Dhawan15}.
%from -15.500 to +125.250d relative to B-band maximum
%SN 2014J \citep{Srivastav16}
\begin{figure}[htb]
\begin{center}
\includegraphics[width=0.9\textwidth]{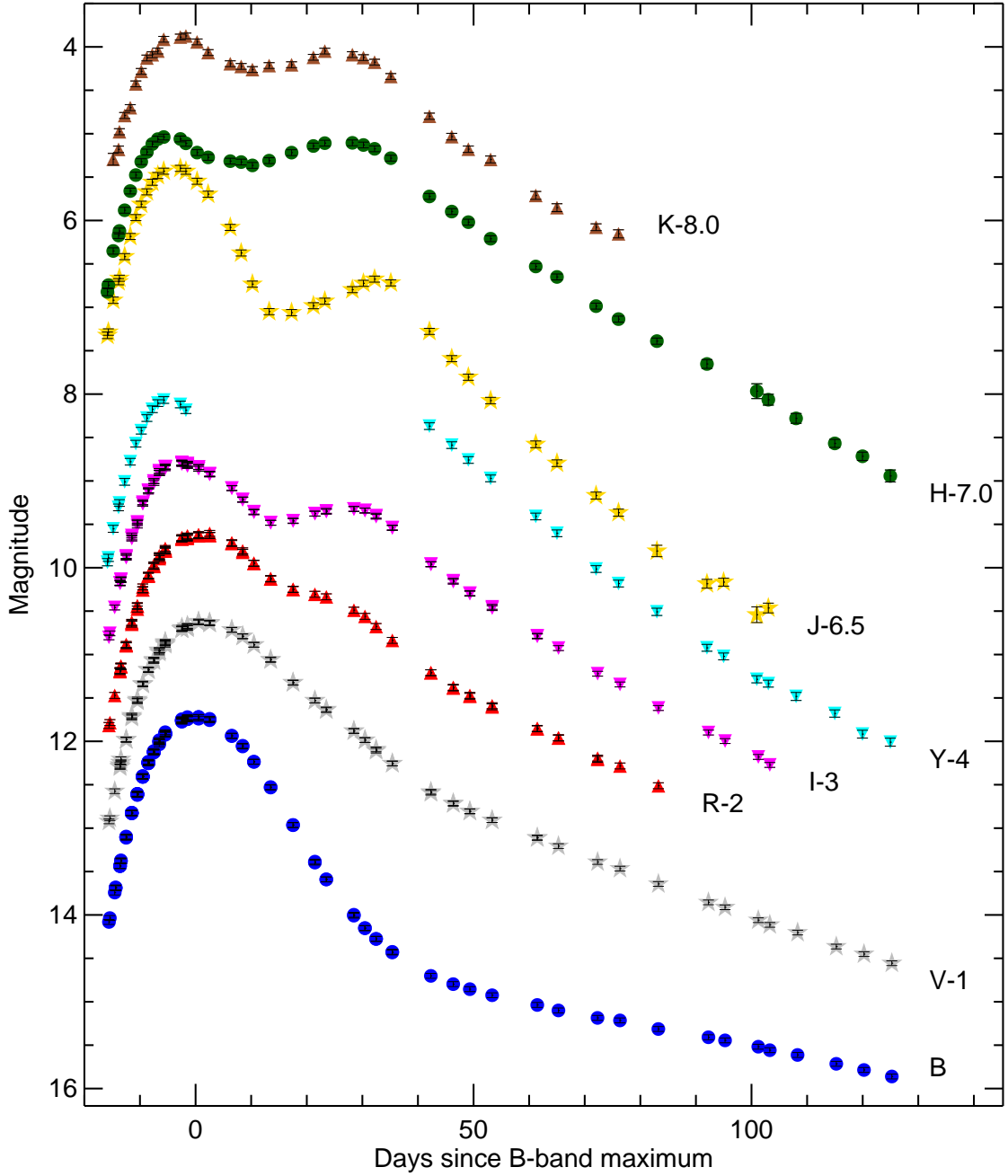}
\caption{The optical and NIR light curves of SN~2017cbv, spanning from $\sim$$-$16 to +125 d with respect to $B$-band maximum. Data for this figure are available online. \label{fig:lc}}
\end{center}
\end{figure}
%\subsection{Optical Light curves}

%%%%%%%%%%%%%%%%%%%%%%%%%%%%%%%%%%%%%%%%%%%%%%%%%%%%%%%%%%
\subsubsection{Light Curve Parameters}

 Gaussian Process Regression (GPR) was applied via the Python module Scikit-learn \citep{scikit11} to estimate the light curve shape parameter $\Delta m_{15}(B)=0.990\pm0.013$ mag and the peak time of $B$-band light curve as $t_B^{max}= 57840.87\pm0.10$ MJD, which are used throughout this paper. The value $\Delta m_{15}(B)$ is the $B$-band magnitude difference between the peak $B_{t=0}=11.710\pm0.006$~mag and 15~days after $B_{t=15}=12.700\pm0.011$~mag. GPR is a nonparametric, Bayesian approach to regression in the area of machine learning \citep{Rasmussen06}. Similarly, we implemented GPR to fit the phases and maximum peak magnitudes of the $VRIYJHK_s$ bands. When possible, we also use the same GPR to deduce the phases and peak magnitudes of the secondary maximum and the minimum magnitude between the two maximums for each of $IYJHK_s$-band light curves. We follow the nomenclature adopted by \citet{Biscardi12} to parameterize the light curves of SN~2017cbv. For X band, at the phases of t1(X), t2(X) and t0(X) relative to the $B$ maximum, the first maximum m1(X), the secondary maximum m2(X) and a minimum m0(X) between the two maximums are reached. The uncertainties of the phases were measured using a  jack-knife procedure. For example, we select N points around the maximum t1, and then start a loop taking one data point out and fitting with the GPR to the rest of the N-1 data points. An array of N measurements of t1 is obtained once the loop is completed. Then we do statistics with this array to get the standard deviation $\sigma_{t1}$ and the final uncertainty of t1 will be $\sigma_{t1} \times \sqrt N$. We repeat the above process to obtain the uncertainties of t1, t2, t0 for each band.  
These light curve parameters, m1, m2, m0, and their times relative to $t_B^{max}$: t1, t2 and t0, as well as the decay rate $\beta$ between 40 and 90 days are tabulated in Tables~\ref{tab:shape} and ~\ref{tab:beta}. As shown in \citet{Elias81}, the $IYJHK_s$ light curves show the first maximum within -2 to -5 d of the $B$-band maximum, which is consistent with the results for SNe Ia with $\Delta m_{15} < 1.8$ \citep{Folatelli10} and is in agreement with the statistical studies in the $YJH$ bands \citep{Dhawan15}. Recent studies suggested that the timing of $i$-band maximum indicates the physical state of the SN Ia explosion \citep{Gonzalez-Gaitan14,Ashall20}. Thus the time of $i$-band maximum can be used to sub-classify Ia \citep{Gonzalez-Gaitan14,Ashall20} and SN~2017cbv locates in the normal sub-type.

\citet{Dhawan15} have also studied the $YJH$-band light curves of 91 SNe Ia from the literature and made an extensive statistical analysis for NIR light curve shape parameters, i.e., t1, t0, t2, and late time decay $\beta$ in these three bands. We found that the phases t1, t0, and t2 for SN~2017cbv in Table~\ref{tab:shape} for each band are consistent with values reported by \citet{Dhawan15}. As predicted by \citet{Kasen06a}, a larger Ni mass of supernova leads to higher temperatures and thus a later $2\rightarrow1$ recombination wave occurring around 7000 K, which tends to delay the secondary maximum. The epoch of secondary maximum t2 was slightly delayed for SN~2017cbv, perhaps indicating a larger Ni mass, when compared with SN~2011fe, which shows an earlier secondary maximum. SN~2011fe has a $\Delta m_{15}=1.18$~mag \citep{Zhang16}, which also indicated a lower Ni mass. 

After the (secondary) maximum, the light curves of SNe Ia usually show a linear decline in magnitude. We also calculated the decay rate $\beta$ for SN 2017cbv in $BVRIYJHK_s$ bands during the phases $40<t<90$ days. The values of the decay rate in different bands are listed in Table~\ref{tab:beta}. Our optical decline rates $\beta$ are consistent with those reviewed by \citet{Leibundgut00}. In the early nebular phase, the NIR-band light curves of SN~2017cbv are found to have faster decay rates than the optical ones, consistent with the results reported by \citet{Dhawan15}. As noted by \citet{Dhawan15}, SNe~Ia tend to have similar late-time decay rates in the NIR bands. At late time, the SN gradually becomes transparent to the $\gamma$ rays produced by radioactive decays. Similar late-time decay rates perhaps suggest a similar internal structure of the explosions, which is also in agreement with the predictions for Chandrasekhar mass models \citep{Woosley07} that produce different nickel masses but with similar radial distributions of iron group elements.

\begin{deluxetable}{cccccccccc}[thb]
\tablewidth{0pt}
%\tablenum{7}
\tablecaption{Light curve parameters of SN 2017cbv. \label{tab:shape}}

% new table
\tablehead{\colhead{Filter }   & \colhead{m($t_B^{max})$} & \colhead{t1$^a$}  & \colhead{m1} & \colhead{t0} & \colhead{m0}  & \colhead{t2} & \colhead{m2} & $A_{MilkyWay}$\\
                               &       mag           &    days          &  mag         &        days     &    mag        &     days        &    mag        & mag \\}
\startdata

%update 2020-7-15
$B$ & $11.710\pm0.006$ & $0.00\pm0.10$ & $11.710\pm0.006$  & ... & ... & ... & ... &      0.615\\
$V$ & $11.643\pm0.007$ & $1.24\pm0.10$ & $11.637\pm0.007$  & ... & ... & ... & ... &      0.453\\
$R$ & $11.607\pm0.008$ & $0.66\pm0.17$ & $11.605\pm0.008$  & ... & ... & ... & ... &      0.358\\
$I$ & $11.840\pm0.008$ & $-2.99\pm0.14$ & $11.793\pm0.007$ & $15.32\pm0.22$ & $12.472\pm0.012$ & $26.98\pm0.11$ & $12.312\pm0.012$ & 0.256\\

$Y$ & ...              & $-4.74\pm0.09$ & $12.050\pm0.020$  & ... & ... & ... & ... &      0.175\\
$J$ & $12.004\pm0.018$ & $-3.57\pm0.19$ & $11.883\pm0.015$ & $16.74\pm0.77$ & $13.613\pm0.024$ & $31.66\pm0.69$ & $13.206\pm0.018$ & 0.122\\
$H$ & $12.180\pm0.018$ & $-4.72\pm0.13$ & $12.027\pm0.016$ & $7.79\pm0.20$ &  $12.343\pm0.016$ & $25.96\pm0.29$ & $12.085\pm0.016$ & 0.078\\
$K_s$&$11.938\pm0.017$ & $-3.07\pm0.27$ & $11.877\pm0.015$ & $12.77\pm0.35$ & $12.252\pm0.017$ & $25.33\pm0.44$ & $12.064\pm0.016$ & 0.052

% slope_Y2,slope_eY2 3.3290121      0.19434440

\enddata
\tablecomments{
$^a$ The t1, t2, t0 are phases of the first maximum m1, the secondary maximum m2, and the minimum m0 between the two maximums, relative to $B$ maximum. \\
}

\end{deluxetable}

\subsubsection{Milky Way Extinction}
The Milky Way extinction in $BVRIJHK_s$ bands toward SN 2017cbv were derived from the dust maps of \citet{SF11} in individual bandpasses -- CTIO $B$, CTIO $V$, CTIO $R$, CTIO $I$, 2MASS $J$, 2MASS $H$, and 2MASS $K_s$, which are similar to the ANDICam filters used in the paper. The extinction values are listed in column 9 of Table~\ref{tab:shape} \footnote{\url{https://irsa.ipac.caltech.edu}}. The Galactic reddening toward SN 2017cbv is $E(B-V)$=0.162~mag \citep{SF11}.   

The $Y$-band extinction $A_Y=0.175$ was estimated by considering the mean $R_V$-dependent extinction law, and we refer to equations (1), (2a), and (2b) in \citet{CCM89}.

\begin{equation}
   <A(\lambda)/A_V>=a(x)+b(x)/R_V
\end{equation}
\begin{equation}
   a(x)=0.574x^{1.61}
\end{equation}
\begin{equation}
   b(x)=-0.527x^{1.61}
\end{equation}
where $x$ is the reciprocal of the $Y$-band central wavelength at $\lambda=1.03\mu$m \citep{Hillenbrand02}, the ratio of total to selective extinction is $R_V=3.1$, and $V$-band extinction toward SN 2017cbv is $A_V=0.453$ \citep{SF11}. 

The Na I D lines in the high resolution spectrum provide an independent measurement of both the Milky Way and host reddening. Burns et al. 2020 published one high resolution spectrum of SN2017cbv using Magellan Inamori Kyocera Echelle \citep[MIKE; ][]{Bernstein03} in their Figure 5. Na I D lines can be seen toward the Milky Way, but no Na I D lines toward the host galaxy. A higher Milky-Way reddening of $E(B-V)$=$0.23\pm0.16$~mag was obtained based on the equivalent width of the Na I D lines \citep{Burns20}. This may further suggest that SN2017cbv suffers some Milky Way reddening, but negligible host reddening. \citet{Ferretti17} also published five high-resolution spectra of SN~2017cbv with the Ultraviolet and Visual Echelle Spectrograph \citep[UVES; ][] {Dekker00} and found low values of Equivalent Width for Na I (D1 and D2) lines, consistent with negligible host reddening. 

\ \par
\subsubsection{Comparison to other SNe Ia}
\begin{figure}[htb]
\begin{center}
\includegraphics[width=0.9\textwidth]{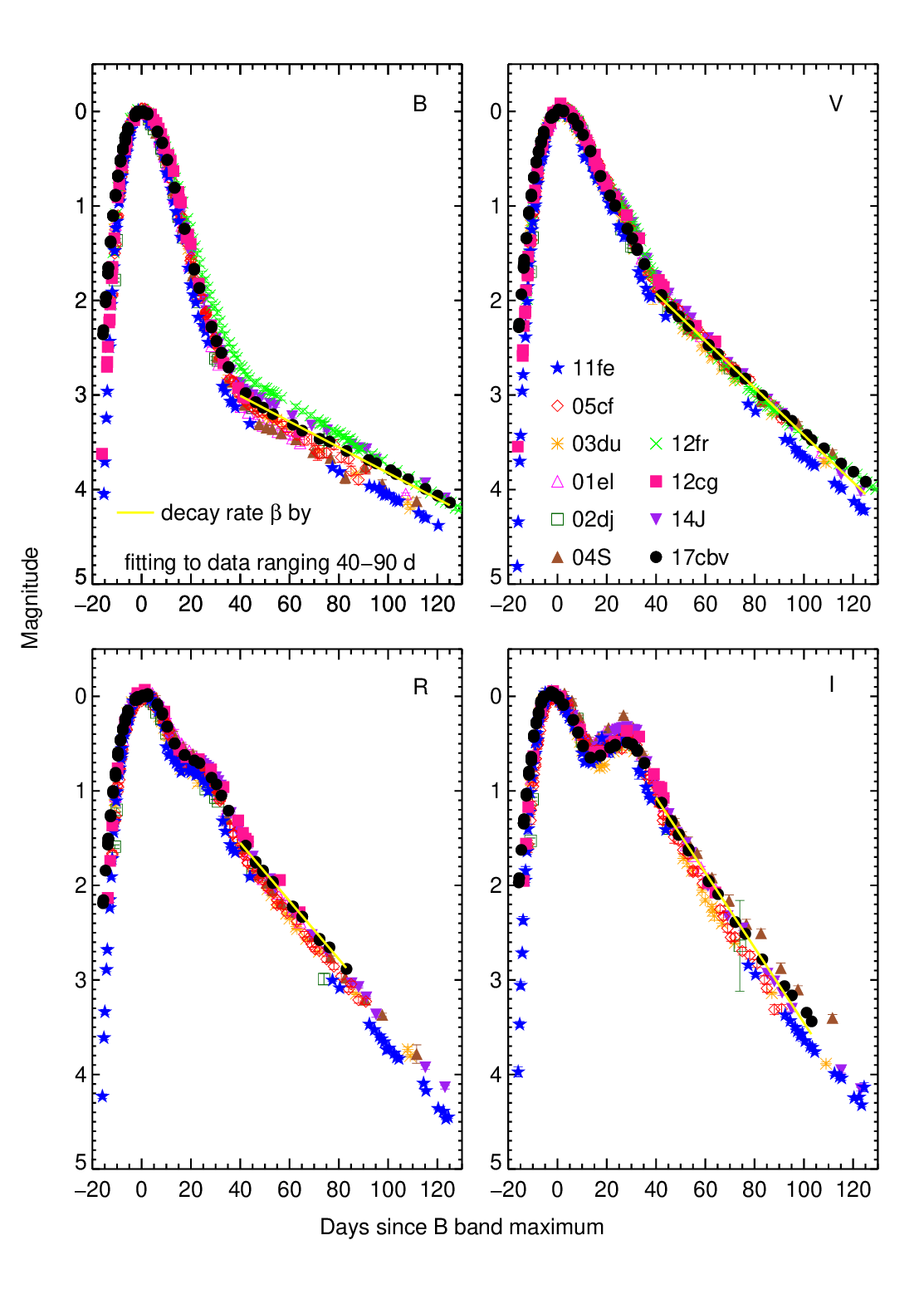}
\caption{Comparison of $BVRI$-band light curves of SN~2017cbv with other well-observed SNe Ia: SNe~2001el \citep{Krisciunas03}, 2002dj \citep{Pignata08}, 2003du \citep{Stanishev07}, 2004S \citep{Krisciunas07}, 2005cf \citep{Wang09b}, 2011fe \citep{Pereira13,Zhang16}, 2012cg \citep{Marion16}, 2012fr \citep{Zhang14,Contreras18}, and 2014J \citep{Foley14,Marion15,Srivastav16,Li19a}. The yellow solid lines mark the decay rate $\beta$ during the interval $t=40-90$ days after $B$ maximum. \label{fig:lcopt}}
\end{center}
\end{figure}

Figure~\ref{fig:lcopt} shows comparisons of the optical light curves of SN~2017cbv with those of well-observed normal SNe Ia, i.e., SN~2001el \citep[$\Delta m_{15}=1.15 $ mag;][]{Krisciunas03}, SN~2002dj \citep[$\Delta m_{15}=1.08$ mag; ][]{Pignata08}, SN~2003du \citep[$\Delta m_{15}=1.02$ mag; ][]{Stanishev07}, SN~2004S \citep[$\Delta m_{15}=1.10$ mag; ][]{Krisciunas07}, SN~2005cf \citep[$\Delta m_{15}=1.07$ mag; ][]{Wang09b}, SN~2011fe \citep[$\Delta m_{15}=1.18$ mag; ][]{Zhang16}, SN~2012cg \citep[$\Delta m_{15}=0.86$ mag; ][]{Marion16}, SN~2012fr \citep[$\Delta m_{15}=0.82$ mag; ][]{Zhang14,Contreras18}, and SN~2014J \citep[$\Delta m_{15}=1.08$ mag; ][]{Foley14,Marion15,Srivastav16,Li19a}. 
The comparison sample have included all available normal SNe~Ia which have been well observed in optical/NIR bands and have similar light curve shapes as SN~2017cbv. It can be seen that the near-maximum-light curves of SN~2017cbv are very similar to the comparison sample. 
 The late-time decay rate during the interval $t=40-90$ days after the peak denoted as $\beta$ here was estimated for the $BVRIYJHK_s$ bands \footnote{https://idlastro.gsfc.nasa.gov/ftp/pro/robust/robust\_linefit.pro}  and the corresponding values were listed in Table~\ref{tab:beta}. The $BVRI$-band late-time decay rates $\beta$ of SN~2017cbv appear relatively slower than or similar to those of corresponding rates of the comparison sample. More details can be seen in Table~\ref{tab:beta} and Figure \ref{fig:lcopt}. During 20 - 90 days after $B$ maximum, the $B$ magnitude of SN~2017cbv falls in between that of SN~2012fr and SN~2011fe, this suggests that these SNe~Ia have different $^{56}$Co hard-gamma ray escaping ratios from the ejecta. %If all the radioactive $^{56}$Co energy was trapped and re-radiated by the ejecta, one would expect all of the SNe Ia to have the same decay rate (the decay of $^{56}$Co) on that phase \citep{Branch92}.
%except SN~2011fe, SN~2014J in $B$ band, and SN~2004S in $I$ band.
% We note that the well-observed SN~2012fr displayed spectroscopic properties similar to SN~2000cx with a long-lasting high-velocity Si II $\lambda 6355$ feature \citep{Li01b} and a normal luminosity \citep{Contreras18}.
%The $\beta$ for comparison sample were also calculated with the same method \footnote{https://idlastro.gsfc.nasa.gov/ftp/pro/robust/robust\_linefit.pro} as the supernova if there are at least 3 data points during $t=40-90$ days. The corresponding $\beta$ values for the comparison sample are also listed in Table~\ref{tab:beta}.

\ \par
\begin{figure}[h]
\begin{center}
\includegraphics[width=0.9\textwidth]{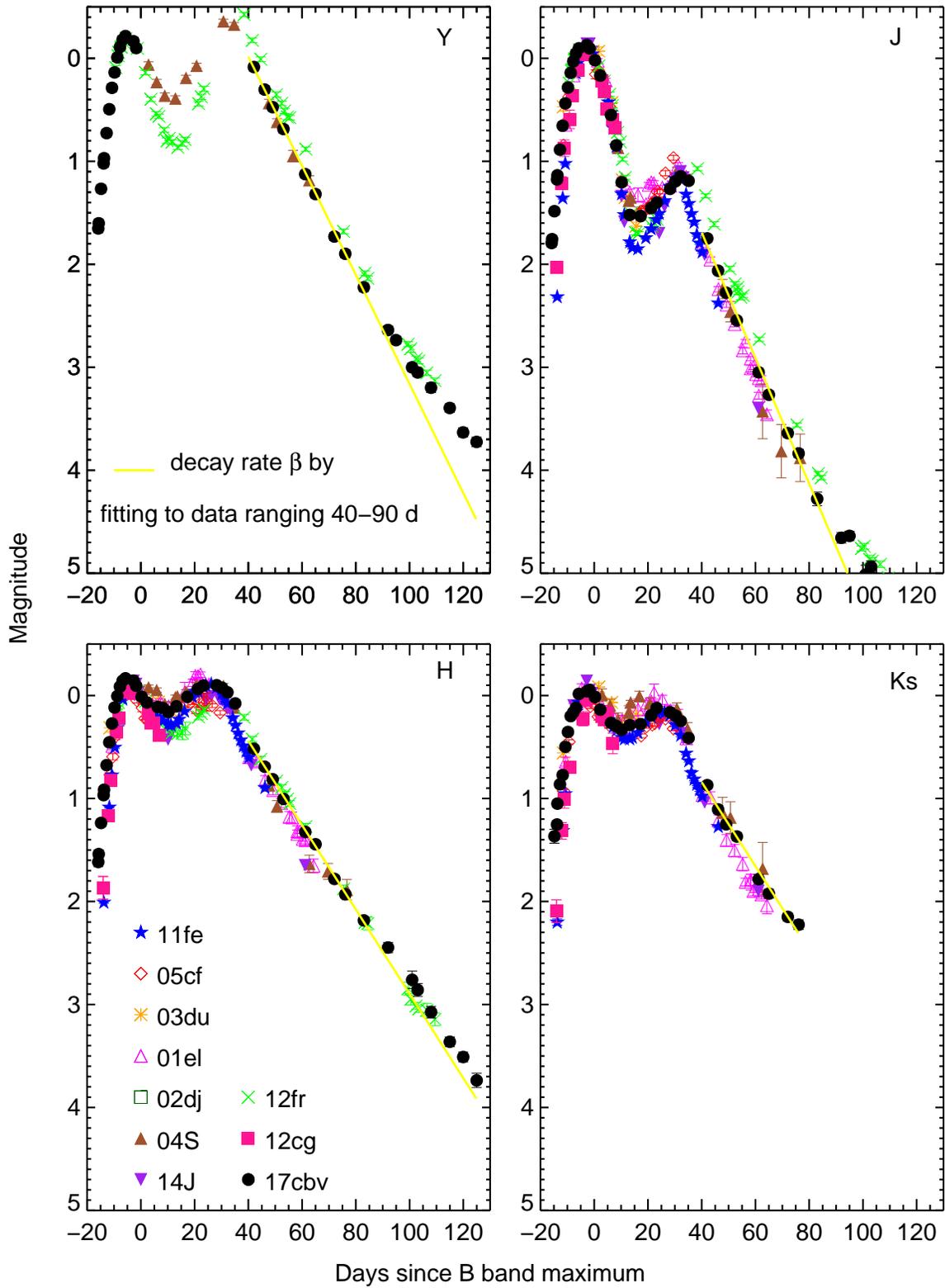}
\caption{Same as Figure ~\ref{fig:lcopt}, but for $YJHK_s$-band light curves. The late-time decay rates $\beta$ of SN~2017cbv are also overplotted in the yellow solid lines. \label{fig:lcnir}}
\end{center}
\end{figure}

In Figure~\ref{fig:lcnir}, the $YJHK_s$-band light curves of SN~2017cbv are compared with those of SNe~2001el, 2002dj, 2003du, 2004S, 2005cf, 2011fe, 2012cg, 2012fr and 2014J. The overall light curves of SN~2017cbv in the $YJHK_s$ resemble the comparison SNe in Figure~\ref{fig:lcnir} and Table~\ref{tab:beta}. One can see that their secondary maximum features show some differences and these variations might be related to the progenitor metallicity, the concentration of iron-group elements and the abundance stratification in SNe Ia \citep{Kasen06a}. After $t>90$~days the light curve decay slope in NIR bands become less steep, this could be influenced by the Co II, like MIR Co II $\lambda~10.5\mu$m time-series variation, flattening out at about day 90-100 \citep{Telesco15}. 

\ \par

\begin{deluxetable}{lcccc}[h]
\tablewidth{0pt}
%\tablenum{7}
\tablecaption{The decay rate $\beta$ of the comparison sample in $BVRI$ bands. \label{tab:beta}}
\tablehead{\colhead{Name }   & \colhead{$\beta_B^a$} & \colhead{$\beta_V$} & \colhead{$\beta_R$} & \colhead{$\beta_I$} \\
           & [mag(100 days)$^{-1}$]&  [mag(100 days)$^{-1}$]& [mag(100 days)$^{-1}$]& [mag(100 days)$^{-1}$] }
\startdata
%SN~2017cbv&  $1.422\pm0.058$ & $2.544\pm0.034$ & $3.114\pm0.035$ & $4.011\pm0.049$\\
SN~2017cbv &  $1.442\pm0.057$ & $2.555\pm0.033$ & $3.139\pm0.037$ & $4.029\pm0.049$\\
SN~2005cf & $1.663\pm0.050$ & $2.639\pm0.036$ & $3.215\pm0.035$ & $4.381\pm0.061$\\
SN~2011fe &$1.397\pm0.004$ & $2.762\pm0.005$ & $3.249\pm0.023$ & $4.236\pm0.047$\\
SN~2012fr &$1.639\pm0.015$ & $2.683\pm0.031$ & ... & ...\\
SN~2014J & $1.334\pm0.024$ & $2.852\pm0.042$ & $3.355\pm0.064$ & $4.248\pm0.099$\\
SN~2001el& $1.493\pm0.097$ & $2.638\pm0.072$ & $3.221\pm0.094$ & $4.244\pm0.059$\\
SN~2004S & $1.608\pm0.114$ & $2.704\pm0.107$ & $3.185\pm0.094$ & $3.773\pm0.064$\\
SN~2003du& $1.707\pm0.033$ & $2.626\pm0.043$ & $3.094\pm0.066$ & $4.577\pm0.127$\\
\hline
           &   &  &  &\\
           & \colhead{$\beta_Y^a$} & \colhead{$\beta_J$} & \colhead{$\beta_H$} & \colhead{$\beta_{K_s}$} \\
%           & [mag(100 days)$^{-1}$]&  [mag(100 days)$^{-1}$]& [mag(100 days)$^{-1}$]& [mag(100 days)$^{-1}$]\\
           &   &   &  & \\
%\hline
SN~2017cbv &$5.288\pm0.053$&$6.033\pm0.121$& $4.092\pm0.046$& $4.034\pm0.173$\\
SN~2012fr  &$5.345\pm0.041$ &$6.167\pm0.110$ &$4.133\pm0.039$ & ...
\enddata
\tablecomments{$^a$ The late-time decay rate $\beta$ of the light curve during the interval $t=40-90$ days relative to $t_B^{max}$.}
\end{deluxetable}

\clearpage
\subsubsection{Color Curves}
\begin{figure}[ht]
\begin{center}
\includegraphics[width=0.9\textwidth]{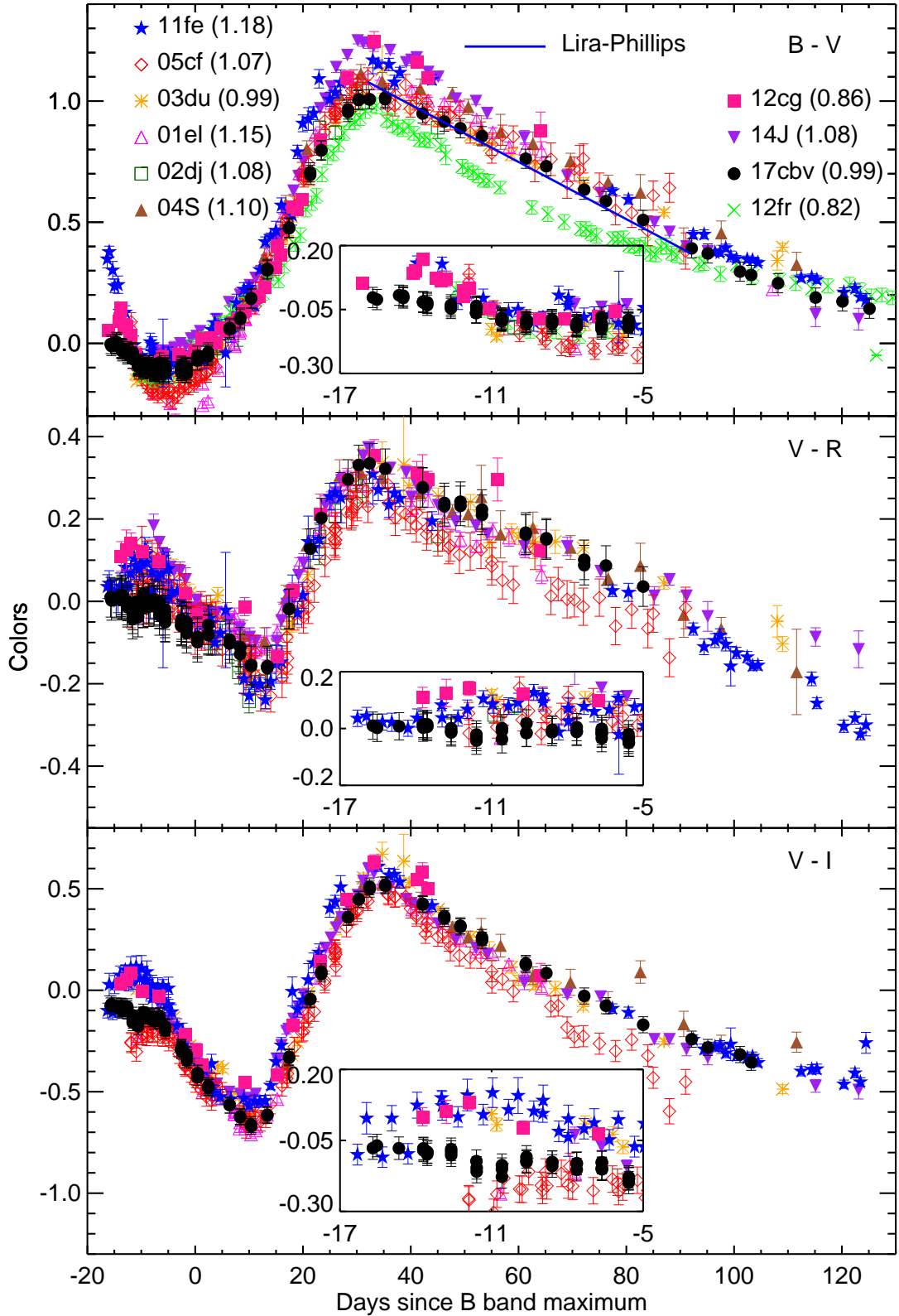}
\caption{$B-V$, $V-R$, and $V-I$ color curves of SN~2017cbv, together with those of SNe~2001el, 2002dj, 2003du, 2004S, 2005cf, 2011fe, 2012cg, 2012fr and 2014J. All of 
the comparison sample have been dereddened. Only the Milky Way extinction for SN~2017cbv was corrected. The blue solid line in $B-V$ panel displays the unreddened Lira-Phillips loci. The data sources are cited in the text, see \S 3.1. The inner panel is a zoom in of the very early phase color curves. The corresponding shape parameter $\Delta m_{15}$ of each supernova is listed in parentheses behind the SN name. \label{fig:copt}}
\end{center}
\end{figure}

\begin{figure}[h]
\begin{center}
\includegraphics[width=0.9\textwidth]{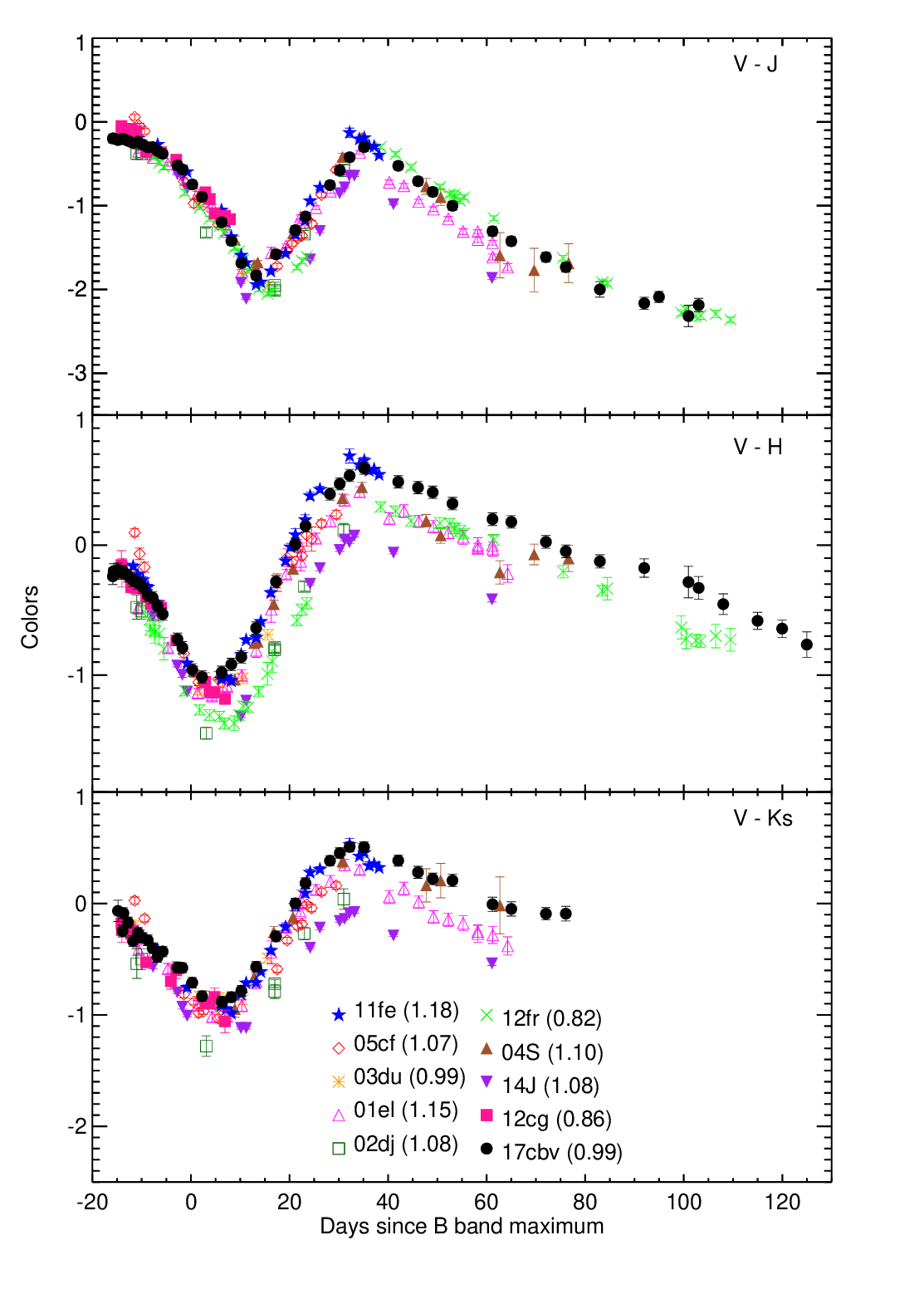}
\caption{$V-JHK_s$ color curves of SN~2017cbv compared with SNe~2001el, 2002dj, 2003du, 2004S, 2005cf, 2012cg, 2011fe, 2012cg, 2012fr, and 2014J. The corresponding shape parameter $\Delta m_{15}$ of each supernova is listed in parentheses behind the SN name.  \label{fig:cvjhk}}
\end{center}
\end{figure}

\begin{figure}[h]
\begin{center}
\includegraphics[width=0.9\textwidth]{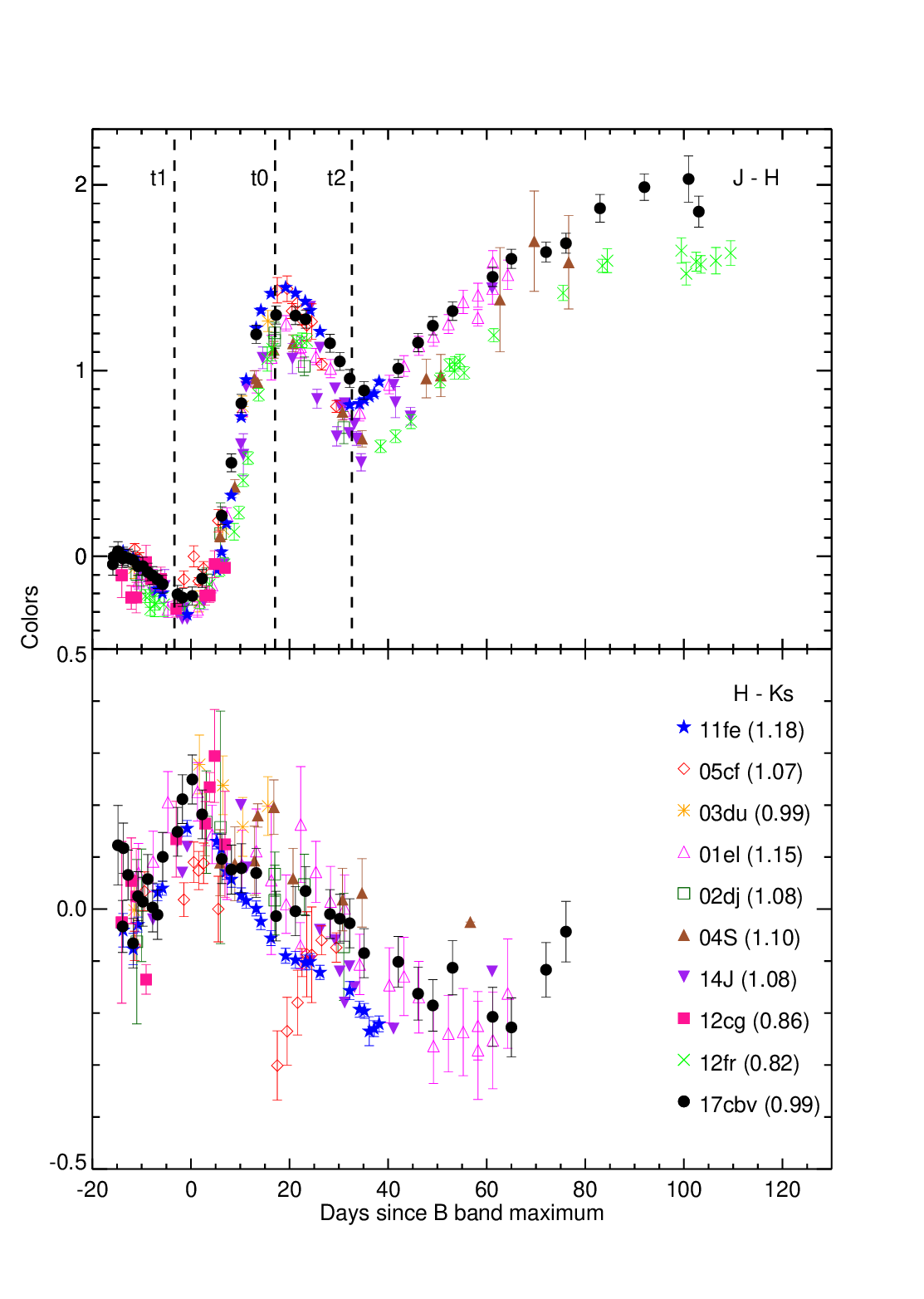}
\caption{$J-H$ and $H-K_s$ color curves of SN~2017cbv, together with the comparison sample SNe 2001el, 2002dj, 2003du, 2004S, 2005cf, 2011fe, 2012cg, 2012fr, and 2014J. The phases t1, t0, and t2 in $J$ band are overplotted in dashed lines in top panel. The corresponding shape parameter $\Delta m_{15}$ of each supernova is listed in parentheses behind the SN name. \label{fig:cjhk}}
\end{center}
\end{figure}

 %It has the bluest color among all the supernovae under comparison.
A comparison of several well-observed SNe is shown in Figures~\ref{fig:copt},~\ref{fig:cvjhk},~\ref{fig:cjhk}. All photometry has been corrected for reddening in the Galaxy and the host galaxies by the values from the corresponding published papers, except for SN~2017cbv for which only the Galactic extinction was corrected. The optical color curves of SN~2017cbv and comparison sample ($B - V$, $V - R$, and $V - I$) are presented in Figure~\ref{fig:copt}. At very early phases $t< -10$~days with respect to $B$-band maximum, the colors of SN~2017cbv are much bluer than the comparison SNe \citep[also see ][]{Hosseinzadeh17,Stritzinger18,Bulla20}. The $B - V$ color of SN~2017cbv stays flat at about $-0.05$~mag soon after explosion and then slowly becomes bluer until day $-5$. Also, it is the bluest SN in colors of $B-V$, $V-R$, and $V-I$ until day $-$11. The $V - R$ color is even bluest among all the comparison SNe~Ia until 10 days after $B$ maximum. The blue colors seen in the early light curves of some SNe~Ia have been interpreted as interactions between SN ejecta and a companion star, serving as evidence in favor of the single degenerate scenario \citep{Brown12,Marion16,Hosseinzadeh17,Dimitriadis19a}. After $B$ maximum light, the color evolution of SN~2017cbv matches well with the comparison sample and at about 30 days past maximum the $Lira-Phillips$ relation \citep[blue solid line; ][]{Phillips99}. For a nearby SNe~Ia sample in the $Lira~law$ regime, \citet{Forster13} claimed that the $B-V$ slope $-0.013~\rm mag~day ^{-1}$ can be used to classify the faster decliners and slower decliners. Faster decliners ($<-0.013 \rm mag~day^{-1}$) have higher equivalent widths (EW) of Na I D lines, redder colors, and lower $R_V$ reddening law at maximum light, suggesting the presence of circumstellar material \citep{Forster13,Wang09a,Wang13,Wang19}, while slower decliners are the opposite. The slope of SN~2017cbv in the Lira phase is $-0.010~\rm mag~day^{-1}$ and thus it is a slower decliner. This is consistent with the very small EW measurements of Na I absorption lines for SN~2017cbv \citep{Ferretti17} and blue $B-V$ color at maximum light. 
\ \par

We also compare the $V-$ NIR color evolution ($V - J$, $V - H$ and $V - K_s$) of SN~2017cbv and the comparison SNe~Ia in Figure~\ref{fig:cvjhk}. We can see that the $V-$ NIR color evolution of SN~2017cbv matches with the comparison sample, especially the well-observed SN~2011fe. The study by \citet{Burns14} derived empirical relations of intrinsic colors $V-J$ and $V-H$ at maximum light relative to light-curve shape parameter $s_{BV}$ in their Table 2. According to the relations for their low reddening sample (LRS), and assuming a $s_{BV}=1.11$ for SN~2017cbv \citep{Burns20}, we obtained intrinsic colors $(V-J)_{\rm host}^{max}=-0.61\pm0.08$~mag and $(V-H)_{\rm host}^{max}=-0.85\pm0.09$~mag. Meanwhile, Table~\ref{tab:shape} gives the observed colors $(V-J)_{\rm host}^{max}=-0.60\pm0.02$~mag and $(V-H)_{\rm host}^{max}=-0.80\pm0.02$~mag at maximum epochs. Thus we can derive $E(V-J)_{\rm host} ^{max}= 0.01\pm0.08$~mag and $E(V-H)_{\rm host}^{max} = 0.05\pm0.09$~mag. This indicates SN~2017cbv suffers neglected host reddening.

%Overplotted are the unreddened $V - $ NIR loci for SNe Ia of middle decliners (with $\Delta m_{15} = 1.0-1.3$ mag) in dashed lines and slow decliners (with $\Delta m_{15} = 0.8-1.0$ mag) in solid lines, respectively \citep{Krisciunas04b}. The $V - J$ loci for slow decliners in the top plot of Figure~\ref{fig:cvjhk} shows the range of uniformity at [$-8$ d, 9.5 d] relative to $B$ maximum in solid line \citep{Krisciunas04b}. Thus $E(V-J)_{\rm host} = 0.116\pm 0.062$~mag was measured for SN 2017cbv. {\bf No $V-J$ loci is available for middle decliners from Table 14 of \citet{Krisciunas04b}.} The dashed lines in the bottom two plots are the loci for middle decliners at [$-9$, 27] days relative to $B$ maximum, while they are solid lines for slow decliners at [$-$8, 27] days \citep{Krisciunas00,Krisciunas03,Krisciunas04b}. Relative to the middle decliners loci, we estimated $E(V-H)_{\rm host} = -0.075\pm0.107$~mag  and $E(V-K_s)_{\rm host} = -0.040\pm0.066$~mag. Relative to the loci of slow decliners, we estimated  $E(V-H)_{\rm host} = 0.286\pm0.060$~mag  and $E(V-K_s)_{\rm host} = 0.233\pm0.106$~mag. SN~2017cbv could be classified as a middle decliner or a slow decliner as its $\Delta m_{15}$ locates at the boundary between the two catalog. {\bf SN~2017cbv tends to be a middle decliner when comparing with the above host reddening estimates with that derived from CMAGIC diagram in S3.2, Phillips intrinsic color and Lira-Phillips relation in \S3.3. } But in any case, it is a normal type Ia based on its color evolution and light curves. 

$J - H$ and $H - K_s$ color evolutions of SN~2017cbv and the comparison SNe~Ia are shown in Figure~\ref{fig:cjhk}. Overplotted are the phase of the first maximum t1$=-3.6$~days, the phase of the secondary maximum t2$=31.7$~days, and the phase of the minimum between the two maximums t0$=16.7$~days in $J$ band relative to $B$-band maximum, as shown by the vertical dashed lines. 
As seen from the top panel of Figure~\ref{fig:cjhk}, $J - H$ color illustrates a pronounced evolution after t1. 
The flux in $J$ band decreases obviously with respect to the $H$ band shortly after t1, probably due to the lack of emission features around 1.2 $\mu$m \citep{Spyromilio94,Hoflich95b,Wheeler98}. This trend holds until t2 when the Fe II $\lambda$ 1.25 $\mu$m emission line forms. Then the $J-H$ color becomes redder again as a result of a faster decline rate in the $J$ band. The bottom panel of Figure~\ref{fig:cjhk} displays the $H - K_s$ plot and SN~2017cbv matches well with the comparison SNe.

%The dashed lines represent the mean color curve for those with $\Delta m_{15} = 1.0-1.3 $ mag; while the solid lines denote the mean loci of the unreddened SNe Ia with $\Delta m_{15} = 0.8-1.0 $ mag \citep{Krisciunas00,Krisciunas03,Krisciunas04b}. No $V-J$ loci is available for middle decliners from \citet{Krisciunas04b}.

%\clearpage
\subsection{Color Magnitude Diagram}
SNe~Ia are assumed to be standard distance candles after a one or two parameter (light-curve shape/color) correction. If they form a one or two parameter group, it is possible to derive distance measurements from their multi-color light curves. \citet{Wang03} studied the color-magnitude relation of SNe~Ia during the first month past maximum and found a linear relation between $B$ and $B-V$ color in SNe~Ia. This linear relation provides distance determinations and dust extinction estimates simultaneously. The color-magnitude intercept calibration (CMAGIC) method provides a tool to obtain accurate distance calibration without data around optical maximum and suggests new observational strategies to estimate accurate distances \citep{Wang03,Conley2006ApJ...644....1C,Wanglifan06,He18}.   

The color-magnitude diagrams of SN~2017cbv are shown in Figure~\ref{fig:cmagic} together with those of SN~2011fe. For the first, second and fourth rows, the first column shows the diagram for the observed magnitudes, the middle column shows the diagram after correction for Milky Way extinction, and the third column shows the color-magnitude diagram after reddening corrections based on various assumptions. Note that on the color-magnitude  diagram (CMD) the two supernovae show genuine differences as indicated by the differences in their CMD shapes. With the high quality NIR data on both SNe, we can use the CMDs to estimate the extinction to these SNe. However, some assumptions need to be made to allow this. In the original CMAGIC construction, \citet{Wang03} used the linear region of the CMD, but there are more features that can be employed by the extensive data on these SNe. Examples are the bluest and the reddest colors on the CMD. The CMDs of $B~vs.~B-K_s$ and $V~vs.~V-K_s$ show a characteristic upside down \enquote{\&} shaped curve. Various aspects of this shape can be used to analyze the properties of the SNe.
 
\subsubsection{$B-V$ color}

 Figure~\ref{fig:cmagic}(a) show the color--magnitude diagram presented by \citet{Wang03,Wanglifan06}. In order to estimate the CMAGIC color excess $E_{BV}(B-V)$ we use the quantity as defined in \citet{Wang03}:
\begin{equation}
    \epsilon(B-V)=(B_{max}-B_{BV})/\beta_{BV}
\end{equation}
where $B_{max}$ are $B$-band maxima, and $\beta_{BV}$ and $B_{BV}$ denote the slope and the value for the intercept at $(B-V)=0$ for the linear region from 5 to 27 days after the $B$-band maximum with the CMAGIC relation \citep{Wang03}:
\begin{equation}
    B=B_{BV}+\beta_{BV}(B-V)
\end{equation}
The CMAGIC color excess for a reddening-free SN $\epsilon_0$ also depends on light curve shape parameter $\Delta m_{15}(B)$ and the linear relationship between them was derived from a low-extinction sample of SNe~Ia \citep{Phillips99,Wang03}:
\begin{equation}
    \epsilon_{0}=(-0.118\pm0.013) + (0.249\pm0.043)(\Delta m_{15}(B)-1.1)
\end{equation}
And the final color excesses of SNe~Ia can be measured based on the equation:
\begin{equation}
    E_{BV}(B-V)=\epsilon(B-V)-\epsilon_0 
\end{equation}

%{\bf LFW: check the question marks}
We fit the $B$ magnitude and $B-V$ color of SNe~2017cbv and 2011fe  between 5 and 27 days and we derived $E_{BV}(B-V)_{11fe}=0.030\pm0.044$~mag, and  $E_{BV}(B-V)_{17cbv}=0.173\pm0.029$~mag, which includes both the contribution from the Milky Way and the SN's host galaxy. We can take SN~2011fe as a reference with no dust extinctions from the Milky Way and its host galaxy \citep{Nugent11,Johansson13,Patat13}. SN~2017cbv has negligible host extinction $E(B-V)_{\rm host}=0.011\pm0.029$~mag after considering the Milky Way extinction toward SN~2017cbv \citep{SF11}. 

\ \par
The distance of the two supernovae can be also derived based on the CMAGIC method by comparing the observed $B_{BV}$ and the absolute value $M_{BV}^{B}$ from empirical relation between the absolute magnitude $M_{BV}^{B}$ $vs.$ $\Delta m_{15}$ in Figure 10 of \citet{Wang03}. Table 3 of \citet{Wang03} gives the fit result to the absolute magnitude $M_{BV}^{B}$ $vs.$ $\Delta m_{15}$ relation for $R_B=3.3$ expressed as
\begin{equation}
    M_{BV}^{B}=(-19.35\pm0.02) + (0.60\pm0.08)(\Delta m_{15}(B)-1.1)
\end{equation}
And the extinction correction $A_{BV}$ is given by Equation 3a of \citet{Wang03} expressed as
\begin{equation}
    A_{BV}=(R_B - \beta_{BV}) E(B-V)
\end{equation}
where $R_B=3.3$, $\beta_{BV}=2.250\pm0.030$, and $E(B-V)_{17cbv}(host)=0.011\pm0.029$~mag. The observable value $B_{BV}$ is the color-magnitude intercept, which is calculated from the intercept of the typical color $B-V=0.6$~mag reported by \citep{Wang03}. To minimize the covariance of the distance estimates to the slope $\beta_{BV}$, \citet{Wang03} defined the following equation to measure $B_{BV}$ as the standard color-magnitude intercept:  
\begin{equation}
    B_{BV}=B_{BV0.6}-1.164
\end{equation}

Equation (14) yields $B_{BV}$ = $11.648\pm0.030$~mag for SN~2017cbv. Thus we obtained the distance modulus $\mu=30.58\pm0.05$~mag ($D=13.1\pm0.3$~Mpc) after applying $M_{BV}^B$, $A_{BV}$, and $B_{BV}$ into equation (17) in \S3.4.1. Based on the same method, we estimated the distance modulus $\mu=29.14\pm0.10$~mag for SN~2011fe, which is consistent with the Cepheid distance modulus of SN~2011fe \citep{Shappee11}.

\subsubsection{$B-K_s$ Color}
\citet{Wang03} also found $B$ magnitudes and the various colors (i.e., $B-R$, $B-I$) are linearly related. Thanks to our well-observed optical and NIR photometric observations, we found that the $B$ magnitude $vs.~B-K_s$ color, and $V$ magnitude $vs.~V-K_s$ color are also linearly related in the phase ranges of $ 5 \le t \le 30$~days and $ t \le -5$~days. The two linear relations have a intersection point for SN~2017cbv (red dashed lines), and for SN~2011fe (black dashed lines), as shown in panels (d) and (j) of Figure~\ref{fig:cmagic}. If we assume that SN~2017cbv and SN~2011fe are intrinsically the same, matching their corresponding colors should give us the comparable interstellar dust extinctions. There is no first principle physical insights to help us to determine which points on the CMDs are  most representative of the intrinsic color of the SN~Ia population. To explore the various possibilities, 
we tested the color excesses $E(B-K_s)$ and $E(V-K_s)$ by matching the reddest color, bluest color, and intersection point between SN~2017cbv and SN~2011fe.

We plot $B~vs.~B-K_s$ diagram in panels (d) to (i) of Figure~\ref{fig:cmagic}. The panels (d), (e), (f) are the observations without extinction corrections of the Milky Way and the host, only corrected of the Milky Way, and corrected of the Milky Way and its host by matching with intersection point of two dashed lines (black for SN~2011fe and red for SN~2017cbv). One line is the linear fit to phase interval $ 5 \le t \le 30$~days relative to $B$-band maximum, and the other is the linear fit to phase range $ t \le -5$ days. Thus we obtained $E(B-K_s)_{\rm host}=-0.05\pm0.05$~mag in panel (f).

Figure~\ref{fig:cmagic}(g) is the same as Figure~\ref{fig:cmagic}(d), except that the solid lines are the interpolation values to $B$ and $K_s$ in the phase range of $-15 \le t \le 60$ days after $B$-band maximum. When matching the bluest color of SN~2017cbv with that of SN~2011fe in panel (h), we obtained $E(B-K_s)_{\rm host}=-0.13\pm0.05$~mag. While matching the reddest color of SN 2017cbv with that of SN 2011fe in panel (i), we obtained $E(B-K_s)_{\rm host}=-0.27\pm0.05$~mag.

Based on these four different assumptions on the uniformity of the intrinsic color, we obtained four different values of $E(B-K_s)$. Among them, matching the reddest color yields the most negative values of $E(B-K_s)\ = \ -0.27\pm0.05$ mag. This value is inconsistent with the CMAGIC estimates. It would imply a significant over correction of the Milky Way reddening if the reddest color is used as the reference point for extinction correction. It is more likely that the difference at the reddest color is intrinsic to the SNe and contrary to what has been employed in the Lira-Phillips relation, the late time color cannot be used as reliable estimates of extinction, at least for these two very well observed SNe.

\subsubsection{$V-K_s$ Color}
\begin{figure}[h]
\begin{center}
\includegraphics[width=0.9\textwidth]{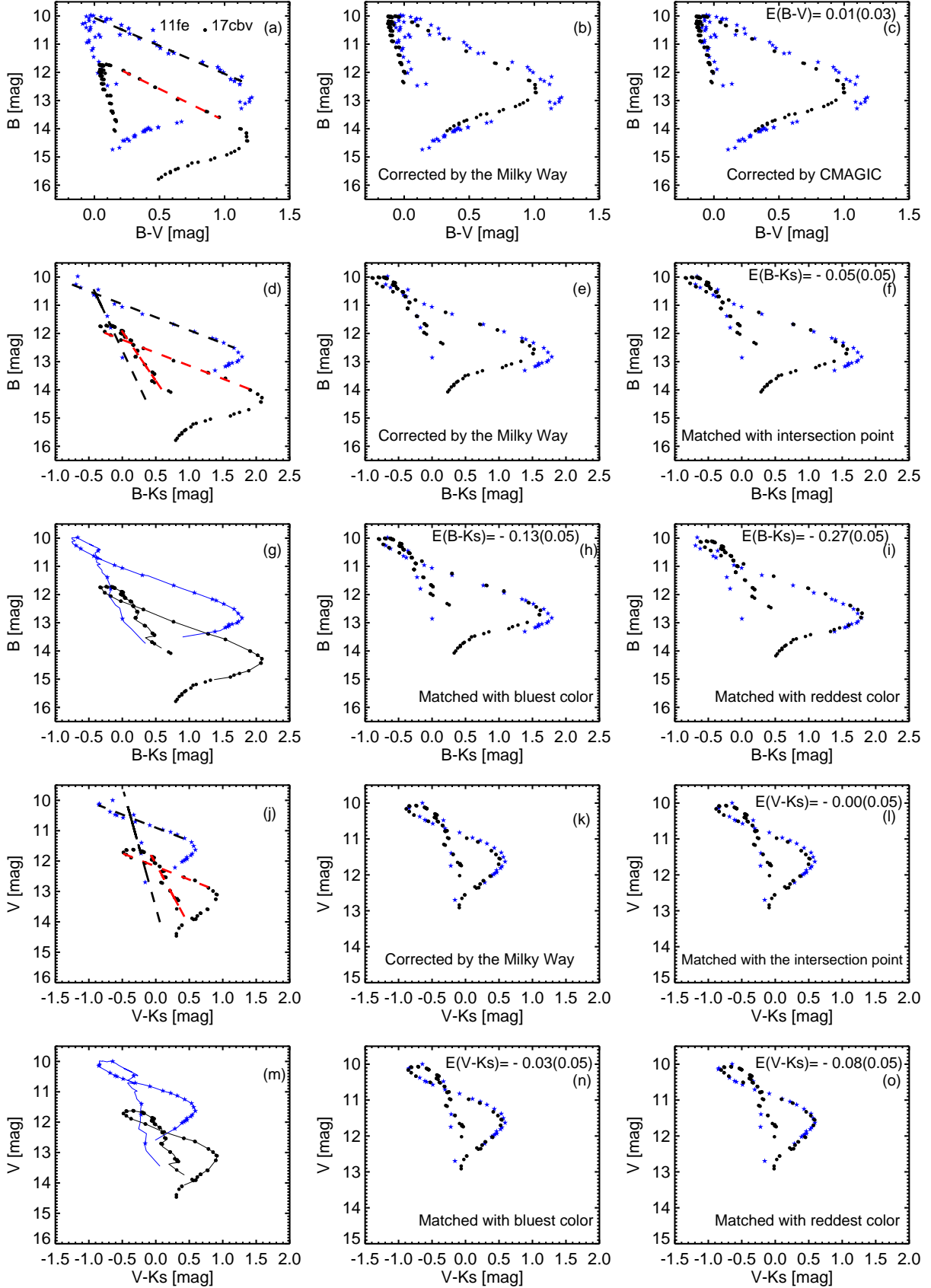}
\caption{Color-magnitude plots of SNe~2017cbv and 2011fe for $B-V$, $B-K_s$, and $V-K_s$ colors, respectively from top to bottom panels. Overplots are (CMAGIC) linear fitting in dashed lines and the interpolation to $BVK_s$ bands of SNe~2017cbv and 2011fe in solid lines. Refer to the text for details.
\label{fig:cmagic}}
\end{center}
\end{figure}

Similar to the previous section, we also plot $V~vs.~V-K_s$ in panels (j) to (o) of Figure ~\ref{fig:cmagic}. By matching with the intersection point of the two lines for SNe~2017cbv and 2011fe, we obtained $E(V-K_s)_{\rm host}=-0.00\pm0.05$~mag in panel (l). By matching with the bluest and reddest colors, we obtained  $E(V-K_s)_{\rm host}=-0.03\pm0.05$~mag in panel (n), and $E(V-K_s)_{\rm host}=-0.08\pm0.05$~mag in panel (o), respectively.

\ \par

We note further that the intrinsic colors $B-K_s$ and $V-K_s$ of SN~2017cbv are bluer than the expected value of SN~2011fe when matching them with the intersection point, the bluest color, and the reddest color: panels (f) and (l), (h) and (n), (i) and (o) in Figure~\ref{fig:cmagic}, respectively. We assumed no host extinction for SN 2011fe based on the reddening analysis and the distance determination on our CMAGIC diagram (in \S 3.2.1) and other work by \citet{Nugent11,Johansson13,Patat13}. 

We also note that the $B-V$ color of SN~2017cbv in panel (c) is bluer than SN~2011fe by $0.18\pm0.07$ mag around $t\sim 32$~days, at this phase, the two SNe have the reddest color. This suggests that SN~2017cbv and SN~2011fe are different in $B$ and/or $V$ bands at some late phases. When we matched the reddest colors $B-K_s$ of SN~2017cbv with SN~2011fe, the estimated color excess $E(B-Ks)$ of host in the panel (i) is different from the early-phase estimates (matching with the bluest color or intersection point) by 2-3 $\sigma$ in  panels (f) and (h). In contrast, the color excess $E(V-K_s)$ of the two SNe by matching their reddest color in panel (o) is comparable with the measurements by matching the bluest color in panel (n) and intersection point in panel (l). This may further suggest that the $B$ magnitudes between SN~2017cbv and SN~2011fe are more different than $V$ magnitudes at these late phases, although $B$ and $V$ magnitudes of SN~2017cbv are brighter than SN~2011fe after $B$-band maximum in Figure~\ref{fig:lcopt}, where their maximum are matched. 
\clearpage

%%%%%%%%%%%host extinction%%%%%%%%%%%%%%%%%%%%%%%%%%%%%%%%%
%\subsection{Host Reddening $E(B-V)$ and Distance of SN~2017cbv}
\subsection{More on Host Reddening $E(B-V)$}
In addition to the insights derived from the CMD, there are several other  mature methods for deriving the host galaxy reddening. 
\subsubsection{Phillips Intrinsic Color}
 
 \citet{Phillips99} compiled a set of unobscured type Ia SNe and derived a relation between their intrinsic pseudo color $(B_{max}-V_{max})_0$ and their 
 decay parameter $\Delta m_{15}(B)$:

 \begin{equation}
 (B_{max}-V_{max})_0=-0.070(\pm0.012)+0.114(\pm0.037)\times(\Delta m_{15}(B)-1.1)
\end{equation}

This relation allows to estimate the host reddening suffered by any normal SN~Ia
by just comparing their measured pseudo color and their intrinsic estimate.
SN~2017cbv, after correcting by the Milky Way reddening, shows a pseudo color $(B_{max}-V_{max})$
that is even bluer than the expected value from \citet{Phillips99}'s relation, or $E(B-V)_{\rm host}=-0.006\pm0.016$~mag. This value correspond to $E(B-K_s)$ of about $-0.019$~mag, which is inconsistent with the value estimated by matching the late time CMD in $B$ vs $B-K_s$, suggesting again that the late time colors of SN~Ia can be substantially different. 
%For that reason, we estimate a zero host reddening from this method.

\subsubsection{Lira-Phillips Relation}
The top panel of Figure~\ref{fig:copt} showed the $B-V$ color evolution curve and the unreddened Lira-Phillips loci was overplotted with a blue solid line. The following relation was derived to describe the intrinsic $B-V$ color evolution in the phase interval $30\le t_V \le 90$d \citep{Lira96,Phillips99}. 

\begin{equation}
    (B-V)_0=0.725-0.0118(t_V - 60)
\end{equation}
Applying the relation to SN 2017cbv, we obtain host extinction $E(B-V)_{\rm host}=0.000\pm0.037$~mag. 

\subsubsection{CMAGIC Diagram}
\citet{Wang03} derived the CMAGIC relation for SNe Ia over the phase interval $5\le t_B\le 27$ days after $B$-band maximum. We applied the CMAGIC relation to SN 2017cbv in \S 3.2 and we derived the the host extinction $E(B-V)_{\rm host}=0.011\pm0.029$~mag. 

 \ \par
 
By averaging the host extinction of SN 2017cbv based on the above three methods, we obtained $E(B-V)_{\rm host}=0.002\pm0.009$~mag. The low host-galaxy reddening is consistent with the facts that the SN exploded at the outskirts of NGC~5643 and no narrow Na I D absorption lines were detected in the low resolution spectra, even in the MIKE spectrum \citep{Burns20}. \citet{Ferretti17} published five high-resolution spectra of SN~2017cbv and found values of Equivalent Width for Na I (D1 and D2) at the lower end of the empirical relation between strength of Na I D absorption $vs.$ reddening \citep{Poznanski12}, consistent with zero reddening. Thus we assume no host galaxy reddening for SN~2017cbv in our study.  

%a total equivalent widths of 13.9 m$\rm \AA$ for Na I D1 and 31.2 m$\rm \AA$ for Na I D2, respectively. This corresponds to 0.02 mag and 0.01 mag for $E(B-V)_{\rm host}$, respectively, following the empirical relation between strength of Na I D absorption and reddening \citep{Poznanski12}. Thus we assume no host galaxy reddening for SN~2017cbv in our study. 

\clearpage

\subsection{Distance of SN~2017cbv}
\subsubsection{NIR-Absolute Calibration}
The effects of extinction are considerably reduced in $JHK_s$ bands and it seems there are relatively constant peak magnitudes in NIR bands \citep{Meikle00,Krisciunas04a,Krisciunas04c,Krisciunas07}. SNe Ia have a more uniform peak luminosity in NIR bands \citep{Krisciunas04a,WoodVasey08,Folatelli10,Phillips12,Matheson12,Avelino19}. The well-sampled NIR photometry of SN~2017cbv can be used to determine the distance modulus toward NGC~5643. For each band, the apparent maximum magnitudes $m1$ and the magnitudes at $t_B^{max}$ are listed in column 3 of Table~\ref{tab:mu} and in columns 4, 2 of Table~\ref{tab:shape}. For each case, we used the following formula to derive the distance modulus $\mu$.

\begin{equation}
\mu =m-M + 5\mathrm{log}(H_0/72) - A_{Milky Way} - A_{Host} +S_{Correction} + K_{Correction}
\end{equation}
\begin{itemize}
    \item $m$ stands for the apparent magnitudes. $M$ represents the absolute NIR magnitudes from these calibration sources, \citet{Krisciunas04a,Mandel09,WoodVasey08,Folatelli10,Burns11,Kattner12} and all assumed Hubble constant $H_{0}$ = 72 $\kmsmpc$ \citep{Freedman01,Spergel07}.
      
    \item The Milky Way extinction $A_{Milky Way}$ towards to SN 2017cbv were adopted with $A_J=0.122$~mag, $A_H=0.078$~mag, $A_{K_s}=0.052$~mag \citep{SF11}. 
    \item Host extinction $A_{Host}$ of SN 2017cbv was assumed to be zero based on the analysis presented in \S 3.3.
    \item $S_{Correction}$ was applied between our $JHK_s$-band magnitudes on 2MASS system and CSP-calibrated magnitude \citep{Contreras10} for calibration sources: \citet{Folatelli10,Burns11,Kattner12}. They are $S_{Correction}(J)=0.005$~mag, $S_{Correction}(H)=-0.038$~mag, and $S_{Correction}(K_s)=0.009$~mag, which are added to CSP-calibrated magnitudes. The remaining calibration sources have been calibrated to 2MASS system \citep{Persson98} and no $S_{Correction}$ are necessary. 
    \item No $K_{Correction}$ has been applied to our photometry due to the close distance of SN 2017cbv. 
\end{itemize}

According to equation~17, the distance modulus to SN~2017cbv ranges from $30.11\pm0.11$~mag ($D=10.5\pm$0.5~Mpc) to $30.41\pm0.19$ mag ($D=12.1\pm$1.1~Mpc) which are shown in Figure~\ref{fig:mu} and listed in Table~\ref{tab:mu}. Note that the uncertainty of each case in Table~\ref{tab:mu} is dominated by the calibration of the absolute NIR peak magnitude. We note that the same absolute calibrations were applied to the $JHK_s$ magnitudes of SN~2011fe by \citet{Matheson12}, yielding a dispersion of 0.31 mag, very similar to the case of SN~2017cbv. 

%From the NIR calibration, the average distance modulus of SN~2017cbv is $\mu =30.24\pm0.09$~mag, marked by the red dashed line in Figure~\ref{fig:mu}. 
%We measured the distance modulus of SN~2017cbv in $J$, $H$, and $K_s$ bands. The average values for $J$, $H$, and $K_s$ bands are $30.24\pm0.08$~mag ($D=11.2\pm0.4$~Mpc), $30.26\pm0.12$~mag($D=11.3\pm0.6$~Mpc), and $30.21\pm0.06$~mag($D=11.0\pm0.3$~Mpc), respectively. 

\begin{deluxetable}{lllll}[ht]
\tiny
\tablewidth{0pt}
%\tablenum{5}
\tablecaption{Derived Distance Moduli $\mu$ of NGC~5643 \label{tab:mu}}
\tablehead{\colhead{Calibration Source}   & \colhead{Filter}  & \colhead{Apparent Magnitude$^{\rm a}$}&\colhead{Absolute Magnitude} & \colhead{Distance Modulus $\mu$}  \\
                                          &                   &                             &                            & \colhead{to NGC 5643 [mag] $^{\rm b}$} \\}
\startdata
\citet{Mandel09}\tablenotemark{c}       & $J$   & 12.004$\pm$0.018 & -18.25 $\pm$ 0.17 & 30.13$\pm$    0.17  \\ 
                                        & $H$   & 12.180$\pm$0.018 & -18.01 $\pm$ 0.11 & 30.11 $\pm$    0.11 \\ 
                                        & $K_s$ & 11.938$\pm$0.017 & -18.25 $\pm$ 0.19 & 30.14 $\pm$    0.19 \\ 
%\tableline
\hline
\citet{WoodVasey08}\tablenotemark{c,d} & $J$   & 12.004$\pm$0.018 & -18.29 $\pm$  0.33 & 30.17 $\pm$     0.33  \\  %2mass
                                       & $H$   & 12.180$\pm$0.018 & -18.08 $\pm$  0.15  & 30.18 $\pm$     0.15 \\ 
                                       &$K_s$  & 11.938$\pm$0.017 & -18.32 $\pm$  0.26 & 30.21 $\pm$     0.26  \\ 
\hline
\citet{Folatelli10}\tablenotemark{c}  & $J$    & 12.004$\pm$0.018 & -18.42 $\pm$  0.18 &  30.30 $\pm$     0.18 \\ 
                                      &  $H$   & 12.180$\pm$0.018 & -18.23 $\pm$ 0.19 &  30.37 $\pm$     0.19  \\ 
                                      & $K_s$  & 11.938$\pm$0.017 & -18.30 $\pm$ 0.27 &  30.18 $\pm$     0.27 \\ 

\hline
\citet{Krisciunas04a}\tablenotemark{e} & $J$  & 11.883$\pm$0.015 & -18.57 $\pm$  0.14 &30.33 $\pm$     0.14   \\ 
                                       & $H$  & 12.027$\pm$0.016 & -18.24 $\pm$  0.18 &30.19 $\pm$     0.18 \\ 
                                       &$K_s$ & 11.877$\pm$0.015 & -18.42 $\pm$  0.12 &30.24 $\pm$     0.12 \\
\hline
\citet{Folatelli10}\tablenotemark{e}  & $J$  & 11.883$\pm$0.015 &  -18.43 $\pm$  0.18 & 30.19 $\pm$     0.18 \\ 
                                      & $H$  & 12.027$\pm$0.016 & -18.42 $\pm$  0.19 & 30.41 $\pm$     0.19  \\ 
                                      &$K_s$ & 11.877$\pm$0.015 &  -18.47 $\pm$  0.27 & 30.29 $\pm$     0.27 \\
\hline 
\citet{Burns11}\tablenotemark{e}      & $J$  & 11.883$\pm$0.015 & -18.44 $\pm$  0.12 &  30.20 $\pm$     0.12  \\
                                      & $H$  & 12.027$\pm$0.016 & -18.26 $\pm$ 0.10  &  30.25 $\pm$     0.10 \\ 
\hline
\citet{Kattner12}\tablenotemark{e,f} &  $J$  & 11.883$\pm$0.015 & -18.57 $\pm$  0.14  & 30.33 $\pm$     0.14 \\ 
                                     &  $H$  & 12.027$\pm$0.016 & -18.42 $\pm$  0.14  & 30.41 $\pm$     0.14  %\\
%\tableline
% {\bf Weighted mean}                 &       &                     & {\bf 30.21 $\pm$     0.04}
\enddata

\tablenotetext{a}{The apparent magnitude in Table~\ref{tab:mu} is the same as that from Table~\ref{tab:shape}.}
\tablenotetext{b}{Distance modulus $\mu$ was derived by combining absolute calibration sources with the apparent magnitudes (see text for details). Only Milky Way extinctions toward to SN 2017cbv were corrected with $A_J=0.122$ mag, $A_H=0.078$ mag, $A_Ks=0.052$ mag \citep{SF11}. We assumed $H_{0}$ = 72 $\kmsmpc$ \citep{Freedman01,Spergel07}.}
\tablenotetext{c}{Fiducial time corresponds to $B$-band maximum brightness.}
\tablenotetext{d}{Using the PAIRITEL subsample only.}
\tablenotetext{e}{Fiducial time corresponds to the first maximum brightness in the given
  filter ($J$, $H$, or $K_s$).}
\tablenotetext{f}{Using subsample 2 of \citet{Kattner12}.}
\end{deluxetable}

\subsubsection{SNooPy Fitting}
SNooPy is a well-established light-curve fitting method to generate template light curves in the CSP natural system and to derive distances to SNe Ia \citep{Burns11}. Applying SNooPy to our $BVRIYJHK_s$-band light curve, we obtained the distance modulus of $\mu=30.46\pm0.08$~mag \citep{Burns14}, or distance $D=12.4\pm0.5$~Mpc. This distance estimate should be an independent measurement of CSP \citep{Burns20} as we have independent data. When we calibrated the optical using the CSP calibration \citep{Burns20}, we established some correlation there. 

\subsubsection{CMAGIC Diagram}

Comparing the measurement of the color-magnitude intercept parameter $B_{BV}$ to its absolute value in Table 3 of \citet{Wang03}, we obtained $\mu=30.58\pm0.05$~mag for SN~2017cbv ($D=13.1\pm0.3$~Mpc).

The color-magnitude diagrams ($B~vs.~B-V$, $B~vs.~B-K_s$, and $V~vs.~V-K_s$ in \S 3.2) in Figure~\ref{fig:cmagic} gives $\Delta \mu_{17cbv}(B)=1.10$~mag relative to SN~2011fe. Assuming a Cepheid distance $\mu_{11fe}=29.04\pm0.19$~mag for SN~2011fe \citep{Shappee11}, we obtained the distance modulus $\mu_{17cbv}=30.14\pm0.19$~mag for SN~2017cbv ($D=10.7\pm0.9$~Mpc). 
\ \par

\subsubsection{Distance of SN 2013aa}
\citet{Burns20} applied three methods to the photometric data of SN~2013aa to estimate its distance modulus. They are $\mu=30.46\pm0.08$~mag from SNooPy fitting \citep{Burns11,Burns14}, $\mu=30.56\pm0.04$~mag from MLCS2k2 fitter \citep{Jha07}, and $\mu=30.62\pm0.04$~mag from the SALT2 algorithm \citep{Guy07}, respectively (assuming $H_0=72~\kmsmpc$). The adopted three methods yield an average estimate of SN~2013aa $\mu=30.55\pm0.08$~mag. SN~2013aa also exploded in the same galaxy with SN 2017cbv, which provides an independent distance determination of SN~2017cbv. SNe~2013aa was discovered by the Backyard Observatory Supernova Survey (BOSS) on 2013 February 13  \citep{Parker13} and classified as a SN Ia \citep{Parrent13}. SN~2013aa is 74 arcsec west and 180 arcsec south from the core of the host galaxy NGC 5643 \citep{Graham17}. 

\ \par
\begin{figure}[htb]
\begin{center}
\includegraphics[width=0.9\textwidth]{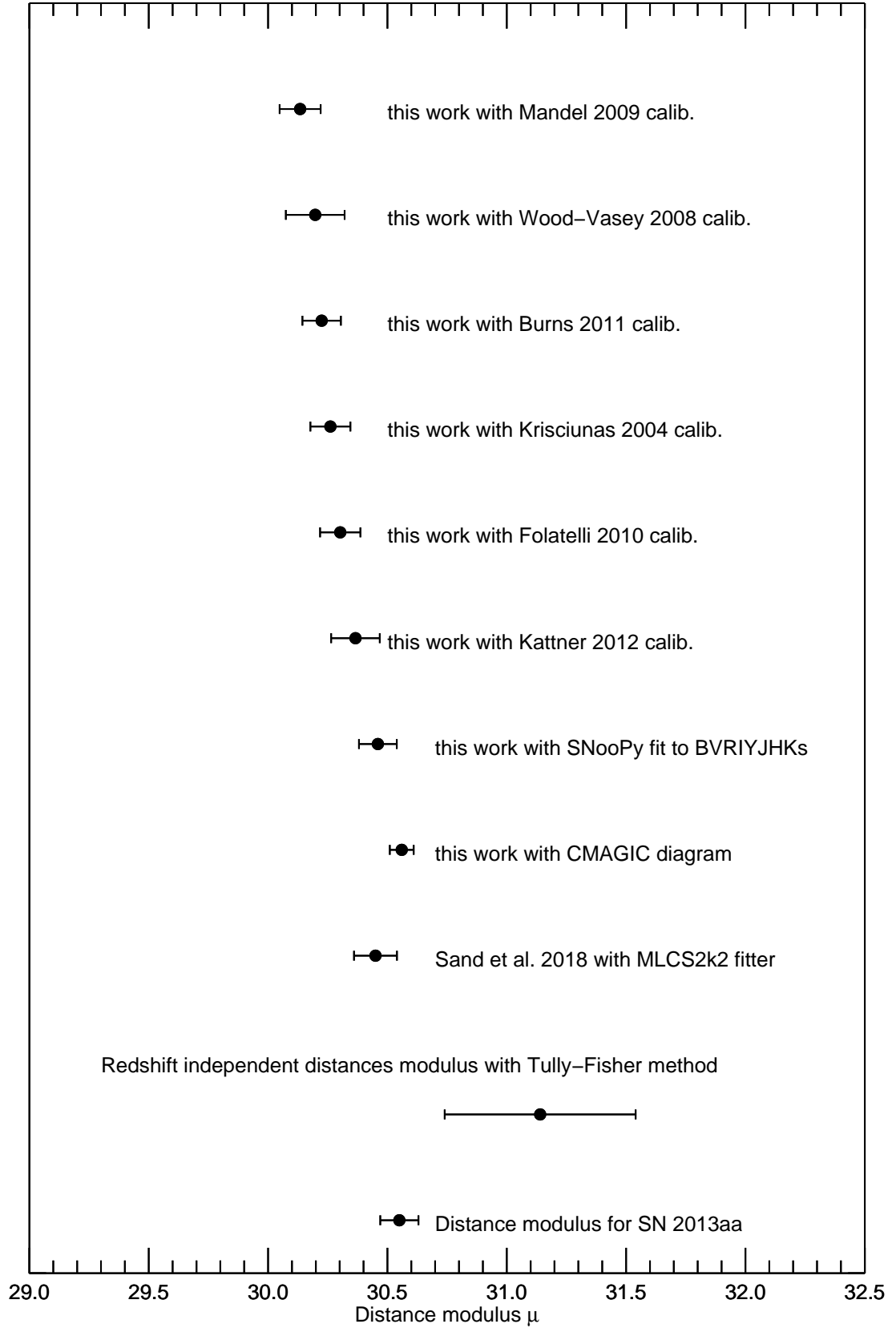}
\caption{Distance moduli toward NGC~5643 from our work on SN~2017cbv with NIR-absolute calibration, SNooPy fit to $BVRIYJHK_s$ light curves, CMAGIC diagram. Other estimates toward NGC~5643 are also listed here for comparison. \citet{Sand18} estimated the distance modulus of SN~2017cbv $\mu=30.45\pm0.09$~mag via MLCS2k2 fit \citep{Jha07} to the light curve obtained by Las Cumbres Observatory's 1~m telescope. CSP II group has measured the distance modulus of SN~2013aa $\mu=30.55\pm0.08$~mag via light curve template fitter \citep{Burns20}. \citet{Bottinelli85} listed the redshift independent distance modulus of NGC~5643 $\mu = 31.14\pm0.40$ mag from Tully-Fisher method. Error bars are 1$\sigma$. \label{fig:mu}}
\end{center}
\end{figure}

In summary, we measured the distance of SN 2017cbv with three methods: NIR-absolute calibration, SNooPy fitting, and the CMAGIC diagram. These derived values in Figure~\ref{fig:mu} are consistent with the results \citep[$\mu = 30.45\pm 0.09$ mag, $D=12.3\pm0.5$ Mpc; ][]{Sand18} made via the MLCS2K2 fitter \citep{Jha07} using independent data, and the value $\mu = 31.14\pm 0.40$ mag ($D=16.9\pm$ 3.1 Mpc) determined from Tully-Fisher method \citep{Bottinelli85,Tully88}. Individually, the distance modulus of SN~2017cbv from NIR-absolute calibration ($\mu =30.11\pm0.11$~mag to $30.41\pm0.19$~mag) are smaller than light-curve template fitters with SNooPy for our data and with MLCS2k2 for independent data \citep{Sand18}, consistent within 2.6$\sigma$. The NIR-absolute calibration values are also smaller than the CMAGIC diagram, consistent within 2.5$\sigma$ except the smallest value ($\mu =30.11\pm0.11$~mag) in $H$ band calibrated with \citet{Mandel09}.

Another SN type Ia, SN~2013aa, exploded in NGC 5643 and provided an independent distance to NGC 5643 \citep{Burns20}, consistent with our measurements. Going forward, we adopt
our CMAGIC results for the distance to this galaxy ($\mu=30.58\pm0.05$~mag) to estimate the quasi-bolometric luminosity and to compare with theoretical models in the following section. 

%Alternatively, another SN~Ia SN~2013aa also exploded in the same host galaxy, which provides an external distance measurement for SN~2017cbv and is completely independent of our photometry. Our measured distance modulus from CMAGIC diagram is close to the external measurement of SN~2013aa. So we prefer to adopt the distance modulus of SN~2017cbv with CMAGIC diagram $\mu=30.58\pm0.05$~mag to estimate the quasi-bolometric luminosity based on our photometry and to compare with theoretical models in the following section. }

% and got a mean value of $30.42\pm0.16$~mag ($D=12.1\pm0.9$~Mpc).
%%%%%%%%%%%%%%%%%%%%%%%%%mu plot%%%%%%%%%%%%%%%%%%%%%%%
%The red dashed line mark the average distance modulus $\mu=30.24\pm0.09$~mag derived from NIR-absolute calibration. 
 
%The distances I get for 13aa are, assuming H_0 = 72 km/s/Mpc:
%\mu = 30.47 +/- 0.08  (SNooPy)
%\mu = 30.56 +/- 0.04  (MLCS2k2)
%\mu = 30.62 +/- 0.04  (SALT2)

\subsection{Bolometric Light Curve}
SN~2017cbv was also observed with the Ultra-Violet Optical Telescope  \citep[UVOT; ][]{Roming05} on board the {\it Swift} satellite \citep{Gehrels04}, spanning from $t=-18.5$ to $t\sim 14$ days relative to the $B$-band maximum light \citep{Hosseinzadeh17}. The UV photometric observations were performed in $UVW2$, $UVM2$, and $UVW1$ filters. \citet{Hosseinzadeh17} published the reduced data and the photometry was performed using the pipeline for the {\it Swift} Optical Ultraviolet Supernova Archive \citep[SOUSA; ][]{Brown14}.

%\subsubsection{Spectral Energy Distribution of SN 2017cbv}
We construct the spectral energy distribution (SED) evolution of SN~2017cbv using the published UVM2 photometry \citep{Hosseinzadeh17} and our optical/NIR-band data, covering wavelengths of 1,800$\AA$ to 25,000$\AA$. We ignored the UVW2 and UVW1 photometry due to their known red leaks \citep{Brown10}. The SED method in SNooPy \citep{Burns11,Burns14} was used to estimate the $uvoir$ quasi-bolometric light curve of SN 2017cbv. The real photometry of each band was matched with the synthetic photometry on the spectral template from \citet{Hsiao07} and then the matched spectral template for each phase was integrated from 1,800$\AA$ to 25,000$\AA$, which we took as the quasi-bolometric luminosity. The light-curve fitting model \enquote{max\_model} in SNooPy  was used to interpolate the missing data points \citep{Prieto06}. For the infrared flux at wavelengths longward of $\lambda_{K_s}$, the Rayleigh-Jeans law is assumed. No host reddening is assumed due to our former analysis in \S 3.3, and the CMAGIC distance modulus of SN~2017cbv was applied to calculate the bolometric luminosity. 

%\subsubsection{Bolometric Light Curve}

Table~\ref{tab:bol} tabulates the quasi-bolometric light curve of SN 2017cbv using SNooPy \citep{Burns11}. The bolometric light curve can be used to estimate the nickel mass synthesized during the explosion using Arnett's rule \citep{Arnett82}. This rule associates the bolometric rise time and the maximum bolometric luminosity $L_{max}$ with the energy deposition $E_{Ni}$, contributed by the radioactive decay chain $^{56}$Ni$\rightarrow ^{56}$Co$\rightarrow^{56}$Fe within the expanding ejecta \citep{Stritzinger05}. The association can be simply expressed as $L_{max}=\alpha E_{Ni}$, where $\alpha$ is the ratio of input to released energy with a value around 1 \citep{Branch92,Hoeflich96,Stritzinger05,Scalzo14a}.

\begin{figure}[htb]
\begin{center}
\includegraphics[width=0.9\textwidth]{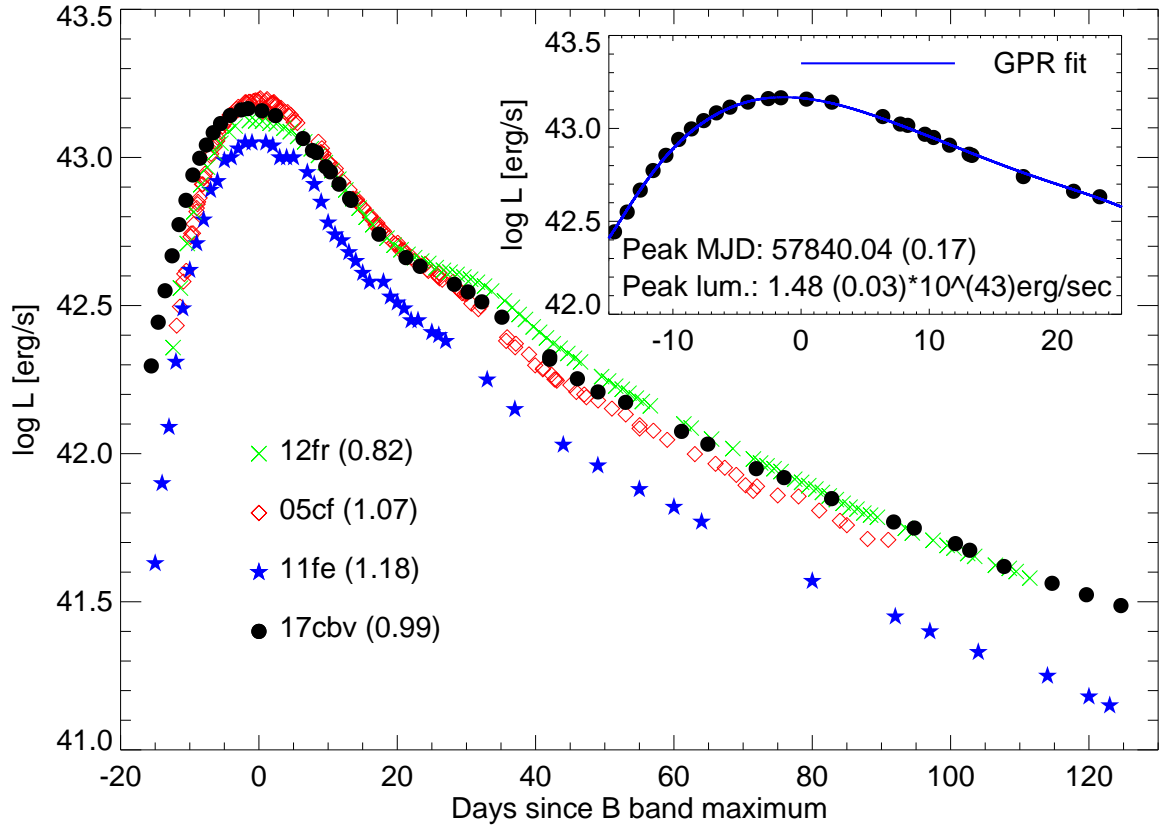}
\caption{UV through NIR quasi-bolometric light curve of SN 2017cbv with the CMAGIC distance modulus $\mu = 30.58$ mag of SN~2017cbv. Overplotted are quasi-bolometric light curves of well-observed normal SNe~2005cf \citep{Wang09b}, 2011fe \citep{Zhang16}, and 2012fr \citep{Contreras18} for comparison. Top right panel shows the Gaussian Process Regression fit to the peak of the bolometric light curve of SN~2017cbv. \label{fig:bol}}

\end{center}
\end{figure}

Figure~\ref{fig:bol} shows the $uvoir$ quasi-bolometric light curve of SN~2017cbv. Overplotted are the bolometric light curves of the normal SNe Ia 2011fe \citep{Zhang16}, 2005cf \citep{Wang09b}, and 2012fr \citep{Contreras18} for comparison, which are the few other SNe~Ia to have been well-observed in UV, optical, and NIR. The GPR fitting was again applied to determine the bolometric rise time and peak luminosity, shown in the top panel of Figure~\ref{fig:bol}. The peak luminosity of SN 2017cbv is $L_{\rm peak}=1.48 \times 10^{43}$ erg s$^{-1}$, or log ($L_{\rm peak}$/erg s$^{-1}$)=43.17 dex, which is consistent with the delayed detonation scenario in the Chandrasekhar-mass models with the peak bolometric luminosity log ($L_{\rm peak}$/erg s$^{-1}$)=42.80 - 43.31 dex \citep{Hoeflich96,Seitenzahl13a,Kromer16, Hoeflich17}. The peak luminosity of SN 2017cbv is also consistent with other models, i.e., sub-Chandrasekhar mass double-detonation models \citep{Fink10,Kromer10}, pulsational-delayed detonation models \citep{Hoeflich96,Dessart14}, and so on.
% which places SN~2017cbv in the parameter space of the delayed detonation scenario in the Chandrasekhar-mass models with the peak bolometric luminosity in the range log ($L_{\rm peak}$/erg s$^{-1}$)=42.80 - 43.31 dex \citep{Hoeflich96,Seitenzahl13a,Kromer16, Hoeflich17}.
\begin{deluxetable}{rlrl}[htb]
\tablewidth{0pt}
\tablecaption{The estimated quasi-bolometric luminosity of SN~2017cbv by adopting the CMAGIC distance modulus of SN 2017cbv; $\mu=30.58$~mag and $H_o=72~\kmsmpc$. \label{tab:bol}}
\tablehead{\colhead{Phase$ ^{a}$} & \colhead{$L_{bol}^b$} &  \colhead{Phase} & \colhead{$L_{bol}$} \\}
%                (d)       & ($10^[43}\mathrm{erg~s}^{-1}$) }
\startdata
$ -15.56 $ &      0.198 & $ 17.34 $ &      0.550 \\ 
$ -14.56 $ &      0.278 & $ 21.26 $ &      0.459 \\ 
$ -13.57 $ &      0.355 & $ 23.30 $ &      0.429 \\ 
$ -12.55 $ &      0.466 & $ 28.24 $ &      0.373 \\ 
$ -11.55 $ &      0.593 & $ 30.23 $ &      0.352 \\ 
$ -10.55 $ &      0.717 & $ 32.22 $ &      0.326 \\ 
$ -9.55 $ &      0.872 &  $ 35.13 $ &      0.289 \\ 
$ -8.56 $ &      0.994 &  $ 42.03 $ &      0.208 \\ 
$ -7.60 $ &      1.101 &  $ 46.04 $ &      0.179 \\ 
$ -6.61 $ &      1.212 &  $ 49.02 $ &      0.162 \\ 
$ -5.55 $ &      1.301 &  $ 53.01 $ &      0.149 \\ 
$ -4.15 $ &      1.386 &  $ 61.10 $ &      0.119 \\ 
$ -2.56 $ &      1.447 &  $ 64.90 $ &      0.108 \\ 
$ -1.58 $ &      1.462 &  $ 71.90 $ &      0.089 \\ 
$  0.42 $ &      1.438 &  $ 75.92 $ &      0.083 \\ 
$  2.40 $ &      1.384 &  $ 82.80 $ &      0.070 \\ 
$  6.37 $ &      1.158 &  $ 91.76 $ &      0.059 \\ 
$  7.74 $ &      1.055 &  $ 94.75 $ &      0.056 \\ 
$  8.32 $ &      1.038 &  $ 100.70 $ &      0.050 \\
$  9.66 $ &      0.930 &  $ 102.76 $ &      0.047 \\
$ 10.33 $ &      0.894 &  $ 107.73 $ &      0.042 \\
$ 11.58 $ &      0.812 &  $ 114.67 $ &      0.036 \\
$ 13.11 $ &      0.726 &  $ 119.63 $ &      0.033 \\
$ 13.33 $ &      0.716 &  $ 124.62 $ &      0.031 
%\tablenotetext{a}{SNe~Ia 2013aa and 2017cbv were exploded in the same host galaxy NGC 5643.} \\
\tablenotetext{a}{Days since $B$-band maximum.} \\
\tablenotetext{b}{ The unit is $10^{43}~\mathrm{erg~s}^{-1}$}
\end{deluxetable}

\clearpage
\subsection{Nickel Mass}
\subsubsection{Nickel Mass from $uvoir$ bolometric light curve}
% SN~2013aa also exploded in the same galaxy with SN~2017cbv, which offers an independent distance determination of SN~2017cbv \citep{Burns20}. 
In calculation of the bolometric luminosity, we adopt the CMAGIC distance modulus of SN~2017cbv $\mu = 30.58\pm0.05$~mag, which is close to the external independent distance determination of SN~2013aa\citep{Burns20}, exploded in the same galaxy. This gives a peak luminosity $L_{\rm peak}= 1.48\pm0.07 \times 10^{43}$ erg s$^{-1}$ and a synthesized nickel mass $M_{Ni} = 0.73\pm0.03$  M$_{\odot}$, which is used throughout this paper. 
\begin{deluxetable}{llll}[htb]
\tablewidth{0pt}
%\tablenum{5}
\tablecaption{$^{56}$Ni mass estimated from $uvoir$ bolometric light curves \label{tab:ni1}}
\tablehead{\colhead{Method} & \colhead{$\mu ^{a}$} & \colhead{$L_{\rm peak}$} & \colhead{Ni mass$^a$} \\
       &         (mag)      & ($10^{43} \mathrm{erg s}^{-1}$) & (M$_{\odot}$)}
\startdata
%Tully-Fisher distance          & 31.14$\pm$0.40 & 2.63$\pm$1.07 & 1.23$\pm$0.51 \\
%\\
%\hline
CMAGIC Distance of SN~2017cbv          & $30.58\pm0.05$ & $1.48\pm0.07$ & $0.73\pm0.03$ 
\tablenotetext{a}{using alpha parameter of 1.0.}
\end{deluxetable}

\subsubsection{Nickel Mass from the phase of the secondary maximum in $YJH$ bands}
\citet{Kasen06a} attributes the secondary maximum of NIR emission to the ionization evolution of iron-group elements in the ejecta and predicted that the secondary maximum should be delayed for larger Fe masses, indicating the phase of the secondary maximum should be a function of the nickel mass in the explosion. This was investigated by recent studies \citep{Jack12,Biscardi12,Dhawan15,Dhawan16}. \citet{Dhawan16} studied the correlation between the phase of the secondary maximum t2 and the bolometric peak luminosity, and they established the $L_{\rm peak}$ vs. t2 relations for $YJH$-band light curves. The well-sampled optical and NIR observations of SN 2017cbv presented in this paper offer a good chance to test the above derived nickel mass using the reddening-free and distance-independent method. The time t2 was measured to be $31.66\pm0.69$ and $25.96\pm0.29$~days from the $J$ and $H$-band photometry and the corresponding peak bolometric luminosities are $1.25\pm0.17$ and $1.11\pm0.27$ $\times 10^{43}$ erg $s^{-1}$, yielding $0.62\pm0.08$ and $0.55\pm0.13$~M$_{\odot}$ nickel masses, respectively. The corresponding value derived from the $Y$ band data is $0.66\pm0.10$ M$_{\odot}$ if we adopted its corresponding of secondary maximum $33.83\pm0.45$~days according to \citet{Wee18}. 

We measured the nickel mass of SN~2017cbv using two methodologies that employ the quasi-bolometric light curve and the phases of the secondary maximum in NIR bands. Those derived values of the nickel mass are listed in Tables~\ref{tab:ni1} and ~\ref{tab:ni2}, which are consistent with each other within 2$\sigma$, and are comparable to that of normal SNe~Ia such as SN~2005cf \citep[$M_{Ni} = 0.78 $ M$_{\odot}$; ][]{Wang09b}, SN~2011fe \citep[$M_{Ni} \sim 0.57 $ M$_{\odot}$; ][]{Zhang16}, SN~2012cg \citep[$M_{Ni} \sim 0.72 $ M$_{\odot}$; ][]{Chakradhari19} and SN~2012fr \citep[$M_{Ni} \sim 0.60 $ M$_{\odot}$; ][]{Contreras18}. Our measured nickel mass is matched with some $M_{ch}$ delayed detonation models and sub-$M_{ch}$ double detonation models discussed in \S4, while the $BV$-band light curves of SN~2017cbv is matched with $M_{ch}$ delayed detonation model 25 with a nickel mass of 0.6~\msun~after $B$ maximum. 

\begin{deluxetable}{lllll}[htb]
%\tiny
\tablewidth{0pt}
%\tablenum{5}
\tablecaption{$^{56}$Ni mass related to the time of the second maximum for $YJH$ bands \label{tab:ni2}}
\tablehead{\colhead{Method$^{a}$} & \colhead{Band} & \colhead{t2} & \colhead{$L_{\rm peak}$} & \colhead{Ni mass$^{b}$} \\
                            &                 &  \colhead{(days)}       & \colhead{($10^{43} \mathrm{erg s}^{-1}$)} & \colhead{(M$_{\odot}$)}}
\startdata
$L_{max}=(0.041\pm0.005)\times t2+(-0.065\pm0.122)$ &$Y$ & $33.83\pm0.45$ & $1.32\pm0.21$ & $0.66\pm0.10$ \\ 
$L_{max}=(0.039\pm0.004)\times t2+(0.013\pm0.106)$ &$J$ & $31.66\pm0.69$  & $1.25\pm0.17$ & $0.62\pm0.08$\\ 
$L_{max}=(0.032\pm0.008)\times t2+(0.282\pm0.174)$ &$H$ & $25.96\pm0.29$  & $1.11\pm0.27$ & $0.55\pm0.13$
\enddata

\tablenotetext{a}{The correlation coefficient values between $L_{ max}$ and $t2$ were taken from Table 2 in \citet{Dhawan16}} \\
\tablenotetext{b}{using alpha parameter of 1.0.}
\end{deluxetable}

\subsection{Is SN~2017cbv a normal type Ia?}
\begin{figure}[htb]
\begin{center}
\includegraphics[width=0.9\textwidth]{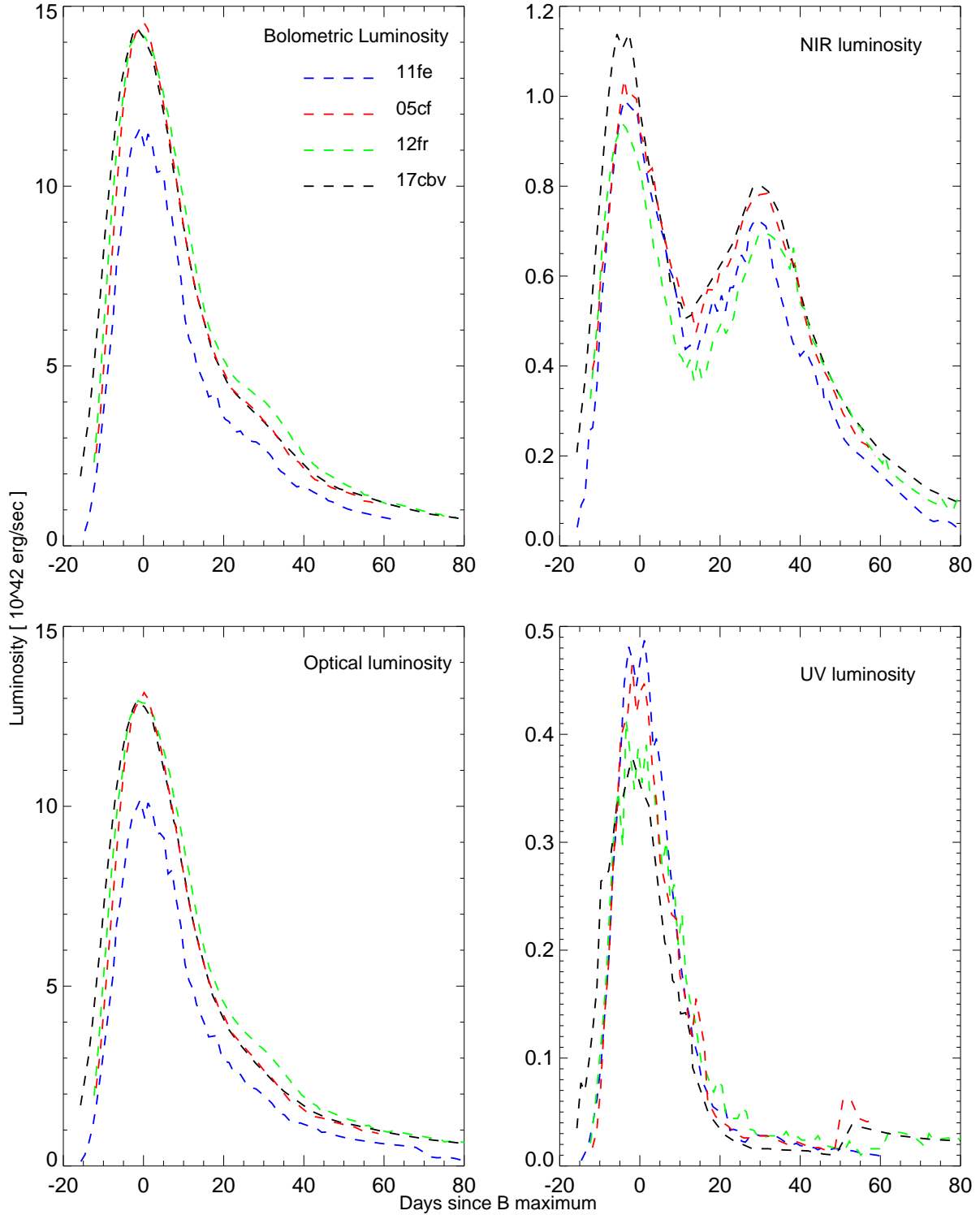}
\caption{The bolometric (1,800\AA-25,000\AA), NIR (9,000\AA-25,000\AA), optical(3,00\AA-9,000\AA), and $UVM2$(1,800\AA-3,000\AA) luminosity of SNe~2005cf, 2011fe, 2012fr, and 2017cbv derived with the same SNooPy method \citep{Burns11}.  \label{fig:flux}}
\end{center}
\end{figure}

Figure~\ref{fig:flux} shows the bolometric luminosity (1,800\AA - 25,000\AA), and the separate contributions of the NIR (9,000\AA - 25,000\AA), optical (3,000 - 9,000\AA), and  UV (1,800 - 3,000\AA) portions of the spectrum for  SNe~2005cf, 2011fe, 2012fr and 2017cbv. All results were derived using the SNooPy module \citep{Burns11,Burns14}. $UVM2$, Johnson-Cousins $UBVRI$, and 2MASS $JHK_s$ filters were applied to the multi-band data of SNe~2005cf \citep{Wang09b} and 2011fe \citep{Zhang16}. $UVM2$ and CSP natual filters $uBgVriYJH$ were applied to multi-band light curves of SN~2012fr \citep{Contreras18}. $UVM2$, ANDICam filters $BVRI$, $Y$, 2MASS $JHK_s$ filters were applied to our data of SN~2017cbv. The CSP natural filter function $Y$ was used for SN~2017cbv as no ANDICam filter function for $Y$-band is available. There are no obvious differences for the bolometric ($\sim 0.1\%$) and NIR luminosities ($\sim 1\%$) if we ignore the $Y$ filter. We adopted the following host extinctions and distance moduli: $E(B-V)_{\rm host}=0.1$~mag and $\mu=32.31$~mag for SN~2005cf \citep{Wang09b};  $E(B-V)_{\rm host}=0$~mag and $\mu=29.04$~mag for SN~2011fe \citep{Zhang16}; $E(B-V)_{\rm host}=0.03$~mag and $\mu=31.27$~mag for SN~2012fr \citep{Contreras18}; $E(B-V)_{\rm host}=0$~mag and $\mu=30.58$~mag for SN~2017cbv. All the Milky Way extinctions toward the four SNe are corrected according to \citet{SF11}. Missing photometry for each epoch was interpolated by the light-curve fitting model \enquote{\rm max\_model} \citep{Prieto06} in SNooPy. For the infrared flux at wavelength longward of the reddest wavelength, a Rayleigh-Jeans law is assumed. An SED-based method is used to estimate the bolometric luminosity by integrating the spectral template for each phase from \citet{Hsiao07} $\times$ a factor, so the synthetic photometry of the template matches the real photometry.  

As showed in Figure~\ref{fig:flux}, SNe~2005cf, 2012fr, and 2017cbv have comparable bolometric luminosity and all of them are brighter than SN~2011fe. The  bolometric light curves of these four SNe are matched with their optical contribution, showing that optical luminosity of a SN~Ia dominates the bolometric luminosity. NIR and UV contributions to the bolometric light curves are generally of less importance than the optical. We note that these four SNe~Ia have comparable NIR and UV luminosities.  

\begin{figure}[htb]
\begin{center}
\includegraphics[width=0.9\textwidth]{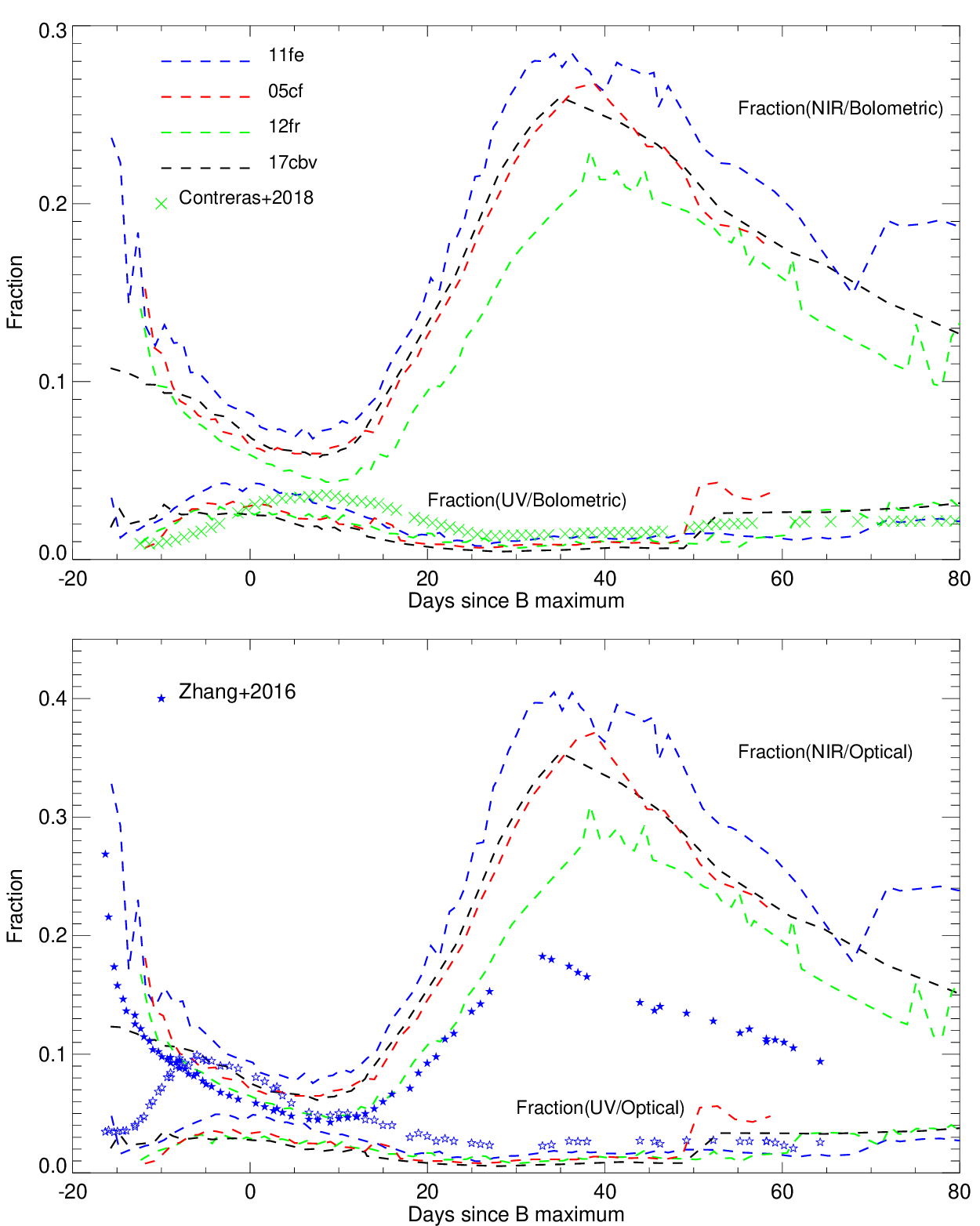}
\caption{Top: the ratio of the NIR (9,000\AA - 25,000\AA) and $UV$ (1,800\AA - 3,000\AA) fluxes to all (1,800\AA-25,000\AA) for SNe~2005cf, 2011fe, 2012fr, and 2017cbv in dashed lines with the same method. The $UV$ fraction of SN~2012fr are overplotted from \citet{Contreras18}. Bottom: the ratio of the NIR (9,000\AA - 25,000\AA) and $UV$ (1,800\AA - 3,000\AA) fluxes to the optical (3,000\AA-9,000\AA). The UV and NIR fraction of SN~2011fe are overplotted from \citet{Zhang16}. \citet{Zhang16} formed spectral energy distribution curves with $UVW2$ and $UVW1$, which are ignored in our analysis due to their red leaks \citep{Brown10}. The difference of NIR fraction of SN~2011fe is caused by difference methods. Refer to the text for details. \label{fig:ratio}}
\end{center}
\end{figure}

Figure~\ref{fig:ratio} illustrates the UV and NIR contributions to the bolometric light curves in the top panel and the optical contribution in the  bottom panel. Overplotted are the flux ratios of SN~2011fe \citep{Zhang16} and SN~2012fr \citep{Contreras18}. The flux ratios measured by SNooPy, \citet{Zhang16} and \citet{Contreras18} point to differences among these SNe, however when the flux ratios are measured with the same methodology as we do here, the flux ratios seem completely consistent. \citet{Contreras18} applied two methods (photometric trapezoidal integration and spectral template fitting) to calculate the luminosity of SN~2012fr based on $UVM2$,$uBgVriYJH$-band magnitudes. The first method calculates the flux from the observed photometry by trapezoidal integration of the fluxes derived from each filter. The UV contribution is made from {\it Swift} $UVM2$ photometry and the NIR contributions at wavelength $\lambda > \lambda_H$ is estimated from the Rayleigh-Jeans law. The second method is similar to the SED method in SNooPy. More detailed steps can be found in the Appendix B of \citet{Contreras18}. The bolometric luminosity from the two methods are consistent at the $\pm5\%$ level \citep[eg., Figure 29 of ][]{Contreras18}. Our flux fraction estimated by the SED method in the SNooPy module is consistent with that of the other two methods, and is much more consistent with the spectral template fitting method (eg., the UV fraction in the top panel of Figure~\ref{fig:ratio}).  

\citet{Zhang16} adopted the photometric direct integration method based on {\it Swift} $UVW2$,$UVM2$,$UVW1$, and $UBVRIJHK_s$-band photometry, which computes effective wavelength of each filter, converts the measured magnitude to flux to generate a spectra energy distribution and thus integrates the derived spectral energy distribution at specific wavelength ranges for bolometric (1,600\AA - 24,000\AA), UV (1,600\AA - 3,200\AA), optical (3,200\AA - 9,000\AA), and NIR (9,000\AA - 24,000\AA) luminosities. Our UV fraction of SN~2011fe is smaller than that of \citet{Zhang16} as we did not use the fluxes of filters $UVW2$ and $UVW1$ due to their red leaks \citep{Brown10}. Our NIR fraction is larger than that of \citet{Zhang16} mainly because we use different methods to calculate them and \citet{Zhang16} did not assume a Rayleigh-Jeans law for the infrared flux at wavelength $\lambda > \lambda_{K_s}$. 

Figures ~\ref{fig:flux} and ~\ref{fig:ratio} illustrate that SN~2017cbv shows similar behaviors as the normal well-observed SNe~Ia SN~2005cf, SN~2011fe, and SN~2012fr. These normal characteristics can also be seen in their light curves (eg., Figures~\ref{fig:lcopt} and ~\ref{fig:lcnir}) and color curves (eg., Figures~\ref{fig:copt},~\ref{fig:cvjhk}, and ~\ref{fig:cjhk}). This behavior is further seen in the NIR spectra from early to late phases relative to $B$-band maximum (eg., Figure~\ref{fig:early} and ~\ref{fig:aftermax}). 
All our multi-band photometry and NIR spectra point to SN~2017cbv as a normal SN Ia. However, the very early colors ($B-V$, $V-R$, and $V-I$) of SN~2017cbv are bluer than the normal comparison sample in Figure~\ref{fig:copt}. Also, the early flux of SN~2017cbv is brighter than both SNe~2005cf and 2012fr with similar bolometric peak, although the difference gradually disappears when close to maximum phase. For example, at the very early phase of SN~2012fr ($t=-11.3$ days), flux ratios of SN~2017cbv over SN~2012fr are 1.7, 1.4, 1.7, and 2.9 for the bolometric flux and NIR, optical and UV contributions, respectively. 

SN 2017cbv shares similar early characteristics (i.e., early light-curve excess and/or blue color) with SN~2012cg \citep{Marion16}, iPTF14atg \citep{Miller18}, MUSSES1604D \citep{Jiang17,Maeda18}, SN~2018oh \citep{Shappee19,Li19b,Dimitriadis19a}, and SN~2019yvq \citep{Miller20}. The early light-curve excess could be the result of either ejecta interaction with its companion \citep{Kasen10,Marion16,Hosseinzadeh17} or vigorous mixing of radioactive $^{56}$Ni in the SN outermost ejecta \citep{Dessart14,Piro16,Miller18,Magee18,Magee20a,Magee20b}. It could also be associated with the He-detonation configuration within a white dwarf with a He layer \citep{Jiang17,Maeda18,Polin19,Tanikawa19}. From the observational perspective, early-color observations of SNe~Ia suggest a \enquote{red} class showing a steep transition from red to bluer colors and a \enquote{blue} class with bluer colors and flatter evolution \citep{Stritzinger18}. Based on a sample of 13 SNe~Ia discovered within 3 days from inferred first light, \citet{Stritzinger18} claimed that events in the \enquote{blue} class are over-luminous and of the Branch Shallow Silicon spectral type, while ones in the \enquote{red} class are more typically related to the Branch Core-Normal or CooL type \citep{Branch06}. \citet{Jiang18} inspected the light curves of 23 young SNe~Ia and drew similar conclusions. Also, \citet{Han20} confirmed the above claim of the distinction between \enquote{blue} and \enquote{red} by adding six events to the sample of \citet{Stritzinger18}. From theoretical perspective, models are expected to have red colors early on when $^{56}$Ni is produced in the high-density innermost regions of ejecta, and later show a transition to blue colors with the photosphere receding into increasing hotter layers. In contrast, the following scenarios predict blue colors soon after explosion, involving the interaction of SN ejecta with a non-degenerate conpanion star \citep{Kasen10}; or unbound material ejected prior to detonation \citep[pulsational-delayed-detonation models, ][]{Dessart14}; or radioactive $^{56}$Ni mixing in the SN outer ejecta \citet{Piro16,Miller18,Magee18,Magee20a,Magee20b,Bulla20}; or ejecta-DOM (disk-originated matter) collision in DD scenario \citep{Levanon17,Levanon19}. 

%\clearpage
\section{Theoretical Perspectives}
From a theoretical perspective, several kinds of explosion models can be expected to generate SNe Ia. Here we examine two popular explosion models: $M_{ch}$ delayed detonation (DDT) scenario, and Sub-$M_{ch}$ Double-detonation scenario. Both scenarios have received much attention as they reproduce aspects of SN~Ia light curves and spectra  \citep{Hoflich95,Hoeflich96,Kasen09,Kromer10,Ropke12,Sim13,Blondin15,Maeda18,Polin19}. Thus both scenarios could make an important contribution to the SN~Ia population \citep{Hachisu08,Mennekens10,Ruiter11,Scalzo14b,Goldstein18}. In detail, \citet{Seitenzahl13b} showed that $\sim 50\%$ of SNe~Ia should explode at or near Chandrasekhar in order to interpret the solar abundance of manganese observed in the Galaxy.  \citet{Scalzo14b} showed the empirical fits to the light curves of the SDSS and SNLS SNe~Ia, which suggests $30\%$ of these events originating from sub-Chandrasekhar scenario.

\subsection{$M_{ch}$ DDT scenario}
In the $M_{ch}$ DDT scenario \citep{Khokhlov91,Hoflich95,Plewa04,Ropke07,Kasen09,Blondin13,Seitenzahl13a,Hoeflich17}, a WD is thought to accrete material from a non-degenerate companion star and explode following carbon ignition near the WD center, which occurs when the WD mass has nearly reached $M_{ch}$. The DDT models involve a transition from deflagration to detonation burning, which is parameterized by a transition density $\rho_{tr}$, regulating the pre-expansion of the white dwarf and determining the abundance structure during the deflagration phase \citep{Hoeflich17}. The DDT scenario predicts some of the basic characteristics of SNe Ia, i.e.,  spherical geometry of SN Ia remnants with layered chemical structures \citep{Rest08,Hoeflich17}, spherical density distributions as suggested by small continuum polarization \citep{Patat12}, and so on.

As mentioned in \S 3.5, the peak luminosity of SN~2017cbv is located in the parameter space of the Chandrasekhar mass delayed detonation models. Here we compared the $BV$-band light curves of SN~2017cbv with three typical DDT models in \citet{Hoeflich17} in Figure~\ref{fig:model}. Models 23 \citep{Hoeflich02}, 23d5 \citep{Hoeflich06,Diamond15}, and 25 \citep{Hoeflich02} are assumed to have solar metallicity and a main sequence mass of 5 \msun, with different central densities ($\rho_c=$2.0, 6.0, 2.0$\times 10^9$~g cm$^{-3}$) and different transition densities ($\rho_{tr}=$2.3, 2.3, 2.5$\times 10^7$~g cm$^{-3}$) at the time of the explosion. The corresponding $^{56}$Ni masses for the three DDT models 23, 23d5, and 25 are 0.56, 0.54, and 0.60 \msun. At phases $t\ge 0$ since the $B$-band maximum, model 25 is more consistent with the $BV$-band data. Compared with model 23, model 25 has a larger transition density $\rho_{tr}$, meaning less deflagration burning and subsequently higher density burning during the detonation phase \citep{Hoeflich17}. Model 25 has log ($L_{peak}$/erg s$^{-1}$)=43.22 dex and provides an absolute brightness $M_V=-19.29$~mag with a rise time of 18.5~days. Model 25 is fainter than observed peak for SN~2017cbv by 0.15~mag, but well within the error bars of distance and reddening uncertainties.

%Our selected DDT models are fainter than observed for SN~2017cbv, perhaps suggesting that a model with slightly higher $^{56}$Ni may be preferred.    

%{\bf In detail, the $\chi^2$ of $B$ magnitude and DDT models are 8.6, 23.2, and 27.6 for models 25, 23, and 25d5, respectively. The corresponding $\chi^2$ of $V$ magnitude and DDT models are 3.9, 11.5, and 25.1.}
%extra energy is contributed by CSM interaction and/or

\subsection{Sub-$M_{ch}$ Double-detonation scenario}
In the sub-$M_{ch}$ double-detonation model \citep{Nomoto80,Taam80,Livne90,Shen09,Kromer10,Woosley11,Fink10,Moll13,Wang17,Blondin17,Shen18a}, the explosion is triggered by the detonation of a helium shell that may be accreted from a helium companion star. This helium detonation then triggers a second detonation in the core. Observational signatures of sub-$M_{ch}$ WD explosions point to two categories of helium shell sizes. Thick shell models predict an early time flux excess caused by the presence of radioactive material in the ashes of the helium shell, but this flux excess will be red due to line blanketing. Thin helium shell models can reproduce the characteristics of many normal SNe~Ia and sub-luminous SN~1991bg-like objects \citep{Shen18b,Polin19}. 

Here we choose double detonation and bare CO detonation in sub-Chandrasekhar mass WDs. For the double detonation scenarios, we take models 3 and 4 of \citet{Fink10,Kromer10} as thick shell models. Model 3 yields 0.55 \msun~of $^{56}$Ni from an initial WD with a 1.025 \msun~CO core and a He shell of 0.055 \msun. In the initial helium-shell detonation, 1.7$\times10^{-3}$ \msun~of $^{56}$Ni, 6.2$\times10^{-3}$ \msun~of $^{52}$Fe, 4.4$\times10^{-3}$ \msun~of $^{48}$Cr have been synthesized close to the ejecta surface. Model 4 yields 0.78 \msun~of $^{56}$Ni from an initial WD with a 1.125 \msun~CO core and a He shell of 0.039 \msun. In the initial helium-shell detonation, 4.4$\times10^{-3}$ \msun~of $^{56}$Ni, 3.5$\times10^{-3}$ \msun~of $^{52}$Fe, 2.2$\times10^{-3}$ \msun~of $^{48}$Cr have been synthesized close to the ejecta surface. As a bare CO detonation, we use model 3c of a detonation of a 1.025 
\msun~that yields 0.55\msun~of $^{56}$Ni \citep{Kromer10}, which represents the extreme case of a thin helium shell model. 

\begin{figure}[htb]
\begin{center}
\includegraphics[width=0.9\textwidth]{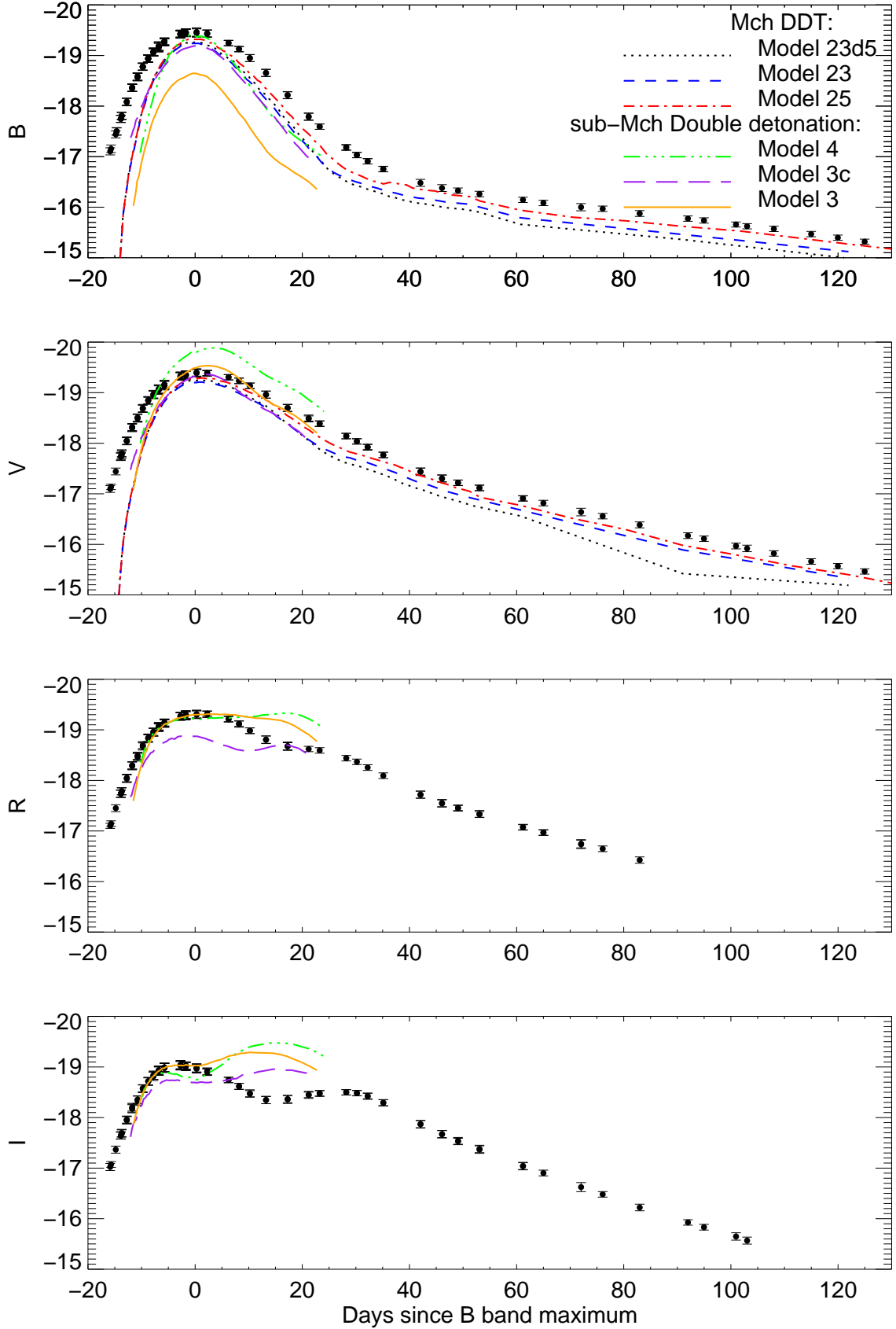}
\caption{The absolute $BVRI$-band light curves of SN 2017cbv. CMAGIC distance modulus $\mu=30.58\pm0.05$~mag and no host extinctions were applied to our $BVRI$-band data. The Milky Way extinctions toward SN~2017cbv are corrected following \citet{SF11}.  Overplotted are the $M_{Ch}$ DDT and sub-$M_{ch}$ double detonation models of normal-bright SNe Ia for comparison. Typical DDT models 23, 23d5, and 25 in \citet{Hoeflich17} are overplotted and they are assumed a progenitor of a solar metallicity and 5 \msun~on the main sequence, with different central densities ($\rho_c=$2.0, 6.0, 2.0$\times 10^9$~g cm$^{-3}$) and different transition densities ($\rho_{tr}=$2.3, 2.3, 2.5$\times 10^7$~g cm$^{-3}$) at the time of the explosion, respectively. Sub-$M_{ch}$ model 4  is composed of a C/O core mass $M_{core}=1.125$~\msun and a shell mass $M_{sh}=0.039$~\msun and model 3 is composed of a C/O core mass $M_{core}=1.025$~\msun and a shell mass $M_{sh}=0.055$~\msun \citep{Kromer10,Fink10}, while sub-$M_{ch}$ model 3c is a bare C/O core with $M_{core}=1.025$~\msun \citep{Kromer10}. \label{fig:model}}
\end{center}
\end{figure}

The models 3, 4 and 3c are available for the optical spectral time series on the Heidelberg Supernova Model Archive \footnote{https://hesma.h-its.org/}. The synthetic photometry for these models in the ANDICam $BVRI$ bands are shown in Figure~\ref{fig:model}. We obtained $M_{B,\rm max}=-19.4$~mag, $M_{V,\rm max}=-19.9$~mag, $M_{R,\rm max}=-19.3$~mag, $M_{I,\rm max}=-19.5$~mag for model 4 and $M_{B,\rm max}=-18.6$~mag, $M_{V,\rm max}=-19.5$~mag, $M_{R,\rm max}=-19.3$~mag, $M_{I,\rm max}=-19.3$~mag for model 3. While for model 3c, they are $M_{B,\rm max}=-19.2$~mag, $M_{V,\rm max}=-19.4$~mag, $M_{R,\rm max}=-18.9$~mag, $M_{I,\rm max}=-19.0$~mag. Those calculated peak values are similar to the ones given by \citet{Kromer10} for model parameters in their Table 1. As showed in Figure~\ref{fig:model}, $V$-band of model 4 is higher than observations of SN~2017cbv after $t \sim -5$ days relative to $B$-band maximum. In contrast, the model 3c  and 3 predict comparable $V$ peaks but fainter $B$ peaks, especially for the model 3 with the faintest $B$-band curve. As the DDT models do, both sub-Chandrasekhar models 3, 4 and 3c predict narrower light-curve profiles (rising and declining) than the observations. The shoulder-phase $R$ band of models 3 and 4 is higher than the observations, while the model 3c is comparable with the observations at the same phase. For the $I$ band, models 3, 4 and 3c are brighter than the observations of SN~2017cbv at shoulder phase. 

The observations of SN~2017cbv are not compatible with sub-Chandrasekhar models due to their brighter $V$-band peak magnitudes for model 4 and due to its much fainter $B$-band magnitudes for model 3, or its red color. We note that the sub-Chandrasekhar bare core detonation model 3c does not predict such red colors as models 3, 4 do, but it expects fainter magnitude at maximum phase in $RI$ bands and brighter magnitudes at the shoulder phase in $I$ band.

\ \par
As showed in Figure~\ref{fig:model}, Chandrasekhar delayed detonation model 25 is more matched with the observations of SN~2017cbv compared with other DDT models, sub-Chandrasekhar double detonation models 3, 4 and bare core detonation model 3c. However, for the early phase, model 25 still does not match with the observations of SN~2017cbv due to its broader rising light curve. As discussed in \citet{Hosseinzadeh17} for the same supernova, extra energy are likely to make a considerable contribution to the early-phase light curve. The early light-curve excess could be the result of either ejecta interaction with its companion \citep{Kasen10,Marion16,Hosseinzadeh17} or vigorous mixing of radioactive $^{56}$Ni in the SN outermost ejecta \citep{Dessart14,Piro16,Miller18,Magee18,Magee20a,Magee20b}. More detailed discussion for the early phase of SN~2017cbv will be presented by Sand et al, in Preparation. Our well-observed data provides valuable reference to model the detailed explosion for SN~2017cbv from very early to late phase.

%%%%%%%%%%%%%%%%%%%%%%%%%%results%%%%%%%%%%%%%%%%%%%%%%%%%%%%
\clearpage

\section{Conclusions}
%Its NIR-band photometry of SN~2011fe just lasted for 60~days \citep{Matheson12}
The optical and NIR-band photometric observations acquired for SN 2017cbv constitute the most complete temporal coverage in eight bands simultaneously among current Ia sample, even for the well-observed SN~2011fe and SN~2012fr. We also present more than 10 time-series NIR spectra with the first observed only 2.3 days past explosion ($-17.6$ days after maximum), the earliest normal Ia target, similar to early-phase NIR spectra only observed for SN~2011fe \citep{Hsiao13}. These make it an ideal reference for comparative investigations of SNe~Ia. Here are our main conclusions.
\begin{itemize}
    \item  SN 2017cbv is a typical normal Ia with decline rate parameter $\Delta m_{15}(B)=0.990\pm0.013$~mag and with peak $B$-band magnitude $M_B^{max}=-19.49\pm0.05$~mag, when taking the distance modulus of SN~2017cbv $\mu=30.58\pm0.05$~mag utilizing the CMAGIC diagram. %with the three methods (the NIR-absolute calibration, SNooPy fitting, and CMAGIC diagram).
    \item The NIR spectra of SN 2017cbv are also well matched with normal SNe Ia, i.e., SN~2011fe, SN~2012fr, and SN~2014J, from the very early phase to about 50 days after $B$-band maximum. 
    \item C I $\lambda~1.0693 \mu$m near 1.03~$\mu$m was likely identified from our NIR spectrum taken at $t\sim$4.4~days via SYNAPPS \citep{Thomas11}, with a clear notch seen in the blue wing of Mg II line. Its blue-shifted velocity is consistent with that of Mg II  $\lambda ~1.0927 \mu$m at the same phase.
    \item A narrow Pa$\beta~ \lambda~1.282 \mu$m was searched for at several phases after maximum light, with no apparent detection, giving a similar hydrogen mass limit of $\sim 0.1 M_{\odot}$ as SN~2014J \citep{Sand16}, based on the red-giant model \citep{Maeda14}. This limit is higher than hydrogen mass predicted by \citet{Pan12} and \citet{Lundqvist13,Lundqvist15}, but comparable to the unbound mass derived from some simulations including a non-degenerate companion star \citep[e.g., ][]{Liu12}. 
    
    \item Thanks to the excellent data set, color-magnitude diagrams $B~vs.~B-V$, $B~vs.~B-K_s$ and $V~vs.~V-K_s$ were used to estimate both host extinction and the SN distance simultaneously.
    \item The host reddening of SN 2017cbv can be neglected based on three methods (Phillips intrinsic color, Lira-Phillips relation, and CMAGIC diagram) and Na I content. 
    
    \item Three methods (NIR-absolute calibration, SNooPy fitting, and CMAGIC diagram) are applied to estimate the distance modulus of SN~2017cbv displayed in Figure~\ref{fig:mu}. These measured values are consistent within the error with external distance determinations of SN~2013aa which exploded in the same host galaxy NGC~5643 \citep{Burns20} and previous light curve-based measurements  \citep{Sand18}.
   
    \item By applying the distance modulus of SN~2017cbv ($\mu=30.58\pm0.05$~mag) using the CMAGIC diagram , the peak bolometric luminosity of SN~2017cbv is estimated as $1.48\pm0.07 \times 10^{43}$ergs$^{-1}$ and the $^{56}$Ni mass synthesized in the explosion is $0.73\pm0.03$  $M_{\odot}$, which is consistent with measurements in Table~\ref{tab:ni2} using the reddening-free and distance-free method via the phases of the secondary maximum of its NIR-band light curves \citep{Dhawan16}. 
    
    %$\mu=30.42\pm0.16$~mag, $D=12.1\pm0.9$~Mpc

     \item The optical/NIR-band light curves and NIR spectra of SN~2017cbv are well matched with normal Ia's around $B$-band peak and thereafter. But it is brighter during the initial early phase, compared with the well-sampled SNe~2011fe, 2005cf and 2012fr. Meanwhile, the early optical photometry of SN~2017cbv is also brighter than $M_{ch}$ DDT models 23d5, 23, 25 \citep{Hoeflich17} and sub-$M_{ch}$ double detonation models 3, 4 and a bare core detonation model 3c \citep{Fink10,Kromer10}, although after $B$ maximum, the $M_{ch}$ model 25 is more matched with $BV$-band light curves than the other models used here. This indicates that $M_{ch}$ DDT explosion with higher transition density (model 25) is preferred for SN~2017cbv, but extra energy is needed to make a considerable contribution for the early-phase light curves when compared with other well-observed normal SNe~Ia and models used here (eg., $M_{ch}$ DDT and sub-$M_{ch}$ double detonation models).

\end{itemize}

\acknowledgements
This work has benefited from the suggestions of an anonymous referee. 
This work is sponsored (in part) by the Chinese Academy of Sciences (CAS), through a grant to the CAS South America Center for Astronomy (CASSACA) in Santiago, Chile.
The CSP-II has been funded by the NSF under grants AST-1613426, AST-1613455, and AST-16133472.
E.Y.H. and C.A. acknowledge the support provided by the National Science Foundation under Grant No. AST-1613472.
 L.G. was funded by the European Union's Horizon 2020 research and innovation programme under the Marie Sk\l{}odowska-Curie grant agreement No. 839090.
 S.G.G acknowledges support by FCT under Project CRISP PTDC/FIS-AST-31546 and UIDB/00099/2020.
 L.W. acknowledge supports from National Science Foundation under grant No.-AST1817099. 
 X.W. acknowledge supports from National Natural Science Foundation of China under grants No. 11633002 and 11761141001. 
M.G. is supported by the Polish NCN MAESTRO grant 2014/14/A/ST9/00121.
Research by D.J.S. is supported by NSF grants AST-1821967, 1821987, 1813708, 1813466, 1908972, and by the Heising-Simons Foundation under grant \#2020-1864.
Support for JLP is provided in part by FONDECYT through the grant 1191038 and by the Ministry of Economy, Development, and Tourism’s Millennium Science Initiative through grant IC120009, awarded to The Millennium Institute of Astrophysics, MAS.
G.P.  acknowledges  support by the Millennium Science Initiative ICN12\_009.
M. S. is a visiting astronomer at the Infrared Telescope Facility, which is operated by the University of Hawaii under contract NNH14CK55B with the National Aeronautics and Space Administration.
Based on observations obtained at the Gemini Observatory under programs GS-2017A-Q-33 (PI: Sand). Gemini is operated by the Association of Universities for Research in Astronomy, Inc., under a cooperative agreement with the NSF on behalf of the Gemini partnership: the NSF (United States), the National Research Council (Canada), CONICYT (Chile), Ministerio de Ciencia, Tecnolog\'{i}a e Innovaci\'{o}n Productiva (Argentina), and Minist\'{e}rio da Ci\^{e}ncia, Tecnologia e Inova\c{c}\~{a}o (Brazil). The data were processed using the Gemini IRAF package. We thank the queue service observers and technical support staff at Gemini Observatory for their assistance.
We have made use of the data with SMARTS/ANDICam and Magellan Baade/FIRE through CNTAC proposal IDs: CN2017A-85, CN2017A-136, CN2017B-61 (PI: Wang). We thank the queue service observers and technical support staff at CTIO and Yale SMARTS team for their assistance. We have also made use of the NASA/IPAC Extragalactic Database (NED) which is operated by the Jet Propulsion
Laboratory, California Institute of Technology, under contract with
the National Aeronautics and Space Administration. This work made use of the Heidelberg Supernova Model Archive (HESMA), https://hesma.h-its.org. 
\  \par
$Facilities$: SMARTS (ANDICam), Magellan Baade (FIRE), Gemini South (FLAMINGOS-2), IRTF (SpeX), NTT (SOFI), DLT40, 2MASS, {\it Swift}

$Software$:
\texttt{firehose} \citep{Simcoe13},
\texttt{IRAF},
\texttt{PSFEx} \citep{Bertin96,Bertin11},
\texttt{SYNAPPS} \citep{Thomas11},
\texttt{SEP} \citep{Barbary18},
\texttt{skycat},
\texttt{xtellcor} \citep{Vacca03}

\clearpage

%\bibliography{/home/lingzhi/Documents/enotes/sne}{}
\bibliography{ms}{}
\bibliographystyle{apj}

\appendix

The $BVRI$-band photometry on ANDICam natural system were listed in long Table~\ref{tab:bvri}. The $YJHK_s$-band photometry on 2MASS system were listed in Table~\ref{tab:yjhk}.

%BVRI table
\LongTables
\begin{deluxetable*}{lrccc}[t]
%\begin{longtable*}{lrccc}[t]
%\rotate
%\tablewidth{0pc}
%\tablewidth{0pt}
%\tablenum{6}
\tablecaption{$BVRI$-band PSF photometry of SN2017cbv. There is a potential 0.02 magnitude systematic error measured from 
aperture photometry. \label{tab:bvri}}
\tablehead{\colhead{MJD }& \colhead{Phase$^a$} & \colhead{filter} &  \colhead{Natural Magnitude$^{b}$}   \\}
\startdata
 57825.17  &   $-$15.70  & $B$ &     14.081 $\pm$     0.019 \\
 57825.33  &   $-$15.54  & $B$ &     14.036 $\pm$     0.019 \\
 57826.21  &   $-$14.66  & $B$ &     13.742 $\pm$     0.019 \\
 57826.37  &   $-$14.50  & $B$ &     13.692 $\pm$     0.019 \\
 57826.37  &   $-$14.50  & $B$ &     13.683 $\pm$     0.019 \\
 57826.37  &   $-$14.50  & $B$ &     13.687 $\pm$     0.019 \\
 57827.16  &   $-$13.71  & $B$ &     13.435 $\pm$     0.019 \\
 57827.16  &   $-$13.71  & $B$ &     13.438 $\pm$     0.019 \\
 57827.16  &   $-$13.71  & $B$ &     13.436 $\pm$     0.019 \\
 57827.31  &   $-$13.56  & $B$ &     13.378 $\pm$     0.019 \\
 57827.32  &   $-$13.55  & $B$ &     13.382 $\pm$     0.019 \\
 57827.32  &   $-$13.55  & $B$ &     13.371 $\pm$     0.019 \\
 57828.26  &   $-$12.61  & $B$ &     13.103 $\pm$     0.019 \\
 57828.26  &   $-$12.61  & $B$ &     13.097 $\pm$     0.019 \\
 57828.27  &   $-$12.60  & $B$ &     13.111 $\pm$     0.019 \\
 57829.26  &   $-$11.61  & $B$ &     12.826 $\pm$     0.019 \\
 57829.26  &   $-$11.61  & $B$ &     12.827 $\pm$     0.019 \\
 57829.27  &   $-$11.60  & $B$ &     12.826 $\pm$     0.019 \\
 57829.27  &   $-$11.60  & $B$ &     12.823 $\pm$     0.019 \\
 57829.27  &   $-$11.60  & $B$ &     12.827 $\pm$     0.019 \\
 57829.27  &   $-$11.60  & $B$ &     12.823 $\pm$     0.019 \\
 57829.27  &   $-$11.60  & $B$ &     12.829 $\pm$     0.019 \\
 57830.27  &   $-$10.60  & $B$ &     12.608 $\pm$     0.019 \\
 57830.27  &   $-$10.60  & $B$ &     12.610 $\pm$     0.019 \\
 57830.27  &   $-$10.60  & $B$ &     12.614 $\pm$     0.019 \\
 57830.27  &   $-$10.60  & $B$ &     12.611 $\pm$     0.019 \\
 57830.27  &   $-$10.60  & $B$ &     12.618 $\pm$     0.019 \\
 57830.27  &   $-$10.60  & $B$ &     12.605 $\pm$     0.019 \\
 57830.27  &   $-$10.60  & $B$ &     12.606 $\pm$     0.019 \\
 57831.24  &    $-$9.63  & $B$ &     12.410 $\pm$     0.019 \\
 57831.24  &    $-$9.63  & $B$ &     12.409 $\pm$     0.019 \\
 57831.24  &    $-$9.63  & $B$ &     12.407 $\pm$     0.019 \\
 57831.24  &    $-$9.63  & $B$ &     12.400 $\pm$     0.019 \\
 57831.24  &    $-$9.63  & $B$ &     12.402 $\pm$     0.019 \\
 57831.24  &    $-$9.63  & $B$ &     12.407 $\pm$     0.019 \\
 57831.25  &    $-$9.62  & $B$ &     12.406 $\pm$     0.019 \\
 57832.26  &    $-$8.61  & $B$ &     12.244 $\pm$     0.019 \\
 57832.26  &    $-$8.61  & $B$ &     12.250 $\pm$     0.019 \\
 57832.26  &    $-$8.61  & $B$ &     12.253 $\pm$     0.019 \\
 57832.26  &    $-$8.61  & $B$ &     12.251 $\pm$     0.019 \\
 57832.26  &    $-$8.61  & $B$ &     12.238 $\pm$     0.019 \\
 57832.26  &    $-$8.61  & $B$ &     12.244 $\pm$     0.019 \\
 57832.26  &    $-$8.61  & $B$ &     12.236 $\pm$     0.019 \\
 57833.23  &    $-$7.64  & $B$ &     12.124 $\pm$     0.019 \\
 57833.23  &    $-$7.64  & $B$ &     12.120 $\pm$     0.019 \\
 57833.23  &    $-$7.64  & $B$ &     12.121 $\pm$     0.019 \\
 57833.23  &    $-$7.64  & $B$ &     12.127 $\pm$     0.019 \\
 57833.23  &    $-$7.64  & $B$ &     12.121 $\pm$     0.019 \\
 57833.24  &    $-$7.63  & $B$ &     12.128 $\pm$     0.019 \\
 57833.24  &    $-$7.63  & $B$ &     12.112 $\pm$     0.019 \\
 57834.22  &    $-$6.65  & $B$ &     11.998 $\pm$     0.019 \\
 57834.22  &    $-$6.65  & $B$ &     12.008 $\pm$     0.019 \\
 57834.22  &    $-$6.65  & $B$ &     12.034 $\pm$     0.019 \\
 57834.23  &    $-$6.64  & $B$ &     11.987 $\pm$     0.019 \\
 57834.23  &    $-$6.64  & $B$ &     12.011 $\pm$     0.019 \\
 57834.23  &    $-$6.64  & $B$ &     11.998 $\pm$     0.019 \\
 57834.23  &    $-$6.64  & $B$ &     12.007 $\pm$     0.019 \\
 57835.29  &    $-$5.58  & $B$ &     11.913 $\pm$     0.019 \\
 57835.29  &    $-$5.58  & $B$ &     11.913 $\pm$     0.019 \\
 57835.29  &    $-$5.58  & $B$ &     11.912 $\pm$     0.019 \\
 57835.29  &    $-$5.58  & $B$ &     11.925 $\pm$     0.019 \\
 57835.29  &    $-$5.58  & $B$ &     11.911 $\pm$     0.019 \\
 57835.29  &    $-$5.58  & $B$ &     11.895 $\pm$     0.019 \\
 57835.29  &    $-$5.58  & $B$ &     11.917 $\pm$     0.019 \\
 57838.28  &    $-$2.59  & $B$ &     11.757 $\pm$     0.019 \\
 57838.28  &    $-$2.59  & $B$ &     11.745 $\pm$     0.019 \\
 57838.28  &    $-$2.59  & $B$ &     11.774 $\pm$     0.019 \\
 57838.28  &    $-$2.59  & $B$ &     11.760 $\pm$     0.019 \\
 57838.28  &    $-$2.59  & $B$ &     11.769 $\pm$     0.019 \\
 57838.28  &    $-$2.59  & $B$ &     11.752 $\pm$     0.019 \\
 57838.28  &    $-$2.59  & $B$ &     11.762 $\pm$     0.019 \\
 57839.27  &    $-$1.60  & $B$ &     11.733 $\pm$     0.019 \\
 57839.27  &    $-$1.60  & $B$ &     11.741 $\pm$     0.019 \\
 57839.28  &    $-$1.59  & $B$ &     11.738 $\pm$     0.019 \\
 57839.28  &    $-$1.59  & $B$ &     11.745 $\pm$     0.019 \\
 57839.28  &    $-$1.59  & $B$ &     11.743 $\pm$     0.019 \\
 57839.28  &    $-$1.59  & $B$ &     11.721 $\pm$     0.019 \\
 57839.28  &    $-$1.59  & $B$ &     11.730 $\pm$     0.019 \\
 57841.29  &     0.42  & $B$ &     11.721 $\pm$     0.019 \\
 57841.29  &     0.42  & $B$ &     11.739 $\pm$     0.019 \\
 57841.29  &     0.42  & $B$ &     11.719 $\pm$     0.019 \\
 57843.27  &     2.40  & $B$ &     11.747 $\pm$     0.019 \\
 57843.27  &     2.40  & $B$ &     11.741 $\pm$     0.019 \\
 57843.27  &     2.40  & $B$ &     11.756 $\pm$     0.019 \\
 57847.26  &     6.39  & $B$ &     11.934 $\pm$     0.019 \\
 57847.26  &     6.39  & $B$ &     11.940 $\pm$     0.019 \\
 57849.21  &     8.34  & $B$ &     12.059 $\pm$     0.019 \\
 57849.22  &     8.35  & $B$ &     12.051 $\pm$     0.019 \\
 57851.24  &    10.37  & $B$ &     12.237 $\pm$     0.019 \\
 57851.24  &    10.37  & $B$ &     12.234 $\pm$     0.019 \\
 57854.25  &    13.38  & $B$ &     12.527 $\pm$     0.019 \\
 57854.25  &    13.38  & $B$ &     12.532 $\pm$     0.019 \\
 57858.28  &    17.41  & $B$ &     12.963 $\pm$     0.019 \\
 57858.28  &    17.41  & $B$ &     12.966 $\pm$     0.019 \\
 57862.21  &    21.34  & $B$ &     13.386 $\pm$     0.019 \\
 57862.21  &    21.34  & $B$ &     13.397 $\pm$     0.019 \\
 57864.26  &    23.39  & $B$ &     13.590 $\pm$     0.019 \\
 57864.26  &    23.39  & $B$ &     13.591 $\pm$     0.019 \\
 57869.22  &    28.35  & $B$ &     14.008 $\pm$     0.019 \\
 57869.22  &    28.35  & $B$ &     13.998 $\pm$     0.019 \\
 57871.21  &    30.34  & $B$ &     14.153 $\pm$     0.019 \\
 57871.21  &    30.34  & $B$ &     14.151 $\pm$     0.019 \\
 57873.22  &    32.35  & $B$ &     14.276 $\pm$     0.019 \\
 57873.22  &    32.35  & $B$ &     14.276 $\pm$     0.019 \\
 57876.13  &    35.26  & $B$ &     14.425 $\pm$     0.019 \\
 57876.13  &    35.26  & $B$ &     14.432 $\pm$     0.019 \\
 57883.07  &    42.20  & $B$ &     14.702 $\pm$     0.019 \\
 57887.09  &    46.22  & $B$ &     14.797 $\pm$     0.019 \\
 57890.08  &    49.21  & $B$ &     14.855 $\pm$     0.019 \\
 57894.09  &    53.22  & $B$ &     14.925 $\pm$     0.019 \\
 57902.21  &    61.34  & $B$ &     15.037 $\pm$     0.019 \\
 57906.02  &    65.15  & $B$ &     15.100 $\pm$     0.019 \\
 57913.05  &    72.18  & $B$ &     15.186 $\pm$     0.019 \\
 57917.09  &    76.22  & $B$ &     15.215 $\pm$     0.019 \\
 57923.99  &    83.12  & $B$ &     15.314 $\pm$     0.019 \\
 57933.00  &    92.13  & $B$ &     15.409 $\pm$     0.019 \\
 57935.99  &    95.12  & $B$ &     15.445 $\pm$     0.019 \\
 57941.97  &   101.10  & $B$ &     15.518 $\pm$     0.019 \\
 57944.04  &   103.17  & $B$ &     15.558 $\pm$     0.019 \\
 57949.04  &   108.17  & $B$ &     15.613 $\pm$     0.019 \\
 57956.00  &   115.13  & $B$ &     15.715 $\pm$     0.019 \\
 57960.98  &   120.11  & $B$ &     15.786 $\pm$     0.019 \\
 57965.99  &   125.12  & $B$ &     15.861 $\pm$     0.019 \\
 57825.17  &   $-$15.70  & $V$ &     13.922 $\pm$     0.018 \\
 57825.33  &   $-$15.54  & $V$ &     13.885 $\pm$     0.018 \\
 57826.21  &   $-$14.66  & $V$ &     13.574 $\pm$     0.018 \\
 57827.16  &   $-$13.71  & $V$ &     13.289 $\pm$     0.018 \\
 57827.16  &   $-$13.71  & $V$ &     13.271 $\pm$     0.018 \\
 57827.16  &   $-$13.71  & $V$ &     13.295 $\pm$     0.018 \\
 57827.32  &   $-$13.55  & $V$ &     13.245 $\pm$     0.018 \\
 57827.32  &   $-$13.55  & $V$ &     13.206 $\pm$     0.018 \\
 57827.32  &   $-$13.55  & $V$ &     13.242 $\pm$     0.018 \\
 57828.27  &   $-$12.60  & $V$ &     12.978 $\pm$     0.018 \\
 57828.27  &   $-$12.60  & $V$ &     12.983 $\pm$     0.018 \\
 57828.27  &   $-$12.60  & $V$ &     12.980 $\pm$     0.018 \\
 57829.27  &   $-$11.60  & $V$ &     12.725 $\pm$     0.018 \\
 57829.27  &   $-$11.60  & $V$ &     12.715 $\pm$     0.018 \\
 57829.27  &   $-$11.60  & $V$ &     12.703 $\pm$     0.018 \\
 57830.27  &   $-$10.60  & $V$ &     12.537 $\pm$     0.018 \\
 57830.28  &   $-$10.59  & $V$ &     12.518 $\pm$     0.018 \\
 57830.28  &   $-$10.59  & $V$ &     12.533 $\pm$     0.018 \\
 57831.25  &    $-$9.62  & $V$ &     12.344 $\pm$     0.018 \\
 57831.25  &    $-$9.62  & $V$ &     12.338 $\pm$     0.018 \\
 57831.25  &    $-$9.62  & $V$ &     12.334 $\pm$     0.018 \\
 57832.26  &    $-$8.61  & $V$ &     12.175 $\pm$     0.018 \\
 57832.26  &    $-$8.61  & $V$ &     12.172 $\pm$     0.018 \\
 57832.26  &    $-$8.61  & $V$ &     12.180 $\pm$     0.018 \\
 57833.24  &    $-$7.63  & $V$ &     12.074 $\pm$     0.018 \\
 57833.24  &    $-$7.63  & $V$ &     12.058 $\pm$     0.018 \\
 57833.24  &    $-$7.63  & $V$ &     12.068 $\pm$     0.018 \\
 57834.23  &    $-$6.64  & $V$ &     11.950 $\pm$     0.018 \\
 57834.23  &    $-$6.64  & $V$ &     11.976 $\pm$     0.018 \\
 57834.23  &    $-$6.64  & $V$ &     11.957 $\pm$     0.018 \\
 57834.23  &    $-$6.64  & $V$ &     11.950 $\pm$     0.018 \\
 57834.23  &    $-$6.64  & $V$ &     11.955 $\pm$     0.018 \\
 57835.29  &    $-$5.58  & $V$ &     11.870 $\pm$     0.018 \\
 57835.29  &    $-$5.58  & $V$ &     11.872 $\pm$     0.018 \\
 57835.29  &    $-$5.58  & $V$ &     11.887 $\pm$     0.018 \\
 57835.29  &    $-$5.58  & $V$ &     11.861 $\pm$     0.018 \\
 57835.29  &    $-$5.58  & $V$ &     11.853 $\pm$     0.018 \\
 57838.28  &    $-$2.59  & $V$ &     11.704 $\pm$     0.018 \\
 57838.28  &    $-$2.59  & $V$ &     11.708 $\pm$     0.018 \\
 57838.28  &    $-$2.59  & $V$ &     11.704 $\pm$     0.018 \\
 57838.29  &    $-$2.58  & $V$ &     11.691 $\pm$     0.018 \\
 57838.29  &    $-$2.58  & $V$ &     11.704 $\pm$     0.018 \\
 57839.28  &    $-$1.59  & $V$ &     11.695 $\pm$     0.018 \\
 57839.28  &    $-$1.59  & $V$ &     11.670 $\pm$     0.018 \\
 57839.28  &    $-$1.59  & $V$ &     11.692 $\pm$     0.018 \\
 57839.28  &    $-$1.59  & $V$ &     11.674 $\pm$     0.018 \\
 57839.28  &    $-$1.59  & $V$ &     11.680 $\pm$     0.018 \\
 57841.29  &     0.42  & $V$ &     11.629 $\pm$     0.018 \\
 57841.29  &     0.42  & $V$ &     11.619 $\pm$     0.018 \\
 57841.29  &     0.42  & $V$ &     11.628 $\pm$     0.018 \\
 57843.27  &     2.40  & $V$ &     11.634 $\pm$     0.018 \\
 57843.27  &     2.40  & $V$ &     11.631 $\pm$     0.018 \\
 57843.27  &     2.40  & $V$ &     11.642 $\pm$     0.018 \\
 57847.26  &     6.39  & $V$ &     11.716 $\pm$     0.018 \\
 57847.26  &     6.39  & $V$ &     11.714 $\pm$     0.018 \\
 57849.22  &     8.35  & $V$ &     11.791 $\pm$     0.018 \\
 57849.22  &     8.35  & $V$ &     11.789 $\pm$     0.018 \\
 57851.24  &    10.37  & $V$ &     11.887 $\pm$     0.018 \\
 57851.24  &    10.37  & $V$ &     11.888 $\pm$     0.018 \\
 57854.25  &    13.38  & $V$ &     12.056 $\pm$     0.018 \\
 57854.25  &    13.38  & $V$ &     12.063 $\pm$     0.018 \\
 57858.28  &    17.41  & $V$ &     12.321 $\pm$     0.018 \\
 57858.28  &    17.41  & $V$ &     12.326 $\pm$     0.018 \\
 57862.21  &    21.34  & $V$ &     12.530 $\pm$     0.018 \\
 57862.21  &    21.34  & $V$ &     12.529 $\pm$     0.018 \\
 57864.26  &    23.39  & $V$ &     12.640 $\pm$     0.018 \\
 57864.26  &    23.39  & $V$ &     12.632 $\pm$     0.018 \\
 57869.22  &    28.35  & $V$ &     12.881 $\pm$     0.018 \\
 57869.23  &    28.36  & $V$ &     12.880 $\pm$     0.018 \\
 57871.21  &    30.34  & $V$ &     12.985 $\pm$     0.018 \\
 57871.22  &    30.35  & $V$ &     12.986 $\pm$     0.018 \\
 57873.22  &    32.35  & $V$ &     13.093 $\pm$     0.018 \\
 57873.22  &    32.35  & $V$ &     13.107 $\pm$     0.018 \\
 57876.14  &    35.27  & $V$ &     13.256 $\pm$     0.018 \\
 57876.14  &    35.27  & $V$ &     13.254 $\pm$     0.018 \\
 57883.08  &    42.21  & $V$ &     13.592 $\pm$     0.018 \\
 57883.08  &    42.21  & $V$ &     13.581 $\pm$     0.018 \\
 57887.10  &    46.23  & $V$ &     13.720 $\pm$     0.018 \\
 57887.10  &    46.23  & $V$ &     13.712 $\pm$     0.018 \\
 57890.08  &    49.21  & $V$ &     13.805 $\pm$     0.018 \\
 57890.09  &    49.22  & $V$ &     13.808 $\pm$     0.018 \\
 57894.09  &    53.22  & $V$ &     13.908 $\pm$     0.018 \\
 57894.09  &    53.22  & $V$ &     13.908 $\pm$     0.018 \\
 57902.21  &    61.34  & $V$ &     14.108 $\pm$     0.018 \\
 57902.21  &    61.34  & $V$ &     14.113 $\pm$     0.018 \\
 57906.02  &    65.15  & $V$ &     14.207 $\pm$     0.018 \\
 57913.06  &    72.19  & $V$ &     14.390 $\pm$     0.018 \\
 57917.09  &    76.22  & $V$ &     14.466 $\pm$     0.018 \\
 57924.00  &    83.13  & $V$ &     14.643 $\pm$     0.018 \\
 57933.00  &    92.13  & $V$ &     14.855 $\pm$     0.018 \\
 57935.99  &    95.12  & $V$ &     14.912 $\pm$     0.018 \\
 57941.97  &   101.10  & $V$ &     15.060 $\pm$     0.018 \\
 57944.04  &   103.17  & $V$ &     15.114 $\pm$     0.018 \\
 57949.04  &   108.17  & $V$ &     15.203 $\pm$     0.018 \\
 57956.00  &   115.13  & $V$ &     15.364 $\pm$     0.018 \\
 57960.98  &   120.11  & $V$ &     15.451 $\pm$     0.018 \\
 57965.99  &   125.12  & $V$ &     15.556 $\pm$     0.018 \\
 57825.17  &   $-$15.70  & $R$ &     13.817 $\pm$     0.027 \\
 57825.34  &   $-$15.53  & $R$ &     13.784 $\pm$     0.027 \\
 57826.21  &   $-$14.66  & $R$ &     13.471 $\pm$     0.027 \\
 57827.16  &   $-$13.71  & $R$ &     13.197 $\pm$     0.027 \\
 57827.16  &   $-$13.71  & $R$ &     13.189 $\pm$     0.027 \\
 57827.16  &   $-$13.71  & $R$ &     13.184 $\pm$     0.027 \\
 57827.32  &   $-$13.55  & $R$ &     13.141 $\pm$     0.027 \\
 57827.32  &   $-$13.55  & $R$ &     13.132 $\pm$     0.027 \\
 57827.32  &   $-$13.55  & $R$ &     13.139 $\pm$     0.027 \\
 57828.27  &   $-$12.60  & $R$ &     12.884 $\pm$     0.027 \\
 57828.27  &   $-$12.60  & $R$ &     12.894 $\pm$     0.027 \\
 57828.27  &   $-$12.60  & $R$ &     12.899 $\pm$     0.027 \\
 57829.27  &   $-$11.60  & $R$ &     12.629 $\pm$     0.027 \\
 57829.27  &   $-$11.60  & $R$ &     12.638 $\pm$     0.027 \\
 57829.27  &   $-$11.60  & $R$ &     12.651 $\pm$     0.027 \\
 57830.28  &   $-$10.59  & $R$ &     12.437 $\pm$     0.027 \\
 57830.28  &   $-$10.59  & $R$ &     12.446 $\pm$     0.027 \\
 57830.28  &   $-$10.59  & $R$ &     12.476 $\pm$     0.027 \\
 57831.25  &    $-$9.62  & $R$ &     12.255 $\pm$     0.027 \\
 57831.25  &    $-$9.62  & $R$ &     12.221 $\pm$     0.027 \\
 57831.25  &    $-$9.62  & $R$ &     12.254 $\pm$     0.027 \\
 57832.27  &    $-$8.60  & $R$ &     12.098 $\pm$     0.027 \\
 57832.27  &    $-$8.60  & $R$ &     12.089 $\pm$     0.027 \\
 57832.27  &    $-$8.60  & $R$ &     12.088 $\pm$     0.027 \\
 57833.24  &    $-$7.63  & $R$ &     11.982 $\pm$     0.027 \\
 57833.24  &    $-$7.63  & $R$ &     11.968 $\pm$     0.027 \\
 57833.24  &    $-$7.63  & $R$ &     11.985 $\pm$     0.027 \\
 57834.23  &    $-$6.64  & $R$ &     11.861 $\pm$     0.027 \\
 57834.23  &    $-$6.64  & $R$ &     11.878 $\pm$     0.027 \\
 57834.23  &    $-$6.64  & $R$ &     11.877 $\pm$     0.027 \\
 57834.23  &    $-$6.64  & $R$ &     11.894 $\pm$     0.027 \\
 57834.23  &    $-$6.64  & $R$ &     11.869 $\pm$     0.027 \\
 57835.29  &    $-$5.58  & $R$ &     11.798 $\pm$     0.027 \\
 57835.29  &    $-$5.58  & $R$ &     11.778 $\pm$     0.027 \\
 57835.29  &    $-$5.58  & $R$ &     11.807 $\pm$     0.027 \\
 57835.30  &    $-$5.57  & $R$ &     11.808 $\pm$     0.027 \\
 57835.30  &    $-$5.57  & $R$ &     11.791 $\pm$     0.027 \\
 57838.29  &    $-$2.58  & $R$ &     11.667 $\pm$     0.027 \\
 57838.29  &    $-$2.58  & $R$ &     11.666 $\pm$     0.027 \\
 57838.29  &    $-$2.58  & $R$ &     11.653 $\pm$     0.027 \\
 57838.29  &    $-$2.58  & $R$ &     11.669 $\pm$     0.027 \\
 57838.29  &    $-$2.58  & $R$ &     11.658 $\pm$     0.027 \\
 57839.28  &    $-$1.59  & $R$ &     11.647 $\pm$     0.027 \\
 57839.28  &    $-$1.59  & $R$ &     11.635 $\pm$     0.027 \\
 57839.29  &    $-$1.58  & $R$ &     11.662 $\pm$     0.027 \\
 57839.29  &    $-$1.58  & $R$ &     11.636 $\pm$     0.027 \\
 57839.29  &    $-$1.58  & $R$ &     11.650 $\pm$     0.027 \\
 57841.29  &     0.42  & $R$ &     11.623 $\pm$     0.027 \\
 57841.29  &     0.42  & $R$ &     11.619 $\pm$     0.027 \\
 57841.29  &     0.42  & $R$ &     11.632 $\pm$     0.027 \\
 57843.28  &     2.41  & $R$ &     11.626 $\pm$     0.027 \\
 57843.28  &     2.41  & $R$ &     11.629 $\pm$     0.027 \\
 57843.28  &     2.41  & $R$ &     11.604 $\pm$     0.027 \\
 57847.26  &     6.39  & $R$ &     11.721 $\pm$     0.027 \\
 57847.27  &     6.40  & $R$ &     11.712 $\pm$     0.027 \\
 57849.22  &     8.35  & $R$ &     11.822 $\pm$     0.027 \\
 57849.22  &     8.35  & $R$ &     11.805 $\pm$     0.027 \\
 57851.24  &    10.37  & $R$ &     11.950 $\pm$     0.027 \\
 57851.25  &    10.38  & $R$ &     11.947 $\pm$     0.027 \\
 57854.25  &    13.38  & $R$ &     12.130 $\pm$     0.027 \\
 57854.25  &    13.38  & $R$ &     12.125 $\pm$     0.027 \\
 57858.28  &    17.41  & $R$ &     12.249 $\pm$     0.027 \\
 57858.29  &    17.42  & $R$ &     12.250 $\pm$     0.027 \\
 57862.21  &    21.34  & $R$ &     12.306 $\pm$     0.027 \\
 57864.27  &    23.40  & $R$ &     12.335 $\pm$     0.027 \\
 57869.23  &    28.36  & $R$ &     12.489 $\pm$     0.027 \\
 57871.22  &    30.35  & $R$ &     12.560 $\pm$     0.027 \\
 57873.22  &    32.35  & $R$ &     12.677 $\pm$     0.027 \\
 57876.14  &    35.27  & $R$ &     12.837 $\pm$     0.027 \\
 57883.08  &    42.21  & $R$ &     13.209 $\pm$     0.027 \\
 57883.08  &    42.21  & $R$ &     13.210 $\pm$     0.027 \\
 57887.10  &    46.23  & $R$ &     13.378 $\pm$     0.027 \\
 57887.10  &    46.23  & $R$ &     13.386 $\pm$     0.027 \\
 57890.09  &    49.22  & $R$ &     13.481 $\pm$     0.027 \\
 57890.09  &    49.22  & $R$ &     13.471 $\pm$     0.027 \\
 57894.09  &    53.22  & $R$ &     13.590 $\pm$     0.027 \\
 57894.10  &    53.23  & $R$ &     13.603 $\pm$     0.027 \\
 57902.21  &    61.34  & $R$ &     13.851 $\pm$     0.027 \\
 57902.21  &    61.34  & $R$ &     13.856 $\pm$     0.027 \\
 57906.03  &    65.16  & $R$ &     13.962 $\pm$     0.027 \\
 57906.03  &    65.16  & $R$ &     13.959 $\pm$     0.027 \\
 57913.06  &    72.19  & $R$ &     14.207 $\pm$     0.027 \\
 57913.06  &    72.19  & $R$ &     14.194 $\pm$     0.027 \\
 57917.09  &    76.22  & $R$ &     14.284 $\pm$     0.027 \\
 57924.00  &    83.13  & $R$ &     14.512 $\pm$     0.027 \\
 57825.17  &   $-$15.70  & $I$ &     13.802 $\pm$     0.020 \\
 57825.33  &   $-$15.54  & $I$ &     13.756 $\pm$     0.020 \\
 57826.21  &   $-$14.66  & $I$ &     13.458 $\pm$     0.020 \\
 57827.17  &   $-$13.70  & $I$ &     13.180 $\pm$     0.020 \\
 57827.17  &   $-$13.70  & $I$ &     13.173 $\pm$     0.020 \\
 57827.17  &   $-$13.70  & $I$ &     13.175 $\pm$     0.020 \\
 57827.32  &   $-$13.55  & $I$ &     13.144 $\pm$     0.020 \\
 57827.33  &   $-$13.54  & $I$ &     13.135 $\pm$     0.020 \\
 57827.33  &   $-$13.54  & $I$ &     13.134 $\pm$     0.020 \\
 57828.27  &   $-$12.60  & $I$ &     12.885 $\pm$     0.020 \\
 57828.27  &   $-$12.60  & $I$ &     12.864 $\pm$     0.020 \\
 57828.27  &   $-$12.60  & $I$ &     12.874 $\pm$     0.020 \\
 57829.27  &   $-$11.60  & $I$ &     12.629 $\pm$     0.020 \\
 57829.28  &   $-$11.59  & $I$ &     12.651 $\pm$     0.020 \\
 57829.28  &   $-$11.59  & $I$ &     12.664 $\pm$     0.020 \\
 57830.28  &   $-$10.59  & $I$ &     12.473 $\pm$     0.020 \\
 57830.28  &   $-$10.59  & $I$ &     12.514 $\pm$     0.020 \\
 57830.28  &   $-$10.59  & $I$ &     12.487 $\pm$     0.020 \\
 57831.25  &    $-$9.62  & $I$ &     12.263 $\pm$     0.020 \\
 57831.25  &    $-$9.62  & $I$ &     12.245 $\pm$     0.020 \\
 57831.25  &    $-$9.62  & $I$ &     12.259 $\pm$     0.020 \\
 57832.27  &    $-$8.60  & $I$ &     12.108 $\pm$     0.020 \\
 57832.27  &    $-$8.60  & $I$ &     12.122 $\pm$     0.020 \\
 57832.27  &    $-$8.60  & $I$ &     12.109 $\pm$     0.020 \\
 57833.24  &    $-$7.63  & $I$ &     12.003 $\pm$     0.020 \\
 57833.24  &    $-$7.63  & $I$ &     12.024 $\pm$     0.020 \\
 57833.24  &    $-$7.63  & $I$ &     12.001 $\pm$     0.020 \\
 57834.23  &    $-$6.64  & $I$ &     11.906 $\pm$     0.020 \\
 57834.23  &    $-$6.64  & $I$ &     11.885 $\pm$     0.020 \\
 57834.24  &    $-$6.63  & $I$ &     11.909 $\pm$     0.020 \\
 57834.24  &    $-$6.63  & $I$ &     11.907 $\pm$     0.020 \\
 57834.24  &    $-$6.63  & $I$ &     11.885 $\pm$     0.020 \\
 57835.30  &    $-$5.57  & $I$ &     11.854 $\pm$     0.020 \\
 57835.30  &    $-$5.57  & $I$ &     11.841 $\pm$     0.020 \\
 57835.30  &    $-$5.57  & $I$ &     11.839 $\pm$     0.020 \\
 57835.30  &    $-$5.57  & $I$ &     11.840 $\pm$     0.020 \\
 57835.30  &    $-$5.57  & $I$ &     11.834 $\pm$     0.020 \\
 57838.29  &    $-$2.58  & $I$ &     11.790 $\pm$     0.020 \\
 57838.29  &    $-$2.58  & $I$ &     11.796 $\pm$     0.020 \\
 57838.29  &    $-$2.58  & $I$ &     11.799 $\pm$     0.020 \\
 57838.29  &    $-$2.58  & $I$ &     11.805 $\pm$     0.020 \\
 57838.29  &    $-$2.58  & $I$ &     11.788 $\pm$     0.020 \\
 57839.29  &    $-$1.58  & $I$ &     11.803 $\pm$     0.020 \\
 57839.29  &    $-$1.58  & $I$ &     11.821 $\pm$     0.020 \\
 57839.29  &    $-$1.58  & $I$ &     11.828 $\pm$     0.020 \\
 57839.29  &    $-$1.58  & $I$ &     11.801 $\pm$     0.020 \\
 57839.29  &    $-$1.58  & $I$ &     11.814 $\pm$     0.020 \\
 57841.30  &     0.43  & $I$ &     11.839 $\pm$     0.020 \\
 57841.30  &     0.43  & $I$ &     11.860 $\pm$     0.020 \\
 57841.30  &     0.43  & $I$ &     11.842 $\pm$     0.020 \\
 57843.28  &     2.41  & $I$ &     11.923 $\pm$     0.020 \\
 57843.28  &     2.41  & $I$ &     11.925 $\pm$     0.020 \\
 57843.28  &     2.41  & $I$ &     11.916 $\pm$     0.020 \\
 57847.27  &     6.40  & $I$ &     12.083 $\pm$     0.020 \\
 57847.27  &     6.40  & $I$ &     12.083 $\pm$     0.020 \\
 57849.22  &     8.35  & $I$ &     12.213 $\pm$     0.020 \\
 57849.23  &     8.36  & $I$ &     12.218 $\pm$     0.020 \\
 57851.25  &    10.38  & $I$ &     12.362 $\pm$     0.020 \\
 57851.25  &    10.38  & $I$ &     12.354 $\pm$     0.020 \\
 57854.25  &    13.38  & $I$ &     12.481 $\pm$     0.020 \\
 57854.26  &    13.39  & $I$ &     12.483 $\pm$     0.020 \\
 57858.29  &    17.42  & $I$ &     12.462 $\pm$     0.020 \\
 57858.29  &    17.42  & $I$ &     12.458 $\pm$     0.020 \\
 57862.21  &    21.34  & $I$ &     12.377 $\pm$     0.020 \\
 57862.22  &    21.35  & $I$ &     12.377 $\pm$     0.020 \\
 57864.27  &    23.40  & $I$ &     12.355 $\pm$     0.020 \\
 57864.27  &    23.40  & $I$ &     12.345 $\pm$     0.020 \\
 57869.23  &    28.36  & $I$ &     12.323 $\pm$     0.020 \\
 57869.23  &    28.36  & $I$ &     12.323 $\pm$     0.020 \\
 57871.22  &    30.35  & $I$ &     12.341 $\pm$     0.020 \\
 57871.22  &    30.35  & $I$ &     12.341 $\pm$     0.020 \\
 57873.23  &    32.36  & $I$ &     12.400 $\pm$     0.020 \\
 57873.23  &    32.36  & $I$ &     12.412 $\pm$     0.020 \\
 57876.14  &    35.27  & $I$ &     12.537 $\pm$     0.020 \\
 57876.14  &    35.27  & $I$ &     12.542 $\pm$     0.020 \\
 57883.09  &    42.22  & $I$ &     12.962 $\pm$     0.020 \\
 57883.09  &    42.22  & $I$ &     12.958 $\pm$     0.020 \\
 57887.10  &    46.23  & $I$ &     13.150 $\pm$     0.020 \\
 57887.10  &    46.23  & $I$ &     13.161 $\pm$     0.020 \\
 57890.09  &    49.22  & $I$ &     13.293 $\pm$     0.020 \\
 57890.09  &    49.22  & $I$ &     13.299 $\pm$     0.020 \\
 57894.10  &    53.23  & $I$ &     13.465 $\pm$     0.020 \\
 57894.10  &    53.23  & $I$ &     13.451 $\pm$     0.020 \\
 57902.21  &    61.34  & $I$ &     13.784 $\pm$     0.020 \\
 57902.22  &    61.35  & $I$ &     13.791 $\pm$     0.020 \\
 57906.03  &    65.16  & $I$ &     13.926 $\pm$     0.020 \\
 57913.06  &    72.19  & $I$ &     14.221 $\pm$     0.020 \\
 57917.09  &    76.22  & $I$ &     14.345 $\pm$     0.020 \\
 57924.00  &    83.13  & $I$ &     14.616 $\pm$     0.020 \\
 57933.00  &    92.13  & $I$ &     14.899 $\pm$     0.020 \\
 57936.00  &    95.13  & $I$ &     14.997 $\pm$     0.020 \\
 57941.98  &   101.11  & $I$ &     15.180 $\pm$     0.020 \\
 57944.04  &   103.17  & $I$ &     15.272 $\pm$     0.020 

\enddata
\tablecomments{$^a$ Relative to the epoch of $B$-band maximum, MJD = 57840.87 from Gauss process regression \citep{Rasmussen06,scikit11}.\\
$^b$ Uncertainty coming from PSFEx photometry was set to 0.01~mag if it is smaller than 0.01~mag. }
% , in units of 0.001 mag, are 1$\sigma$. }
\end{deluxetable*}
%\end{longtable}

\clearpage

%yjhk table
\begin{deluxetable}{lrcccc}[ht]
%\rotate
%\tablewidth{0pc}
\tablewidth{0pt}
%\tablenum{6}
\tablecaption{$YJHK$-band PSF photometry of SN2017cbv. \label{tab:yjhk}}
\tablehead{\colhead{MJD }& \colhead{Phase$^a$}  & \colhead{$Y$(mag)}  & \colhead{$J$(mag)} & \colhead{$H$(mag)} & \colhead{$K_s$(mag)}  \\}
%MJD & Phase & $B$ & $V$ & $R$ & $I$ & $Y$ & $J$ & $H$ & $K$ \\
%update on 2019-7-22
\startdata
%57825.17 & $-$15.70 & 13.937$\pm$0.042 & 13.822$\pm$0.038 & 13.822$\pm$0.044 & 13.380$\pm$0.076 \\
57825.17 & $-$15.70 & 13.937$\pm$0.042 & 13.822$\pm$0.038 & 13.822$\pm$0.044 &  \nd \\ 
57825.33 & $-$15.54 & 13.887$\pm$0.041 & 13.785$\pm$0.037 & 13.745$\pm$0.041 &  \nd  \\ 
57826.21 & $-$14.67 & 13.552$\pm$0.040 & 13.418$\pm$0.036 & 13.351$\pm$0.036 & 13.294$\pm$0.068 \\ 
57827.16 & $-$13.71 & 13.305$\pm$0.039 & 13.205$\pm$0.035 & 13.173$\pm$0.034 & 13.180$\pm$0.037 \\ 
57827.32 & $-$13.55 & 13.254$\pm$0.039 & 13.165$\pm$0.035 & 13.121$\pm$0.034 & 12.978$\pm$0.035 \\ 
57828.27 & $-$12.60 & 13.011$\pm$0.038 & 12.916$\pm$0.035 & 12.882$\pm$0.034 & 12.790$\pm$0.034 \\ 
57829.27 & $-$11.60 & 12.779$\pm$0.038 & 12.683$\pm$0.035 & 12.660$\pm$0.033 & 12.700$\pm$0.034 \\ 
57830.28 & $-$10.60 & 12.571$\pm$0.038 & 12.466$\pm$0.035 & 12.477$\pm$0.033 & 12.426$\pm$0.033 \\ 
57831.25 & $-$9.62 & 12.422$\pm$0.038 & 12.313$\pm$0.035 & 12.322$\pm$0.033 & 12.282$\pm$0.033 \\ 
57832.26 & $-$8.61 & 12.276$\pm$0.038 & 12.169$\pm$0.035 & 12.211$\pm$0.033 & 12.127$\pm$0.033 \\ 
57833.24 & $-$7.63 & 12.175$\pm$0.038 & 12.059$\pm$0.035 & 12.119$\pm$0.033 & 12.090$\pm$0.033 \\ 
57834.23 & $-$6.64 & 12.103$\pm$0.038 & 11.981$\pm$0.035 & 12.062$\pm$0.033 & 12.047$\pm$0.034 \\ 
57835.29 & $-$5.58 & 12.068$\pm$0.038 & 11.932$\pm$0.035 & 12.038$\pm$0.033 & 11.912$\pm$0.033 \\ 
57838.28 & $-$2.59 & 12.119$\pm$0.038 & 11.899$\pm$0.035 & 12.061$\pm$0.033 & 11.886$\pm$0.033 \\ 
57839.28 & $-$1.59 & 12.185$\pm$0.038 & 11.933$\pm$0.035 & 12.111$\pm$0.033 & 11.873$\pm$0.033 \\ 
57841.29 & 0.42 &  \nd  & 12.048$\pm$0.035 & 12.218$\pm$0.033 & 11.942$\pm$0.033 \\ 
57843.27 & 2.40 &  \nd  & 12.197$\pm$0.035 & 12.273$\pm$0.033 & 12.065$\pm$0.033 \\ 
57847.26 & 6.39 &  \nd  & 12.579$\pm$0.035 & 12.316$\pm$0.033 & 12.194$\pm$0.033 \\ 
57849.22 & 8.35 &  \nd  & 12.874$\pm$0.035 & 12.327$\pm$0.033 & 12.225$\pm$0.033 \\ 
57851.24 & 10.37 &  \nd  & 13.233$\pm$0.035 & 12.365$\pm$0.033 & 12.261$\pm$0.033 \\ 
57854.25 & 13.38 &  \nd  & 13.550$\pm$0.035 & 12.310$\pm$0.033 & 12.215$\pm$0.033 \\ 
57858.28 & 17.41 &  \nd  & 13.561$\pm$0.035 & 12.218$\pm$0.033 & 12.206$\pm$0.034 \\ 
57862.21 & 21.34 &  \nd  & 13.482$\pm$0.035 & 12.142$\pm$0.033 & 12.120$\pm$0.033 \\ 
57864.26 & 23.39 &  \nd  & 13.430$\pm$0.035 & 12.109$\pm$0.033 & 12.048$\pm$0.033 \\ 
57869.22 & 28.35 &  \nd  & 13.297$\pm$0.035 & 12.106$\pm$0.033 & 12.089$\pm$0.033 \\ 
57871.21 & 30.34 &  \nd  & 13.224$\pm$0.035 & 12.130$\pm$0.033 & 12.123$\pm$0.033 \\ 
57873.22 & 32.35 &  \nd  & 13.175$\pm$0.035 & 12.174$\pm$0.033 & 12.176$\pm$0.033 \\ 
57876.14 & 35.27 &  \nd  & 13.219$\pm$0.035 & 12.282$\pm$0.033 & 12.341$\pm$0.033 \\ 
57883.08 & 42.21 & 12.369$\pm$0.038 & 13.778$\pm$0.035 & 12.723$\pm$0.034 & 12.798$\pm$0.035 \\ 
57887.10 & 46.23 & 12.588$\pm$0.038 & 14.091$\pm$0.035 & 12.897$\pm$0.034 & 13.034$\pm$0.038 \\ 
57890.08 & 49.21 & 12.759$\pm$0.038 & 14.305$\pm$0.036 & 13.019$\pm$0.034 & 13.179$\pm$0.037 \\ 
57894.09 & 53.22 & 12.971$\pm$0.039 & 14.574$\pm$0.036 & 13.210$\pm$0.034 & 13.297$\pm$0.040 \\ 
57902.21 & 61.34 & 13.410$\pm$0.039 & 15.078$\pm$0.039 & 13.530$\pm$0.034 & 13.711$\pm$0.045 \\ 
57906.02 & 65.15 & 13.603$\pm$0.039 & 15.296$\pm$0.039 & 13.650$\pm$0.034 & 13.852$\pm$0.045 \\ 
57913.06 & 72.18 & 14.015$\pm$0.039 & 15.669$\pm$0.040 & 13.987$\pm$0.035 & 14.077$\pm$0.039 \\ 
57917.09 & 76.22 & 14.182$\pm$0.039 & 15.866$\pm$0.041 & 14.136$\pm$0.035 & 14.154$\pm$0.047 \\ 
57924.00 & 83.13 & 14.508$\pm$0.042 & 16.307$\pm$0.066 & 14.389$\pm$0.036 &  \nd  \\ 
57933.00 & 92.13 & 14.923$\pm$0.039 & 16.685$\pm$0.051 & 14.653$\pm$0.050 &  \nd  \\ 
57936.00 & 95.13 & 15.021$\pm$0.039 & 16.666$\pm$0.046 & \nd &  \nd  \\ 
57941.97 & 101.10 & 15.285$\pm$0.041 & 17.042$\pm$0.090 & 14.966$\pm$0.086 &  \nd  \\ 
57944.04 & 103.17 & 15.335$\pm$0.040 & 16.964$\pm$0.056 & 15.064$\pm$0.062 &  \nd  \\ 
57949.04 & 108.17 & 15.483$\pm$0.046 &  \nd  & 15.280$\pm$0.056 &  \nd  \\ 
57956.00 & 115.13 & 15.680$\pm$0.042 &  \nd  & 15.568$\pm$0.046 &  \nd  \\ 
57960.98 & 120.11 & 15.917$\pm$0.045 &  \nd  & 15.716$\pm$0.048 &  \nd  \\ 
57965.99 & 125.12 & 16.010$\pm$0.046 &  \nd  & 15.943$\pm$0.069 &  \nd  
\enddata
\tablecomments{$^a$ Relative to the epoch of $B$-band maximum, $t_{B}^{max}$ = 57840.87 MJD from Gauss process regression \citep{Rasmussen06,scikit11}.}
%$^b$ Uncertainties, in units of 0.001 mag, are 1$\sigma$. }
\end{deluxetable}

\end{document}